# mRNA active transport in oocyte-early embryo: 3D agent-based modeling

## Активный транспорт материнских мРНК в ооците-раннем эмбрионе дрозофилы: подходы агентного 3D моделирования

*Marat A. Sabirov[1,ǂ], Ekaterina M. Myasnikova[1], Alexander V. Spirov[1,†]*

*[1]Sechenov Institute of Evolutionary Physiology and Biochemistry of the Russian Academy of Sciences (IEPhB RAS), St-Petersburg, Russia*

*[ǂ] Current address: … … …*

*[†]Corresponding author*

## Graphical Abstract

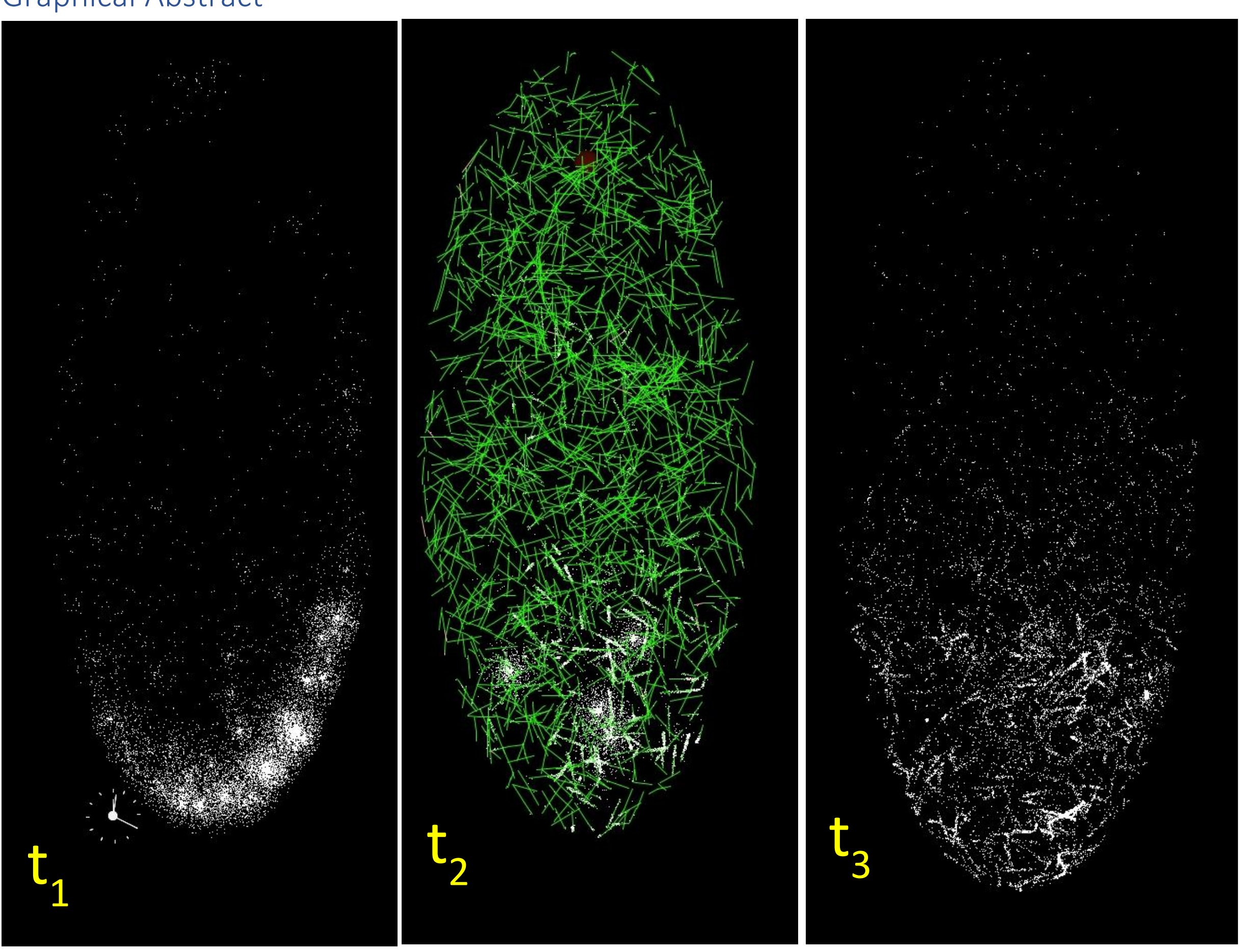

**The 3D agent-based model reproduces active transport by molecular motors (+adaptors) over an undirected network of short microtubules (MT) in syncytial Drosophila embryo.**

## Annotation
Axes of polarity (and primary morphogenetic gradients) are established in the oocyte - early embryo through active transport and localization of maternal factors. It is the oocyte - syncytial embryo of Drosophila (*D. melanogaster*) that is a model object for studying the molecular machinery of such transport systems. The attention of researchers is focused on the processes of formation, maintenance, and functioning of active transport systems of maternal mRNAs and proteins that are key for early Drosophila embryogenesis. Here we develop an approach for agent-based 3D modeling of the key components of transport by molecular motors (by elements of the cytoskeleton) of the Drosophila oocyte-syncytial embryo. The models were developed using Skeledyne software developed by Odell and Foe [Odell and Foe, 2008]. We start with the results of modeling transport along oriented microtubule (MT) bundles in the oocyte. This is a model of transport systems in the Drosophila oocyte, where three maternal mRNAs (bicoid (bcd), oskar, and gurken) that are key to embryonic polarity are transported along their oriented MT bundles. Then we consider models of oriented MT networks in the volume of a cell (oocyte) generated by a single microtubule organization center (or a pair of the centers). This model reproduces the formation of bcd mRNA intrusions deep into the cytoplasm in the head half of the early syncytial embryo. Finally, we consider models for the active transport of bcd mRNA in a syncytial embryo along a randomized network of many short MT strands. In conclusion, we consider the prospects for the implementation of cytoplasmic fountain flows in the active transport model.

## Резюме
Оси полярности и первичные морфогенетические градиенты устанавливаются в ооците – зиготе – раннем эмбрионе посредством механизмов активного транспорта и локализации матернальных факторов. Именно ооцит - синцитиальный эмбрион дрозофилы (*D.melanogaster*) является модельным объектом для исследования молекулярной машинерии таких систем активного транспорта. Внимание исследователей сосредоточено на процессах становления, подержания и функционирования транспортных систем для материнских мРНК в раннем эмбриогенезе дрозофилы. Мы в этой статье развиваем подход по агентному 3D моделированию ключевых компонент систем транспорта молекулярными моторами по элементам цитоскелета развивающегося ооцита – синцитиального эмбриона дрозофилы. Модель разработана на основе программного обеспечения Skeledyne, разработанного Оделом и Фоу [Odell and Foe, 2008]. Мы начинаем с результатов моделирования транспорта по ориентированным пучкам микротрубочек (МТ) в ооците. Это модель транспортных систем в ооците дрозофилы, где три ключевых для полярности эмбриона материнских мРНК (бикоид, bcd, оскар и гуркен) транспортируются по своим ориентированным пучкам МТ. Затем мы рассматриваем модели ориентированных сетей МТ в объеме всей клетки, генерируемых единственным центром организации микротрубочек (или их парой). Такая модель воспроизводит процессы формирования интрузий мРНК бикоида (в комплексе с фактором Staufen) вглубь цитоплазмы в головной половине раннего синцитиального эмбриона. Наконец, мы рассматриваем модели активного транспорта мРНК bcd в синцитиальном эмбрионе по неориентированной сети из множества небольших нитей МТ. В заключение мы рассматриваем перспективы имплементации фонтанных токов цитоплазмы в модели активного транспорта.

# 1. Введение: роль транспорта материнских мРНК в раннем эмбриогенезе

Ряд материнских мРНК и белков посредством специальных молекулярных механизмов транспортируются (из питающих клеток яичника) и затем локализуются в определенных областях ооцита-зиготы (преимущественно в определенных областях кортекса). Эта локализация в итоге определяет положение передне-задней и дорзовентральной осей будущего эмбриона. Важность процессов локализации мРНК для оогенеза-раннего эмбриогенеза продемонстрирована для модельных объектов биологии развития: гидроидных, двукрылых насекомых, асцидий, костных рыб и амфибий [Medioni et al., 2012[1]].

Транспорт таких мРНК (и белка) осуществляется преимущественно молекулярным моторами по элементам цитоскелета (микротрубочкам, МТ, и микрофиламентам, МФ). Активный транспорт макромолекул по элементам реорганизующегося цитоскелета критичен для оогенеза и эмбриогенеза всех модельных объектов биологии развития, заканчивая млекопитающими [Kloc, Etkin, 2005[2]; Vazquez-Pianzola, Suter, 2012[3]]. Наиболее изучены системы активного транспорта на примере ооцитов дрозофилы и ксенопуса.

Именно оогенез-ранний эмбриогенез дрозофилы в силу уникальной изученности и удобности этого экспериментального объекта стал во многом прототипическим для сопоставления с более сложными и менее удобными для исследований объектами биологии развития, вплоть до млекопитающих и человека. В частности, детальные исследования активного транспорта мРНК молекулярными моторами по МТ в оогенезе дрозофилы позволили обобщить их в виде парадигмы оскар (Oskar Paradigm). Здесь мРНК оскар (в сравнении с мРНК бикоид) это типичный и наиболее исследованный представитель мРНК, транспортируемый из питающих клеток и локализуемый в ооците посредством молекулярной системы Bic-D/Egl/Dyn (см обзор [Vazquez-Pianzola, Suter, 2012]). Локализация таких мРНК в ооците определяет в итоге становление осей полярности ооцита – зиготы. Эта молекулярная система используется и для транспорта и локализации мРНК не только в ооците, но и в нейронах эмбриона дрозофилы.

Мы в этой статье развиваем подход по агентному 3D моделированию ключевых компонент систем транспорта молекулярными моторами по элементам цитоскелета развивающегося ооцита–зиготы-синцитиального эмбриона дрозофилы.

## 1.1. Агентное 3D моделирование ооцита-раннего эмбриона

Зрелый ооцит, зигота, как и типичная клетка, – организованы в трех пространственных измерениях и эта 3D организация принципиальна и критична для разворачивающихся процессов [St Johnston, Ahringer, 2010[4]; Medioni et al., 2012[5]; Li, Albertini, 2013[6]]. Соответствующие процессы организации / реорганизации цитоскелета, активного транспорта и направленных токов цитоплазмы разворачиваются в объеме клетки, и они методически чрезвычайно сложны для детального количественного анализа. Современные экспериментальные подходы (прежде всего, биофотоника нормы и мутантов) позволяют наблюдать такие процессы в реальном времени на живом объекте с разрешением на уровне индивидуальных молекул [Lasko, 2012[7]; Little et al., 2015[8]]. Однако нарабатываемые сырые данные еще не дают количественной картины этих ключевых процессов. Для этих целей требуется разрабатывать (исходя из уже наработанных данных) компьютерные 3D модели, численные тесты с которыми позволяют в итоге оценить процессы количественно в 3х пространственных измерениях [Castro-González et al., 2012[9]; Mogilner, Odde, 2011[10]; Mogilner, Manhart, 2016[11]]. Заключения из численных экспериментов с моделями далее верифицируются экспериментально.

Для анализа процессов в объеме клетки все чаще используют подход, известный как агентное 3D моделирование (agent-based 3D modeling) [Odell and Foe, 2008[12]; Wordeman et al., 2016[13]; Mogilner and Manhart, 2016].

Агентные модели — это вычислительные модели, имитирующие действия и взаимодействия автономных агентов (либо отдельных молекул, либо коллективных молекулярных ансамблей). Ключевое понятие, на котором основаны такие модели, заключается в том, что множество агентов взаимодействуют в соответствии с простыми правилами и эти правила легко закодировать в компьютерной программе, а взаимодействия легко смоделировать. Однако, несмотря на свою простоту, эти правила и взаимодействия могут в итоге давать сложное эмерджентное поведение [Mogilner and Manhart, 2016].

Агентное 3D-моделирование клеток позволяет, например, явно анализировать динамику цитоскелета и активного транспорта в цитоплазме. При этом, основные молекулярные компоненты клеточных систем активного транспорта моделируются в явном виде как агенты в объеме клетки, с конкретными координатами, скоростям и другими параметрами, включая параметры химической кинетики. А именно, мы развиваем конкретную упрощенную модель клетки, предназначаемую для анализа сопряжения процессов в сетях МТ и МФ, и мы демонстрируем ее функциональные возможности. Модели развиты нами на основе программного обеспечения Skeledyne, разработанного Гарретом Оделлом и Викторией Фоу [Odell and Foe, 2008]. -

## 2. Активный транспорт мРНК в ооците-раннем эмбрионе дрозофилы

Мы в этой публикации сосредоточимся на моделировании ключевых событий транспорта материнских мРНК в развивающимся ооците (как самом изученном модельном объекте биологии развития) и в раннем синцитиальном эмбрионе.

### 2.1. Активный транспорт материнальных мРНК в развивающимся ооците

Созревающий ооцит (в сравнении с ранним синцитиальным эмбрионом) дрозофилы характеризуются хорошо изученными системами активного транспорта, которые можно детально моделировать как отдельности. Прежде всего, это системы активного транспорта материнальных мРНК (оскар, бикоид, gurken) молекулярными моторами (прежде всего, кинезин и динеин) по ориентированным пучкам МТ в раннем, среднем и позднем оогенезе (и в их взаимодействии с МФ).

Обширные исследования транспорта мРНК в ооцитах привели в итоге к формулировке трех основных конкурирующих моделей организации и функционирования сетей МТ, которые ответственны за транспорт грузов мРНК в ооците [Clark et al., 1994[14]; Clark et al., 1997[15]; Theurkauf and Hazelrigg, 1998[16]; Cha et al., 2001[17]; Cha et al., 2002[18]; Januschke et al., 2006[19]; Zimyanin et al., 2008[20]; Lasko, 2020[21]]. В первой, самой ранней модели полагали, что МТ поляризованы вдоль переднезадней оси, так что минус-концы расположены антериорно, а плюс-концы направлены в постериорном направлении [Clark et al., 1994[22]; Clark et al., 1997[23]; Micklem et al., 1997[24]]. Во второй модели полагали, что МТ нуклеируются кортексом ооцита, что приводит к тому, что плюс-концы МТ направлены к центру, радиально [Cha et al., 2002; Serbus et al., 2005[25]]. Третья модель предполагает, что неполяризованная сеть МТ осуществляет транспорт вдоль специфически ориентированных, биохимически и функционально различных субпопуляций МТ [Parton et al., 2011[26]].

Исходя из того, что ни одна из этих трех моделей не стала общепризнанной, как и ни одна не отвергнута большинством исследований, мы находим вполне резонным исследовать каждую из них средствами агентного 3D моделирования. В перспективе это даст возможность сопоставить поведение этих моделей с накапливающимися количественными данными по транспорту РНК в ооците.

## 2.2. Активный транспорт мРНК бикоида в раннем эмбрионе

Здесь мы кратко охарактеризуем что известно о поведении мРНК бикоида в раннем эмбриогенезе дрозофилы.

В зрелой зиготе мРНК bcd закреплена в переднем кортикальном слое яйца. Вслед за оплодотворением mRNA-содержащие частицы высвобождаются в цитоплазму. Вслед за этим запускаются сложные процессы перераспределения этой мРНК в самом раннем эмбриогенезе [Alexandrov et al., 2018[27]]. Можно выделить минимум три стадии перераспределения мРНК бикоида.

Во-первых, это перераспределение частиц, содержащих mRNA, в течение самых первых циклов деления дробления на стадии пре-бластодермы (до девятого цикла или 3-й стадии, примерно через 70 мин, 25 ° C) [Little et al., 2011[28]]. Детальный анализ данных, предполагает согласованную полномасштабную реорганизацию ранней головной области зародыша [Spirov et al., 2009[29]; Little et al., 2011].

Вторым событием является дальнейшее расширение градиента мРНК в постериорном направлении [Spirov et al., 2009]. Оно длится от стадии позднего дробления и до начала 14 деления дробления.

Третье ключевое событие — это базо-апикальное перераспределение мРНК bcd в начале 14 цикла [Spirov et al., 2009].

Мы можем рассматривать эти наблюдаемые перераспределения и реорганизации мРНК-содержащего материла как идентификацию некоторых ключевых стадий в формировании источника градиента белка Bcd (сайта продукции Bcd). Этими стадиями являются: формирование (1), последовательное укрупнение и усиление (2) и, наконец, разборка (3) пространственно распределенного источника белка бикоид.

В силу важности процессов перераспределений мРНК бикоида в самом раннем эмбриогенезе мы рассмотрим их подробнее ниже.

### 2.2.1. Процессы активного транспорта bicoid

В самом раннем эмбрионе, со стадии 1 деления дробления (и в связи с прохождением волны активации [Weil et al., 2010[30]]) формируются и начинают действовать две взаимосвязанные системы активного транспорта мРНК бикоида и белка бикоид (и возможно других компонент).

**Система сердцевинного транспорта:** это система (активного) транспорта из апикального кортекса внутрь, в сердцевину (core) головной части раннего эмбриона [Little et al., 2011; Fahmy et al., 2014[31]; Ali-Murthy and Kornberg, 2016[32]]. (Мы будем именовать ее «сердцевинной»). При этом, мРНК бикоида прокрашиваются как интрузии, проникающие вглубь цитоплазмы и могут быть двойными. Молекулярные детали этой системы мало охарактеризованы. мРНК бикоида здесь транспортируется в комплексе с фактором Staufen (Stau). Авторы недавних публикаций отмечают явную морфологическую обособленность этой системы [Little et al., 2011; Fahmy et al., 2014; Ali-Murthy and Kornberg, 2016]. При этом, Fahmy с соавторами настаивают, что эта система – основана на сети МТ и транспорт осуществляется молекулярными моторами. Они полагают, что это именно та сеть МТ в сердцевине цитоплазмы, которая ответственна за процессы осевой (аксиальной) экспансии (axial expansion) ядер на стадии 2- 4 деления дробления (эта сеть была описана в [Baker et al., 1993[33]]). Наше понимание процессов становления этой системы приведено на Рис 1.

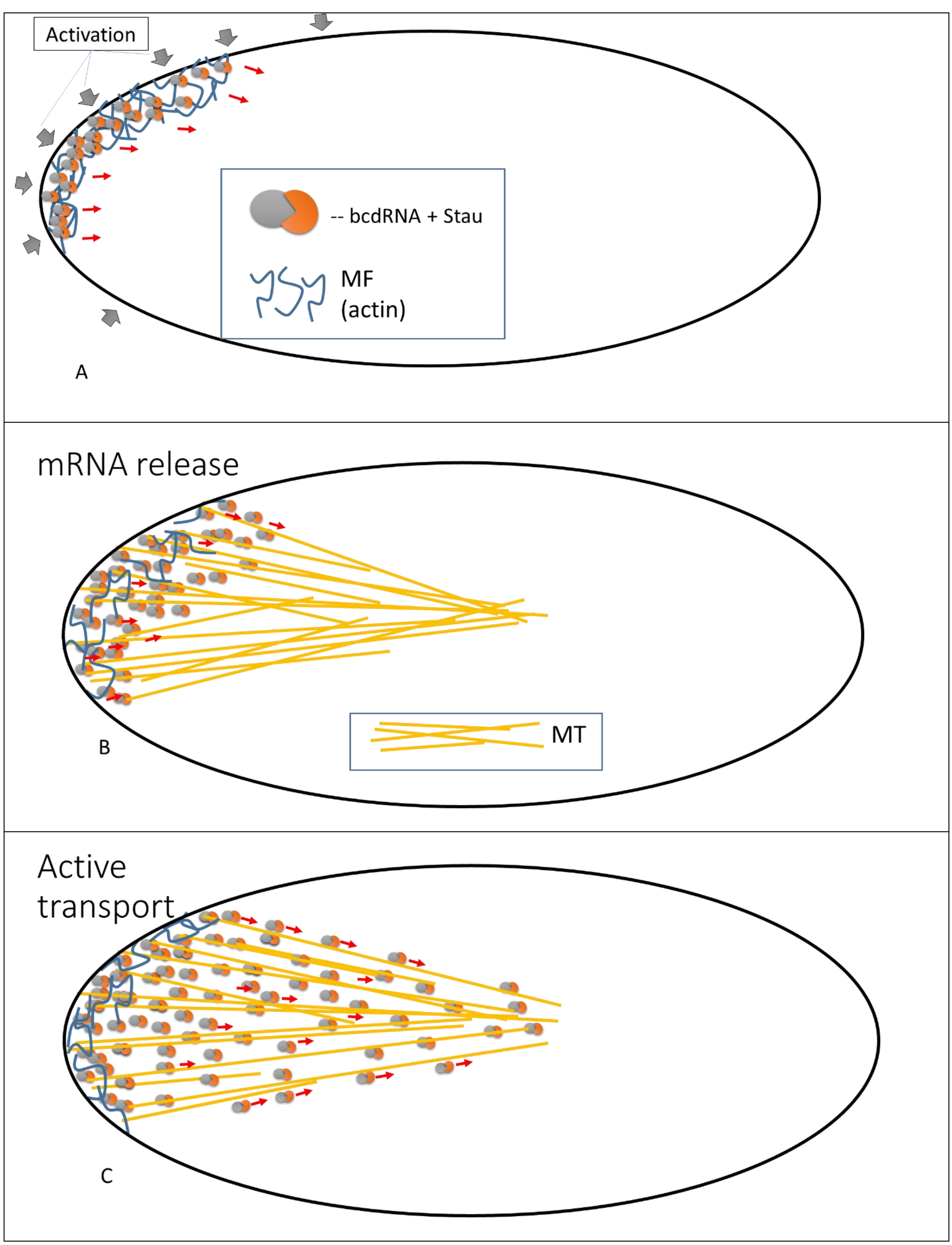

*Рисунок 1. СердцевТранспСист Наша схема становления сердцевинной транспортной системы в самом раннем эмбрионе (и неоплодотворенной яйцеклетке) от события активации (A), за которым следует mRNA release (B) и разворачивания сердцевинного транспорта (C).*

Отметим так же, что хорошо изученная система транспорта и локализации мРНК Vg1 в ооците ксенопуса весьма напоминает организацию и работу сердцевинной системы дрозофилы [Farley, Ryder, 2008[34]]. А именно, пучок МТ связывает центр клетки (где ядро) с вегетативным кортексом. Вначале мРНК Vg1 (в комплексе со Stau) собирается в пределах веретеновидной структуры из МТ и endoplasmic reticulum (EPR), локализованной между ядром и вегетативным кортексом. Вскоре эти комплексы с мРНК активно транспортируются к вегетативному кортексу, где они специфически ассоциируются с МФ кортикального цитоскелета [Yisraeli et al., 1990[35]].

**Система кортикального транспорта:** это система транспорта в кортексе головной части эмбриона. (Мы будем именовать ее «кортикальной»). Кортикальная система активно транспортирует мРНК бикоида из сайта ее компактной локализации (в головном кортексе) в передне-заднем направлении, по периплазме [Spirov et al., 2009; Fahmy et al., 2014; Ali-Murthy and Kornberg, 2016; Cai et al., 2017[36]]. Согласно Fahmy с соавторами [Fahmy et al., 2014] эта система на основе сети МТ и включает тубулин αTub67C, который, известно, поставляется клетками яичника и преимущественно составляет длинные МТ в ооците и раннем эмбрионе. Молекулярным мотором для транспорта служит здесь фактор ncd (из семейства кинезинов), а адаптором для образования специфического комплекса с мРНК бикоида и с мотором - фактор Stau [Fahmy et al., 2014].

Помимо транспорта мРНК бикоида, кортикальная система также осуществляет активный транспорт (молекулярными моторами?) белка бикоид ([Cai et al., 2017]; сравни [Ali-Murthy and Kornberg, 2016]). В настоящее время неясно какие факторы могут быть адапторами для специфического связывания белка бикоид с моторами и это требует дальнейшего изучения. Затем, согласно наблюдениям ([Lucchetta et al., 2008[37]] Figure S7A–C), цитоскелетные элементы, например, актин, могут осуществлять депонирование молекул белка бикоид в окрестности ядра синцитиального эмбриона (тем самым обеспечивая устойчивость морфогенетического градиента бикоид к возвратно-поступательным токам цитоплазмы при каждом митотическом цикле). Все это нуждается в дальнейшем анализе.

Имеются основания полагать, что эти две транспортные системы, сердцевинная и кортикальная, взаимодействуют [Fahmy et al., 2014; Ali-Murthy and Kornberg, 2016]. Для нашего рассмотрения важно также то, что нуль-мутации тубулина αTub67C и сильные мутации ncd приводят к отсутствию кортикального транспорта (по кортикальной системе), но явно усиливают транспорт мРНК сердцевинной системой, так что интрузия этой РНК доходит до середины раннего эмбриона (наш рисунок 5B-C; [Fahmy et al., 2014] Fig.4C,F). То есть, эти две транспортные системы находятся в, своего рода, конкурентных отношениях, так что инактивация первой системы (кортикальной) приводит к существенно более активному транспорту мРНК бикоида второй.

Как продемонстрировали Cai с соавторами и недавно Баумгартнер [Cai et al., 2019[38]; Baumgartner, 2024[39]], механизмы транспорта мРНК бикоида много более комплексны, чем принято думать.

В нашей статье мы развиваем агентные модели для анализа этих сложных процессов в объеме раннего эмбриона дрозофилы. Наша цель здесь – очертить такой набор конкретных моделей, тогда как детали и подробные результаты тестов с ними будут изложены в последующих публикациях.

# 3. Методы и подходы

Полуколичественные и качественные данные мы берем из опубликованных источников. Количественные данные по мРНК bcd и по gfp-bcd были получены как описано ниже.

## 3.1. Количественные данные по мРНК бикоида в зиготе-раннем эмбрионе

Эмбрионы фиксировались теплом и обрабатывались методом FISH на мРНК бикоида, модифицированной так, чтобы достичь высокой чувствительности (как описано в [Spirov et al., 2009]). Далее эмбрионы подвергались послойному конфокальному сканированию. Затем, полученные сагиттальные изображения подвергались количественной обработке так, чтобы получить серии профилей экспрессии.

## 3.2. Данные на живых развивающихся эмбрионах gfp-bcd

We use a transgenic fly line in which GFP (Green Fluorescent Protein) has been fused onto the Bcd protein (gift of Dr. Wieschaus, [20]). With these flies, we can follow developmental dynamics in live embryos. FPs are very sensitive to bleaching and live embryos are very sensitive to photo-damage. To minimize limitations of the approach we use scanning the mid-sagittal plane with a very small pinhole (0.5 airy) and low resolution (512*512 or even 256*256 pixels) minimizes bleaching [Golyandina et al., 2012[40]]. This allows us both to achieve 1 frame/sec. time resolution, allowing us to observe such processes as Bcd redistribution during mitosis.

## 3.3. Агентное 3D моделирование

Мы развиваем конкретные 3D агентные модели средствами пакета *Skeledyne* [[41]], который доступен согласно General Public License. Этот пакет программ был разработан Гарретом Оделлом (Garrett M. Odell) в соавторстве с Викторией Фоу (Victoria E. Foe) (Университет Вашингтона в Сиэтле) для моделирования клетки (включая зиготу) и предназначался изначально для исследований молекулярных деталей и механики клеточных делений [Odell & Foe, 2008; Wordeman et al., 2016; Mogilner, Manhart, 2016]. Имплементированный в *Skeledyne* подход предполагает, что каждая единица макромолекулярных ансамблей, реализуемых в модели, описывается в явном виде как отдельность со своими характеристиками в объеме клетки. Это, например, отдельные микротрубочки, отдельные актиновые филаменты, связанные молекулярные моторы (динеин и кинезин), молекулы грузов в специфических комплексах с адапторами для их активного транспорта, и т.д. При этом, в модели отслеживается динамика и взаимодействия всех этих единиц в объеме клетки. Развиваемые агентные модели позволяют в явном виде воспроизвести 3D геометрию МТ и МФ, их размеры, локализацию, ориентацию, перемещение в цитоплазме, рост и деградацию.

В описании подхода к моделированию авторы особо подчеркивают использование приближения цитодоменов для моделирования процессов в собственно цитоплазме. А именно, цитоплазма (весь материал внутри клетки, помимо других агентов) представлена как плотная упаковка сферических тел заранее заданного радиуса. В зависимости от соотношения размеров, в клетке могут быть заданы от сотен до многих тысяч цитодоменов. По клетке цитодомены перемещаются как целое, и они не деформируемы и не способны проникать друг в друга. При этом, каждый цитодомен несет в себе свой набор концентраций растворимых факторов. В рамках подхода внутри цитодоменов нет пространственной размерности и это значимо ускоряет моделирование.

### 3.3.1. Ооцит и ранний синцитиальный эмбрион дрозофилы

Ооцит в наших агентных моделях представлен как сфероид. Ранний синцитиальный зародыш дрозофилы мы моделируем как вытянутый эллипсоид. Поверхность модели механически жесткая и непроницаемая. Размеры моделей (в мкм) по возможности приближены к реальным, если не исследуются проблемы скейлинга.

В модели эмбриона имплементированы множественные центриоли (предполагается, что они относятся к ядрам дробящегося и синцитиального эмбриона, но сами ядра не имплементированы, поскольку мы не связываем с ними в модели каких-либо функций). Число центриолей в модели определяется моделируемой стадией раннего эмбриогенеза. Исследуется нами активный транспорт молекулярными моторами по пучкам МТ центриольных веретен и по рандомным сетям небольших множественных МТ в цитоплазме. Для моделирования ключевых фаз самого раннего эмбриогенеза (раннее дробление) имплементированы единичные (или парные) центриоли, подразумеваемые как центриоли пронуклеусов. На этих же фазах может быть имплементирована рандомная сеть небольших актиновых филаментов со своими плюс-концевыми моторами.

# 4. Результаты и обсуждение

Ниже мы опишем модели активного транспорта для нескольких сценариев в созревающем ооците и развивающемся синцитиальном эмбрионе.

## 4.1. Активный транспорт в ооците

В этом разделе мы охарактеризуем модели, описывающие основные черты трех гипотез активного транспорта моторами по микротрубочкам в ооците (обсуждались в подразделе «2.1. Активный транспорт матернальных мРНК в развивающимся ооците»).

### 4.1.1. Ориентированная сеть МТ в объеме ооцита

Здесь мы рассмотрим общую ситуацию, когда ориентированная вдоль главной оси клетки и нуклеированная единственной центриолью сеть МТ обеспечивает транспорт изначально равномерно распределенного груза к одному из полюсов клетки (как иллюстрирует Рисунок 2). Такая модель напоминает ранние концепции активного транспорта в ооците дрозофилы, когда предполагалось, что МТ высоко-поляризованы вдоль передне-задней оси (минус-концы преимущественно локализованы антериорно, а плюс-концы – направлены постериорно), тем самым обеспечивая эффективное концентрирование груза на одном полюсе клетки [Clark et al., 1994; Clark et al., 1997]. Это соответствует первой модели из раздела «2.1. Активный транспорт матернальных мРНК в развивающимся ооците».

Мы продемонстрируем как пучки МТ формируют в итоге стабильные паттерны (локальные скопления груза) и отметим от каких параметров этот процесс зависит. Локальные скопления груза стабилизируются тогда, когда все процессы, в которые вовлечен груз (компоненты его активного транспорта по пучку МТ и процессы свободной диффузии груза в цитоплазме) приходят в равновесие.

Во всех этих случаях резонная рабочая гипотеза, что процессы активного транспорта стремятся (или достигают, или, по крайней мере, могут достигнуть) стационарного состояния со стационарными паттернами интересующих нас грузов. В рамках анализируемой модели стационарный паттерн грузов представляет собой его скопление на концах МТ, достигающее стационарного равновесия.

Модель (для большей реалистичности) включает две стадии: стадия быстрого роста центросомных МТ до достижения ими такой длины, что они простираются от одного полюса клетки к другому (1) и стадия активного транспорта, когда МТ больше не растут (2) и осуществляется только активный транспорт до достижения динамического равновесия. Начальная и конечная картины роста МТ приведены на Рис 2B-C, соответственно. Начальное и стационарное распределение груза приведены на Рис 2D-E, соответственно.

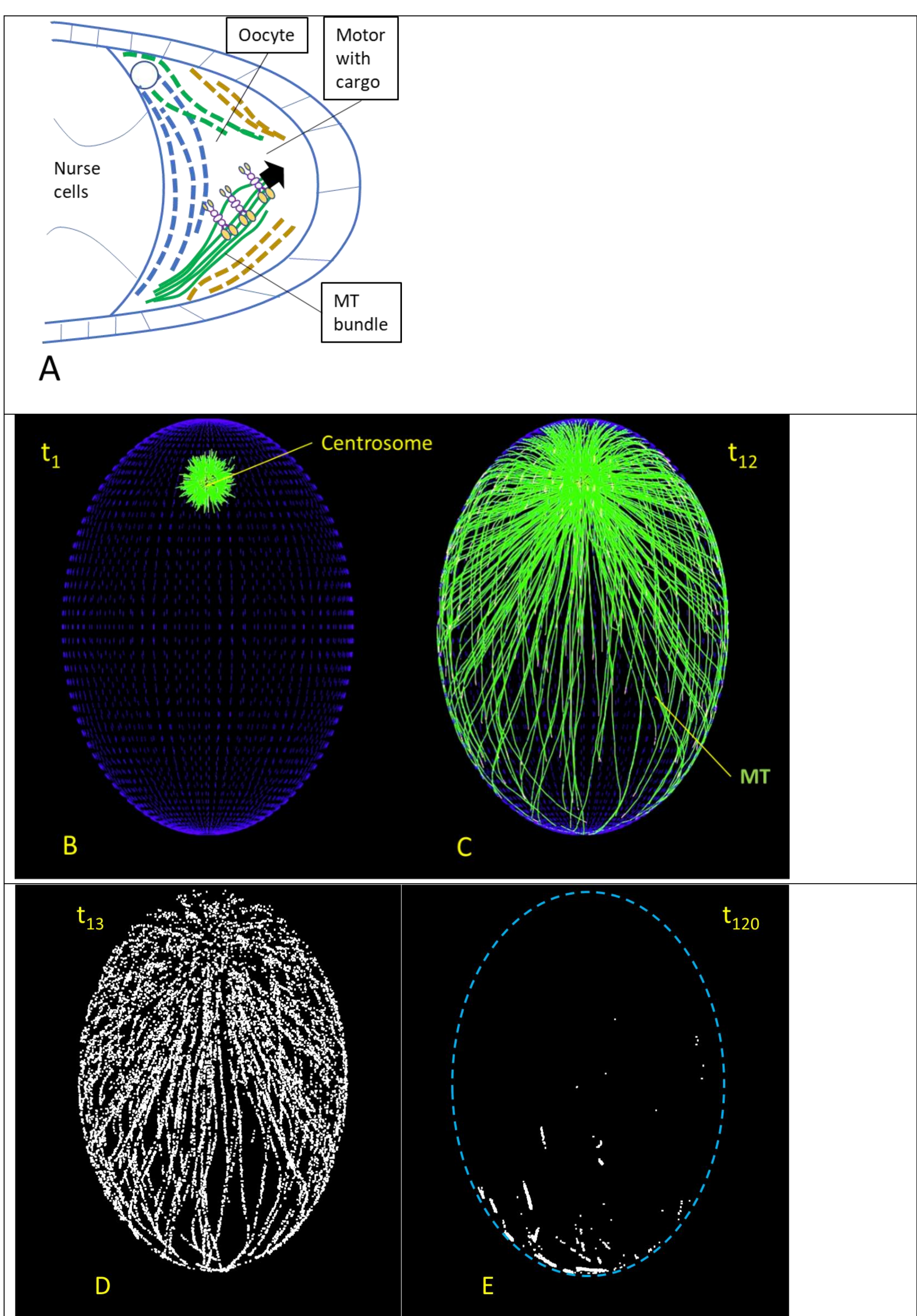

*Рисунок 2. Ориентированная вдоль главной оси клетки и нуклеированная (единственной) центриолью сеть МТ обеспечивает транспорт изначально равномерно распределенного груза к одному из полюсов клетки. (A)*

*Схема высоко-ориентированной сети ооцита дрозофилы (*согласно [Clark, 1994] с изменениями*). (B-E) 3D агентная модель сети МТ, ориентированной вдоль главной оси клетки. Клетка -- элипсоид с осями 266*380 мкм. (B-C) Стадия роста сети МТ, нуклеированной единственной центриолью; $t_2$ – начало роста сети МТ (2я секунда времени модели), $t_{12}$ – остановка роста сети МТ (12я секунда времени модели). (D-E) Стадия активного транспорта груза по ориентированным МТ; $t_{13}$ – начало транспортного процесса; груз распределен по всей клетке (3я секунда времени модели), $t_{120}$ – стационарное скопление груза у полюса клетки (120я секунда времени модели).*

Как можно видеть, рассматриваемая здесь модель системы активного транспорта эффективно концентрирует и удерживает груз на одном из полюсов клетки (Рисунок 2).

Серии экспериментов с моделями таких клеток разных размеров показали, что для поддержания неизменным времени формирования сети (скейлинговость) основной параметр — это скорость полимеризации МТ. Для поддержания времени достижения стационарного распределения груза на 2й стадии моделирования основной параметр — это скорость мотора по МТ. Более подробно эта модель исследуется на предмет масштабируемости в последующей публикации [Sabirov, Spirov, in preparation[42]].

### 4.1.2. Система радиально ориентированных кортикальных МТ ооцита

В этом подразделе мы остановимся на агентном 3D моделировании в рамках второй модели из раздела «2.1. Активный транспорт матернальных мРНК в развивающимся ооците». В этой модели, предложенной в статьях [Cha et al., 2002; Serbus et al., 2005], МТ нуклеируются кортексом ооцита, что приводит к тому, что плюс-концы МТ направлены к центру клетки.

Рисунок 3 иллюстрирует агентную модель активного транспорта мРНК оскар моторами кинезина по ориентированным МТ в ооците. В течение 8й стадии оогенеза МТ нуклеируются по всему кортексу ооцита и кинезин транспортирует мРНК оскара в сердцевину клетки [Cha et al., 2002].

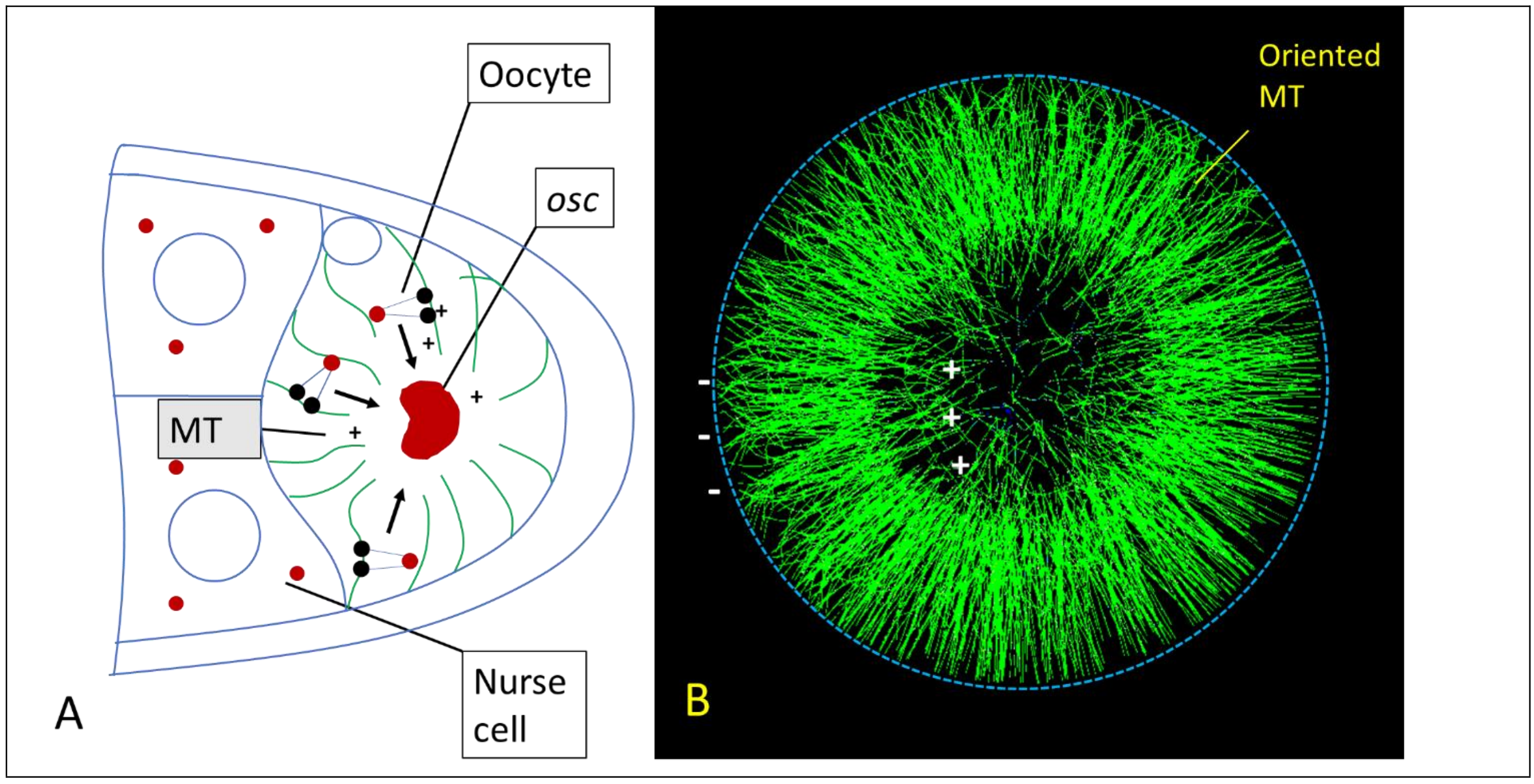

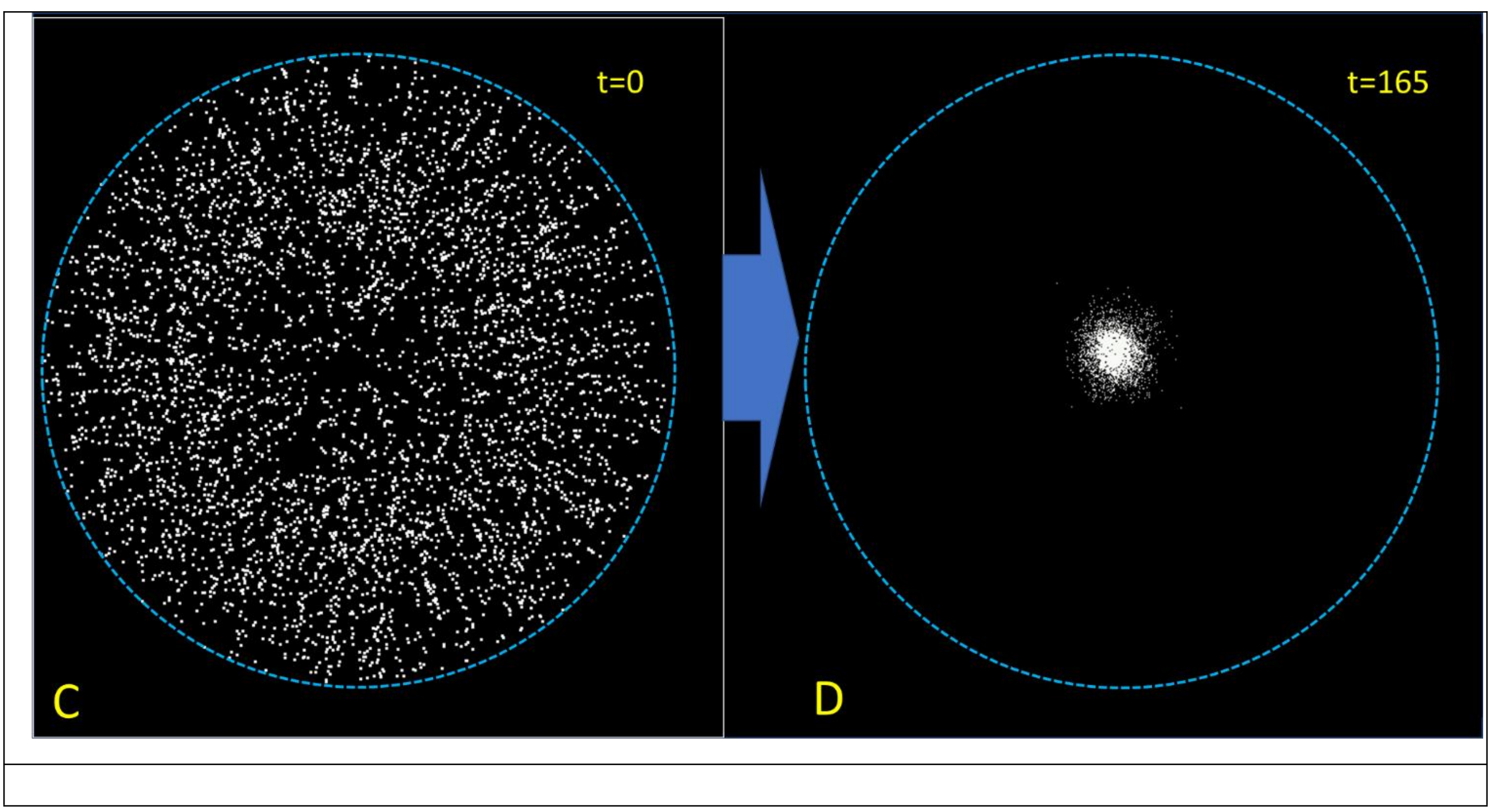

*Рисунок 3. Модель активного транспорта мРНК оскар моторами кинезина по ориентированным МТ в оогенезе (стадия 8) и ее реализация 3D агентным моделированием. (А) В течение 8 стадии МТ нуклеируются по всему кортексу ооцита и кинезин транспортирует мРНК оскара в сердцевину клетки (согласно [Cha et al., 2002] с изменениями). Слева – две питающие клетки, ооцит справа; частицы оскара – красные кружки; кинезин – черные овалы; МТ – зеленые; стрелки указывают направление активного транспорта к середине ооцита; красным в сердцевине ооцита передано скопление оскар; плюсы отмечают плюс-концы МТ. (В) 3D агентая модель клетки, кортекс которой нуклеирует МТ (ориентированные плюс-концом к сердцевине), а кинезин образует комплекс с мРНК оскар и способен транспортировать его по МТ в плюс-направлении. (С) Изначально комплексы кинезин-оскар распределены в цитоплазме клетки равномерно. (D) Со временем комплекс кинезин-оскар перераспределяется в сердцевину клетки и удерживается там активным транспортом.*

В модели Рис 3 задано 5 тыс МТ одинаковой длины (длиной 90% от радиуса клетки) и 6 тыс специфических комплексов молекул кинезина в комплексе с оскар. Клетка задана сферической формы радиусом 200 мкм и с длиной МТ в 192 мкм. Изначально комплексы кинезин-оскар распределены в цитоплазме клетки равномерно (Рисунок 3С). (D) Со временем комплекс кинезин-оскар перераспределяется в сердцевину клетки и удерживается там активным транспортом (результат прогона модели в 100 сек модельного времени приведен на Рис 3D).

Эта модель хорошо описывает предполагаемую динамику перераспределения мРНК оскар. В силу ее простоты она позволяет исследовать от каких параметров молекулярного транспорта и как именно меняется скорость достижения равновесия пространственного распределения груза и как можно достичь скейлинговости.

Мы показали, что даже в случае упрощенных условий наших моделей активного транспорта масштабируемость требует четко синхронизированных изменений сразу нескольких его параметров. При этом, одни параметры должны меняться пропорционально изменению радиуса клетки (например, скорость молекулярного мотора), другие – пропорционально изменению поверхности клетки (например, количество микротрубочек в пучке), третьи – пропорционально изменению объема клетки (например, количество грузов в клетке).

Более подробно эта модель исследуется на предмет масштабируемости в последующей статье [Sabirov, Spirov, in preparation].

Развитие этой модели включением МФ рассматривается ниже в подразделе «4.3. Роль актиновых филаментов в раннем транспорте».

#### *4.1.2.1. К более реалистичной модели транспорта мРНК оскар*

Наконец, приведем наше развитие исходной простой модели с радиально ориентированными нитями МТ. Здесь мы вводим асимметрию в процесс нуклеации МТ кортексом, как иллюстрирует Рисунок 4А.

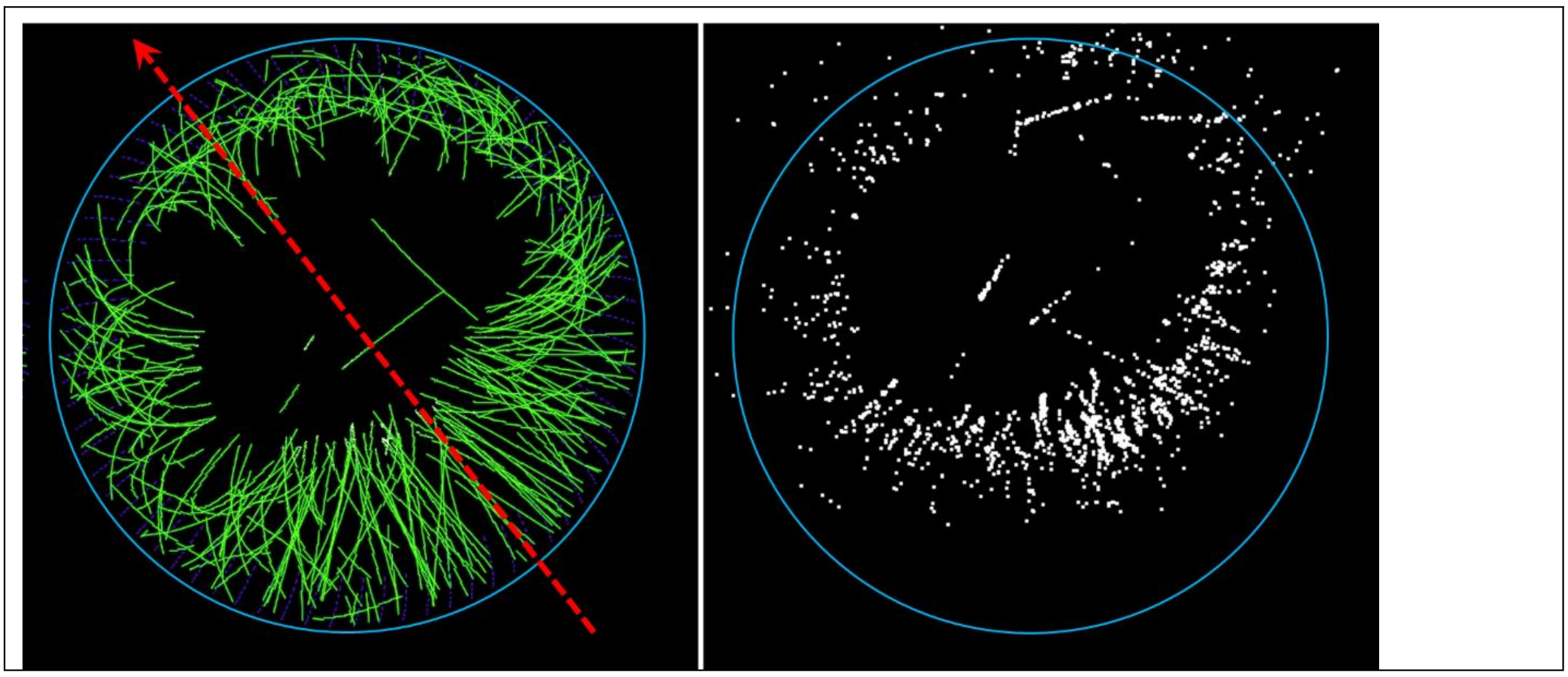

*Рисунок 4. Асимметричное расположение нуклеированных кортексом МТ приводит к асимметричному распределению груза. Это напоминает соответствующую модель транспорта osk в созревающем ооците. Преимущественное направление асимметричной организации МТ сети (ось симметрии) указано стрелкой.*

Такая асимметрия приводит к асимметрии достигаемого стационарного распределения груза (Рисунок 4В). Мы полагаем, что такой подход с внесением асимметрии в пространственное распределение и ориентацию нитей МТ в кортексе имеет перспективы для дальнейшего развития модели транспорта osk.

### 4.1.3. Активный транспорт по отдельным ориентированным пучкам МТ в ооците

Мы этом подразделе мы продолжаем моделирование активного транспорта в ооците. Это будет соответствовать третьей модели из раздела «2.1. Активный транспорт матернальных мРНК в развивающимся ооците». Выбранная модель сходна с транспортными системами в ооците дрозофилы 11й стадии, где три ключевых для полярности эмбриона матернальных мРНК (бикоид, оскар и гуркен) транспортируются по своим ориентированным пучкам МТ [Cohen, 2002[43]; St Johnston, 2005[44]]. К тому же она напоминает транспортную систему ооцита ксенопуса (транспорт Vg1 мРНК), только транспорт идет в противоположном направлении.

Конкретно мы рассмотрим стабильный единичный пучок МТ, транспортирующий груз к центру клетки (Рисунок 5А-С). Транспорту по единственному пучку резонно сопоставить транспорт по паре таких пучков, расположенных симметрично и транспортирующих груз к центру клетки (Рисунок 5D-F).

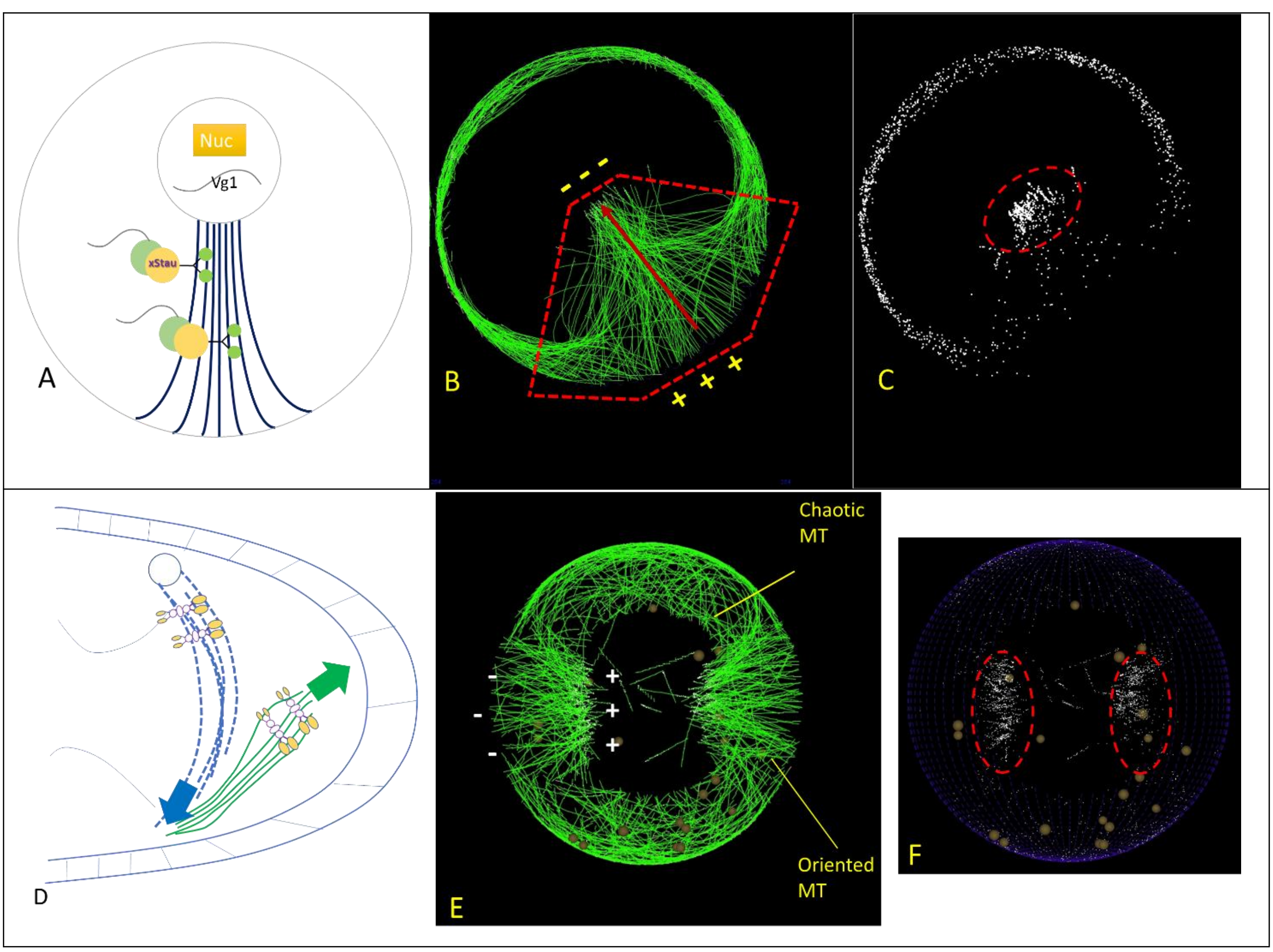

*Рисунок 5. Примеры пучков МТ, в цитоплазме ооцита и соответствующие им 3D агентные модели. Пучки нуклеируются кортексом и обычно ориентированы радиально, плюс-концами МТ к центру клетки. (A-C) Ориентированный пучок МТ в ооците ксенопуса, по которому идет активный транспорт матернальной Vg1 мРНК, и модель, отображающие некоторые ключевые характеристики такого транспорта. (A). Общая схема системы транспорта Vg1 мРНК ооцита ксенопуса: пучок ориентированных к кортексу плюс-концами МТ протянут от ядра к кортексу и мРНК транспортируется моторами к кортексу. (B). Единственный ориентированный пучок в нашей 3D агентной модели, нуклеированный кортексом и ориентированный плюс-концами к центру клетки. (C). Устойчивые агрегации груза в районе плюс-концов пучков МТ модели (B), очертанные пунктирным эллипсом. (D-F) Ориентированные пучки МТ в ооците дрозофилы (стадия 11), по которым идет активный транспорт матернальных мРНК, и модели, отображающие некоторые ключевые характеристики этого транспорта. (D) Общая схема локализации двух ориентированных пучков МТ в цитоплазме ооцита дрозофилы 11 стадии. Направления транспорта по этим пучкам указаны стрелками (голубая и зеленая). (E) Симметричная пара ориентированных пучков в нашей 3D агентной модели. Пучки нуклеированы кортексом и ориентированы радиально плюс-концами к центру клетки. (F) Устойчивые агрегации груза в районе плюс-концов пучков МТ модели (E), очертанные пунктирными эллипсами. Коричневый маленькие сферы – желточные вакуоли.*

Транспорт по паре (или большем числе ориентированных пучков) можно развивать как простую модель активности нескольких ориентированных пучков МТ в ооците дрозофилы, как на Рис 5D-F. По этим пучкам идет активный транспорт матернальных мРНК bcd, oscar и gurken в мид-оогенезе (см обзор [Cohen, 2002]).

В силу простоты этой конкретной модели, она позволяет исследовать характеристики моделируемого процесса в существенных деталях. Это такие параметры, как длина МТ и плотность их пучков, концентрации грузов (и их адапторов), как и параметры взаимодействий моторов и МТ. Все эти вопросы в связи с проблемой скейлинговости будут рассмотрены подробнее в статье-компаньоне [Sabirov, Spirov, in preparation].

## 4.2. Активный транспорт в раннем эмбрионе

В этом разделе мы будем рассматривать несколько агентных моделей, описывающих некоторые ключевые процессы активного транспорта в развивающемся синцитиальном эмбрионе дрозофилы. Отдельно остановимся на одном из самых загадочных событий перераспределений мРНК bcd в самом раннем развитии [Spirov et al., 2009; Little et al., 2011; Fahmy et al., 2014; Ali-Murthy and Kornberg, 2016; Cai et al., 2017]. К тому же нечто сходное наблюдается у неоплодотворенных яиц [Cai et al., 2017]. Это то, что мы кратко описали в разделе «2.2. Активный транспорт мРНК бикоида в раннем эмбрионе». Какова роль того перераспределения мРНК бикоида можно только догадываться. Загадкой остаются и молекулярная машинерия этих процессов. Хотя у нас есть некоторые основания рассматривать это как активный транспорт молекулярными моторами по сетям МТ (и, возможно, по сетям МФ).

### 4.2.1. Транспорт bcd по сердцевинной системе раннего эмбриона

Здесь исследуем простые модели активного транспорта моторами по сетям (пучкам) МТ. Мы представим нашу конкретную версию сердцевинной транспортной системы, описанной в разделе «2.2.1. Процессы активного транспорта bicoid». Как отмечалось, по ряду методических проблем и немногочисленности публикаций наши знания по динамике цитоскелета и молекулярных моторов на этой самой ранней стадии эмбриогенеза весьма фрагментарны и ограничены.

Мы для начала возьмем за основу гипотезу, что частицы, содержащие bcd, активно транспортируются по пучкам МТ, напоминающим гигантские звезды (asters) пронуклеусов в самом начале этой фазы, как иллюстрирует Рисунок 6А. Рисунок 6 схематически воспроизводит форму и направленность таких пучков МТ, тогда как схемы некоторых описанных аномалий этой системы активного транспорта иллюстрирует Рисунок 8.

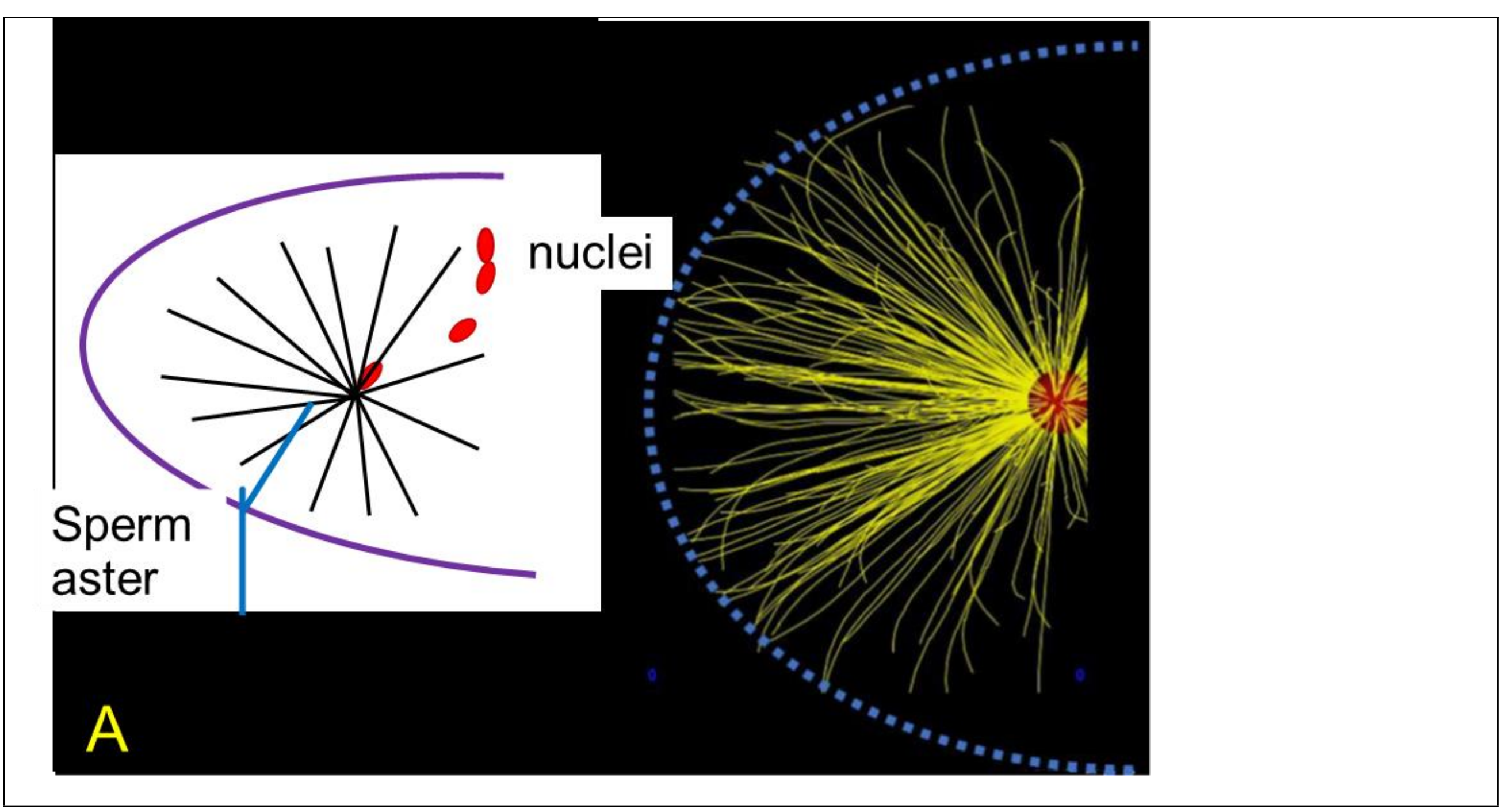

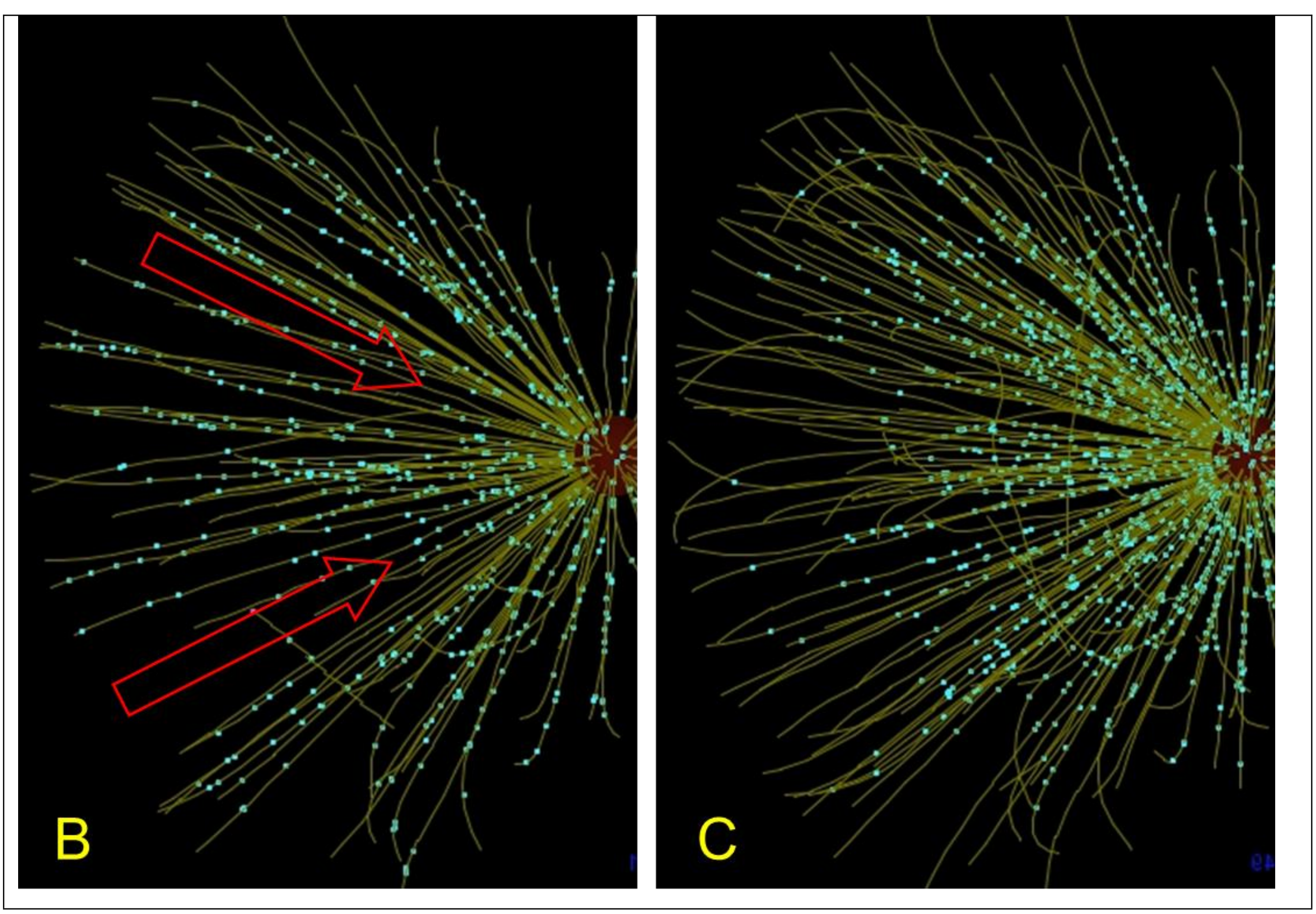

*Рисунок 6. Наша схема становления сердцевинной транспортной системы в самом раннем эмбрионе (и неоплодотворенной яйцеклетке). В качестве своего рода вдохновляющего примера мы возьмем известные представления об организации и поведении the sperm aster на этой стадии раннего эмбриогенеза (A). (B-C) Агентная модель сердцевинной транспортной системы на основе центриоли («sperm aster»). Груз транспортируется с периферии в сердцевину головной части модели раннего эмбриона, что особенно заметно, если сравнить периферию и центр. МТ -- коричневатые, груз – ярко-голубые.*

Для конкретной модели средствами Skeledyne мы имплементировали одиночную центриоль в головной части модели эмбриона и задали процесс асимметричного роста МТ из этой центриоли так, что основной пучок растет в направлении головного (антериорного) кортекса (Рисунок 6B). А по достижении пучком пределов плазматической мембраны рост замедляется и сходит на нет. Типично такие нити МТ в модели не превышают 180-200 мкм (Протяженность интрузий на имиджах статьи [Little et al., 2011] можно оценить в ~75 мкм.)

Когда нити МТ достигают кортикального слоя, где депонированы комплексы грузов с моторами, они начинают связываться с плюс-концами МТ. Эта начальная фаза транспорта в модели иллюстрируется Рис 6B. Начинается процесс активного транспорта моторами в сердцевину головной части раннего эмбриона. Грузы распределяются по всем МТ, напоминая те веретенные или конусовидные интрузии, описанные экспериментаторами (сравни Рисунок 6C). По достижении некоторого времени (зависящего от нескольких параметров и больше всего от скорости мотора) большая часть груза оказывается в сердцевинной части эмбриона. (Отметим, что на Рис 6 визуализированы только грузы в контакте с МТ – груз, теряющий МТ и становящийся растворимой компонентой, не виден.)

Анализ как опубликованных, так и наших неопубликованных данных свидетельствует, что моделируемые здесь процессы имеют характерное время менее часа. От оплодотворения до завершения 6 цикла проходит чуть больше 50 мин (при t=25). При этом Little с соавторами [Little et al., 2011] приводят типичные картины интрузий (включая двойные) для 4го (примерно 35 мин после оплодотворения) и 6го циклов. Наши данные (включая опубликованные) подтверждают типичность интрузий в 6м цикле.

Поведение этой модели зависит от целого ряда параметров и прежде всего от параметров, характеризующих мотор динеин. (Отметим в скобках, что в литературе имеются экспериментальные оценки скорости молекул динеина при движении вдоль МТ [Reck-Peterson et al., 2006[45]]). Дальнейшая работа в этом направлении предполагает более подробное и реалистическое моделирование поведения интрузий, включая последствия обработки химическими агентами (сравни [Cai et al., 2017]).

#### *4.2.1.1. Взаимодействие с кортикальными актиновыми филаментами*

Исходя из известных свойств сети кортикальных актиновых филаментов мы принимаем их неподвижными, тогда как те центриольные МТ свободно растут по направлению к антериорному кортексу. Эти плюс-концевые моторы специфически связываются одновременно с МТ как с грузом и с МФ, как с рельсами, и начинают активное движение. В результате именно нити МТ начинают все больше деформироваться, образуя в итоге изгибы, петли и спирали. Эти процессы иллюстрирует Рисунок 7.

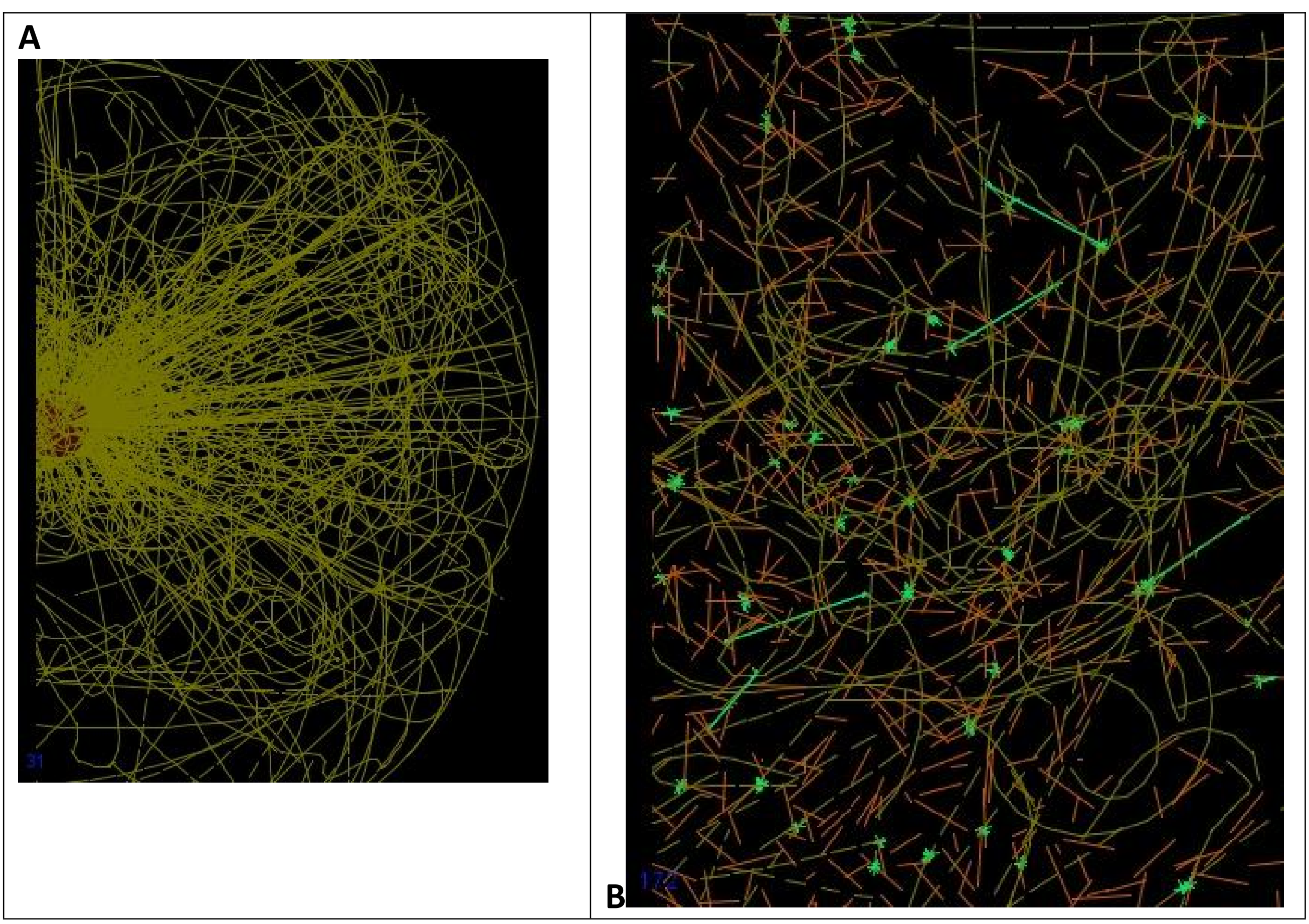

*Рисунок 7. Активная деформация центриольных МТ взаимодействием с неподвижными МФ посредством плюс-концевых молекулярных моторов. (А) Вид сети, образованной пучком центриольных МТ в результате взаимодействия с сетью МФ, МТ – коричневатые. (Сравни с общим видом такого пучка без взаимодействия, как на Рис 6). (Диаметр клетки 360 мкм.) (В) Участок сети (А) при большем увеличении: хорошо видны петли и спирали нитей МТ, как и короткие хаотично разбросанные МФ (красно-коричневые). Ярко-зеленые астериски — это плюс-концевые моторы, связанные и с МТ, и с МФ. Такое взаимодействие приводит к нарастанию деформаций в растущем пучке МТ: они образуют изгибы, петли и спирали. (Радиусы таких петель составляют около 10 мкм). Это отличает рассматриваемую сеть МТ от хорошо ориентированных пучков предыдущего численного эксперимента (Рисунок 6).*

Несмотря на существенно более сложную геометрию рассматриваемой сети (пучка), динамика транспорта груза не сильно изменилась (не показано).

Несколько заложенных в пакет Skeledyne параметров контролируют детали взаимодействия моторов с нитями МТ и МФ. Наши численные эксперименты показали, что основной параметр, контролирующий процесс curving’а это *MaxMotorsPerActinFilament*. Мы полагаем, что и эта модель имеет перспективы дальнейшего развития.

### 4.2.2. Двойные интрузии

В рамках развиваемого здесь семейства моделей одно из простейших объяснений наблюдаемых двойных интрузий это имплементация не единственного пучка МТ, но пары таких пучков от двух центриолей. Рисунок 8 иллюстрирует эту версию модели.

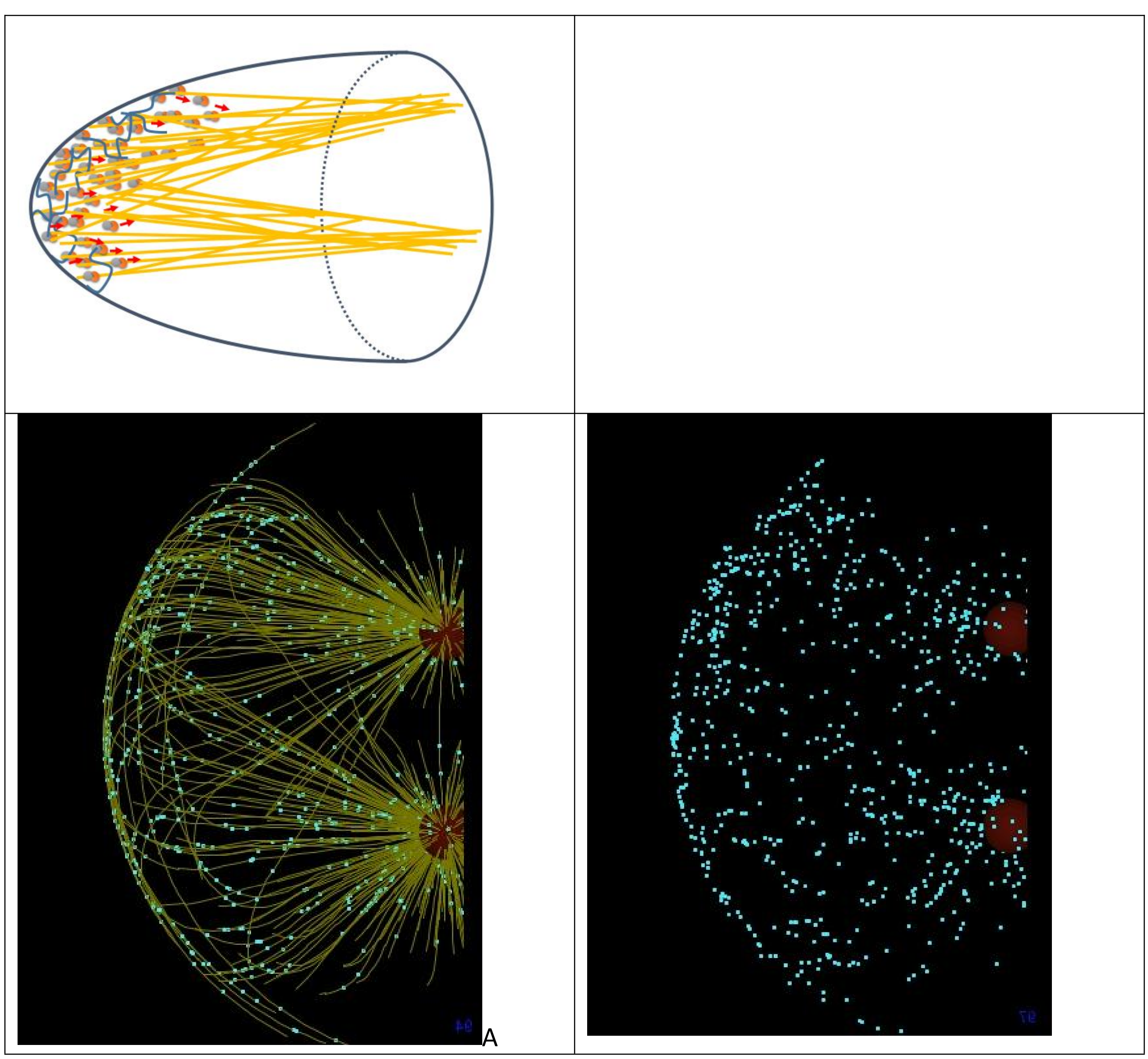

*Рисунок 8. Моделирование активного транспорта bcd плюс-концевыми моторами по паре пучков МТ, сгенерированных двумя центриолями в головной части раннего эмбриона. (A) Наша схема двойной интрузии сердцевинной транспортной системы. (B) Агентная модель пара пучков МТ, сгенерированных двумя центриолями. (Сравни Рисунок 6.) Груз – голубые точки, нити МТ – светло-коричневые, пара сфер -- центриоли. (C) Начало активного транспорта bcd в модели воспроизводит парные интрузии наблюдений экспериментаторов (смотри текст).*

Полученные результаты моделирования (Рисунок 8) весьма напоминают реальные конфокальные изображения ([Little et al., 2011; Fahmy et al., 2014; Ali-Murthy and Kornberg, 2016]).

Результаты этого подраздела демонстрируют перспективность развития таких моделей и в этом конкретном направлении. К сожалению, скудность экспериментальных данных на сей момент ограничивает такую работу. Требуются новые количественные экспериментальные данные для дальнейших численных экспериментов.

## 4.3. Роль актиновых филаментов в раннем транспорте

В этом подразделе мы (в тех же общих рамках подхода) продолжаем моделировать активный транспорт моторами по ориентированным МТ, но уже во взаимодействии с МФ кортекса.

Зависимое от МТ распределение МФ наблюдалось в различных типах клеток [Goode et al., 2000[46]]. Исследования экстрактов яиц Xenopus показывают, что МФ могут перемещаться по МТ в присутствии белков, связывающих МТ [Waterman-Storer et al., 2000[47]]. Во время целлюляризации синцитиальной бластодермы дрозофилы МФ перемещаются вдоль микротрубочек центриолей к их плюс-концам [Foe et al., 2000[48]]. Как предполагают Фой (Foe) с соавторами, такой активный транспорт МФ (и миозина) в определенные области кортекса клетки (где в итоге формируется цитокинетическая борозда) может объяснить детали цитокинеза вообще и целлюляризации в эмбриогенезе дрозофилы в частности (Рисунок 9).

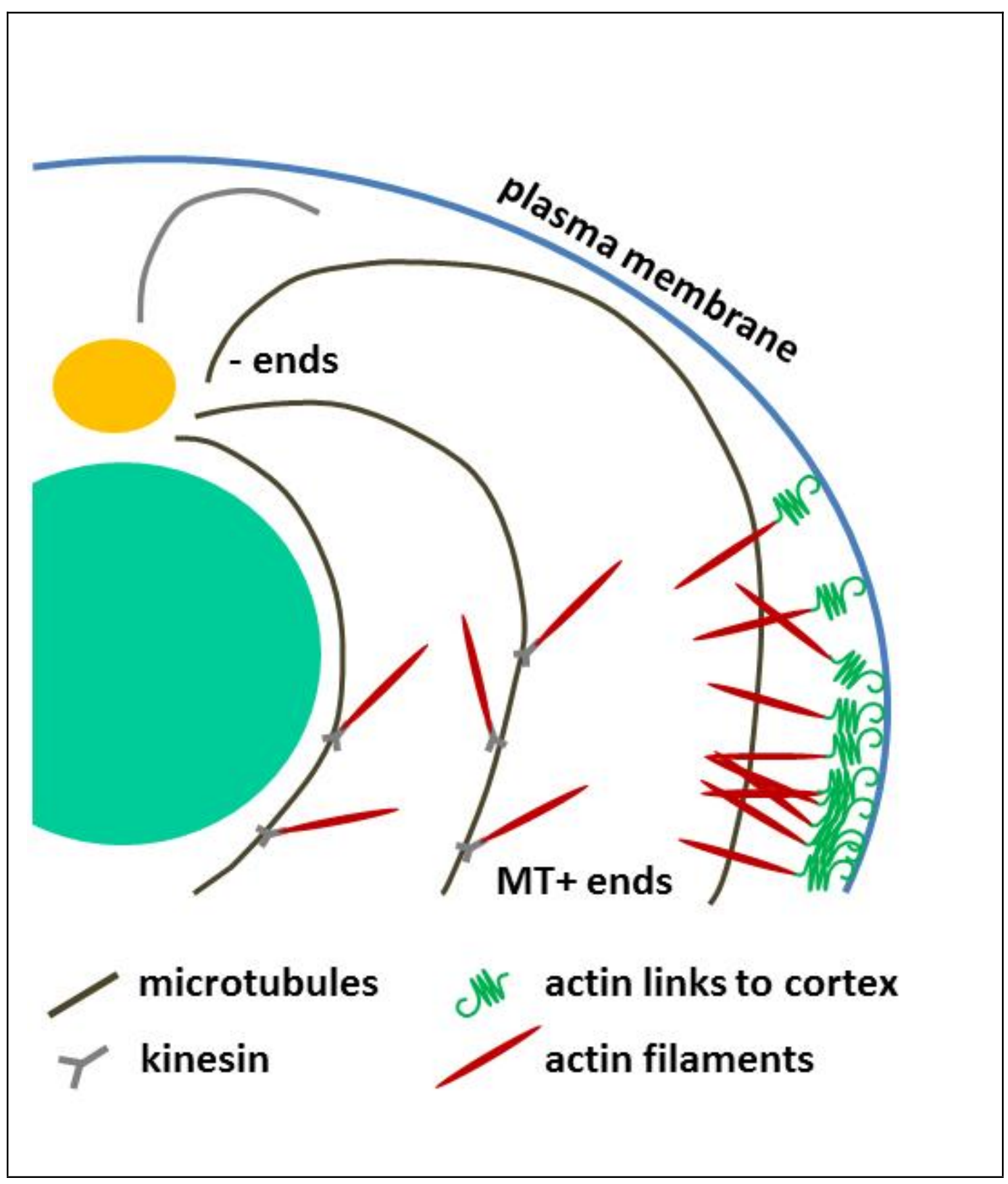

*Рисунок 9. Предполагаемый активный транспорт актиновых филаментов (и миозина) по МТ, нуклеированным парой центросом, в определенные области кортекса клетки. Это может объяснить процессы паттернинга при цитокинезе вообще и целлюляризации в эмбриогенезе дрозофилы в частности. Согласно Foe et al., 2000 с изменениями.*

Пакет *Skeledyne* предоставляет возможности имплементировать процессы активного транспорта актиновых филаментов по МТ посредством plus-end моторов. Однако сколько-нибудь детальный анализ этих механизмов на уровне моделей так и не был опубликован, насколько нам известно.

Наша задача здесь развить простую 3D агентную модель взаимодействия сетей МТ и МФ с рассмотрением процессов активного транспорта молекулярными моторами. На этом этапе нашего проекта мы не ставили целью имплементировать модель какого-либо известного конкретного процесса клеточной биологии. Однако модель имплементирует в упрощенном виде некоторые известные и предполагаемые процессы активного транспорта в клетке (сравни [Foe et al., 2000]). В частности, общая архитектура сети МТ и центростремительный транспорт напоминает процессы транспорта груза мРНК osk мотором кинезином в ооците дрозофилы [Cha et al., 2002].

В рамках пакета Skeledyne конкретно задается следующая модель. Моделируется клетка сферической формы (диаметром 400 мкм). Изначально задается в цитоплазме пул (числом в 5 тысяч) комплексов мотора кинезина с молекулярным грузом (для этих тестов мы не конкретизируем какие именно молекулы играют роль груза). Кинезин с заданной вероятностью может потерять этот молекулярный груз. В модели изначально задается неориентированная сеть (небольших) МФ (Рисунок 10В). Число этих филаментов – 5 тысяч, длина каждого филамента 16 мкм.

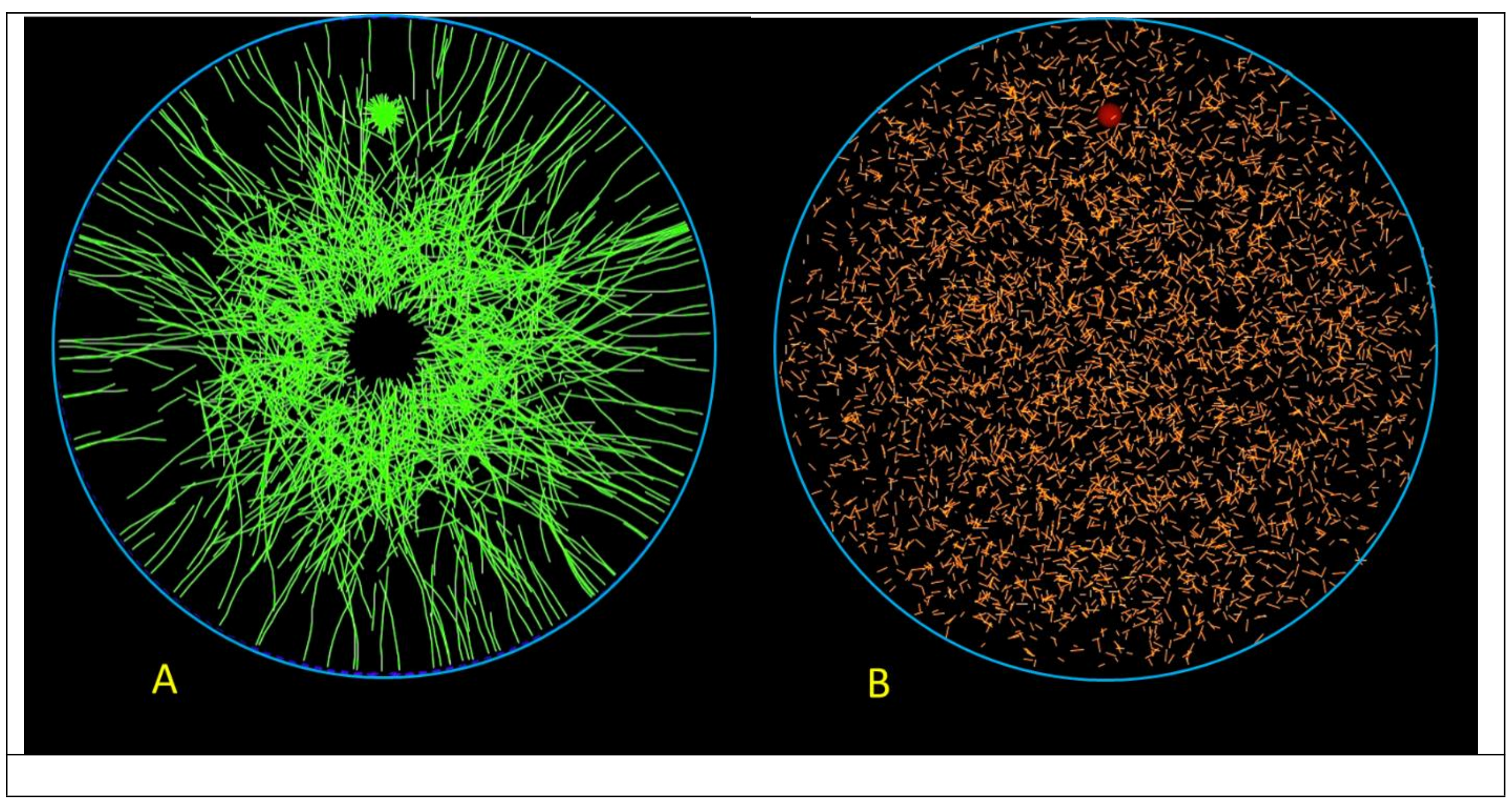

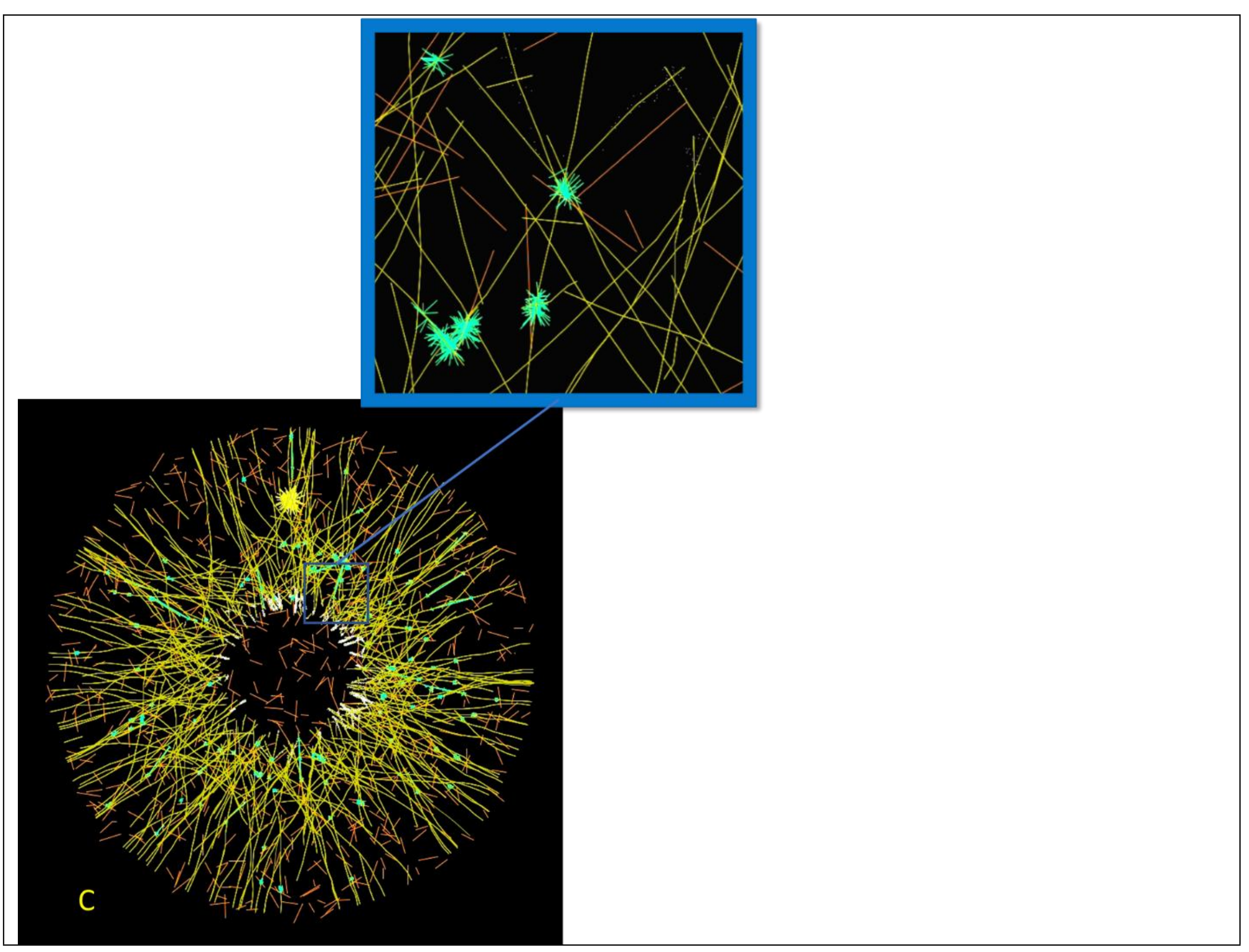

*Рисунок 10. Общая агентная модель активного транспорта молекулярными моторами по ориентированной сети МТ во взаимодействии с МФ. (А) Высоко-ориентированная сеть МТ, нуклеированных кортексом и направленных плюс-концами в сердцевину клетки. (Радиус клетки – 380 мкм; длина МТ – 266 мкм.) (В) Неориентированная сеть МФ во всем объеме цитоплазмы. (С) Визуализация всех агентов этой модели по достижении стационарного состояния транспортных процессов. Участок сердцевинной цитоплазмы приведен на врезке при большем увеличении. Плюс-концевой молекулярный мотор обеспечивает как транспорт молекулярного груза (скопления белых точек вдоль плюс-концевых частей МТ), так и транспортировку МФ (ярко-зеленые астериски визуализируют такие моторы). МТ -- зеленые, МФ -- коричневые.*

Далее, в модели изначально задается сеть МТ, нуклеированных кортексом и ориентированных плюс-концами к центру клетки (Рисунок 10А). Число микротрубочек – 5 тысяч, длина каждого МТ 220 мкм. В таких условиях кинезин с грузом связывается с МТ и транспортирует этот груз к плюс-концу.

В модели также задано, что молекулы кинезина, потерявшие молекулярный груз, способны присоединить в качестве нового груза МФ и транспортировать их по МТ к «плюс»-концу (центростремительно).

С такими условиями модель гоняется такое время, которое соответствует 600 секундам биологического времени. В результате модель типично достигает стационарного распределения кинезина с молекулярным грузом и моторов, несущих МФ (Рисунок 10С). И те и другие комплексы скапливаются в сердцевинной цитоплазме.

Ряд параметров модели мы систематически меняли, так что в итоге эта конкретная модель исследовалась в нескольких десятках прогонов. Конкретно мы меняли ключевые параметры, определяющие плотность и динамику сети МТ (1), плотность и динамику сети МФ (2), поведение plus-end моторов, транспортирующих молекулярный груз (3), и поведение моторов, транспортирующих актиновые филаменты (4).

Имплементированная здесь пространственная организация цитоскелета позволяет не только наблюдать, но и количественно оценивать динамику транспортных процессов (см Рисунок 11С).

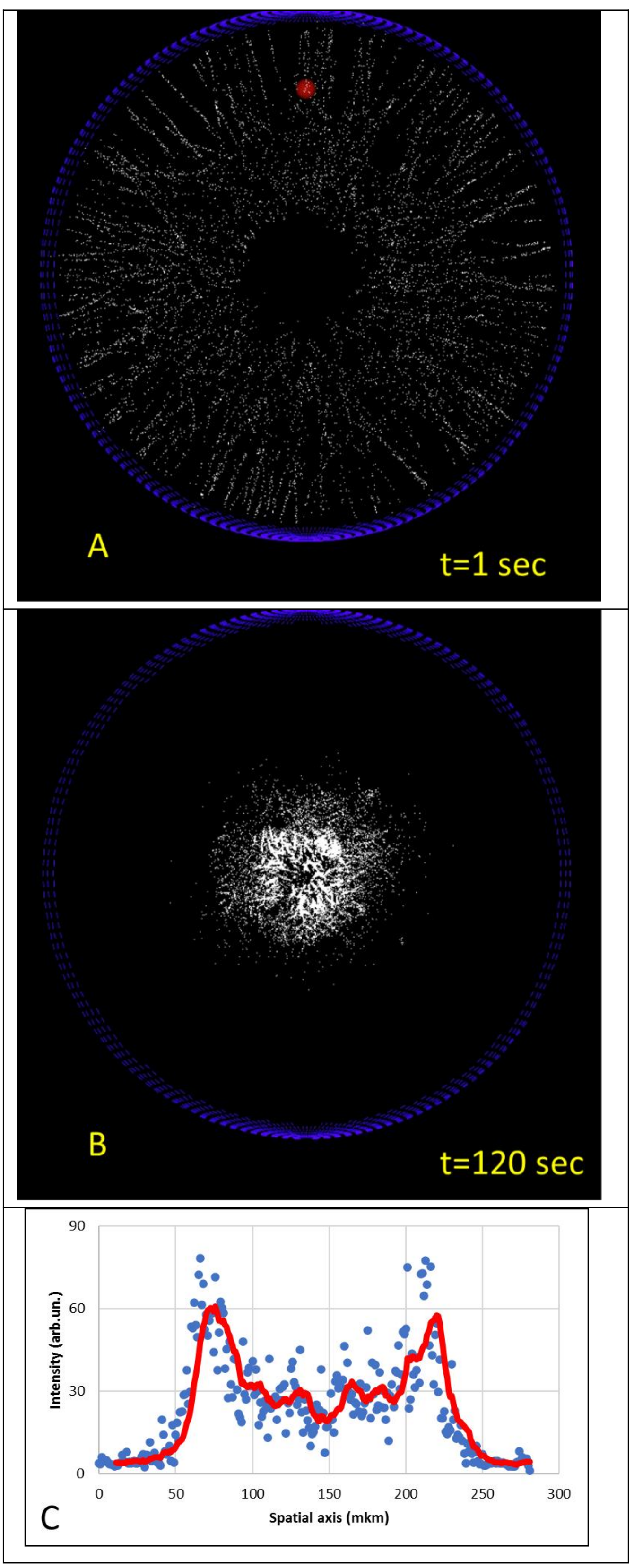

*Рисунок 11. Активный транспорт молекулярного груза мотором кинезином (белые точки – комплексы груз-кинезин) по ориентированной сети МТ (как на Рис 10А), при радиусе клетки в 400 мкм и длине МТ – 198 мкм, приводит к стационарному накоплению груза на плюс-концах микротрубочек в сердцевине модели клетки. (А)*

*Начальное состояние, 1я секунда моделирования. (B) Стационарное состояние транспортного процесса, 120я секунда моделирования. (C) Профиль стационарного распределения груза имиджа (B), полученный оцифровкой полоски, вырезанной из такого имиджа; исходные данные – голубые кружки, сглаживание скользящим средним – красные линии.*

Наши эксперименты in silico продемонстрировали как кинезин транспортирует молекулярный груз по высоко-ориентированной сети МТ к центру клетки до достижения стационарного состояния (Рисунок 11). Стационарное состояние здесь это скопление груза в сердцевине клетки на плюс-концах ориентированных МТ (Рисунок 11B; сравни Рисунок 10C). Ему соответствует некоторое повышение концентрации свободного кинезина в сердцевине клетки (не показано). Точная форма этого стационарного скопления определяется, в первую очередь, длинной МТ (смотри Рисунок 10A). График на Рис 11C показывает профиль такого конкретного скопления.

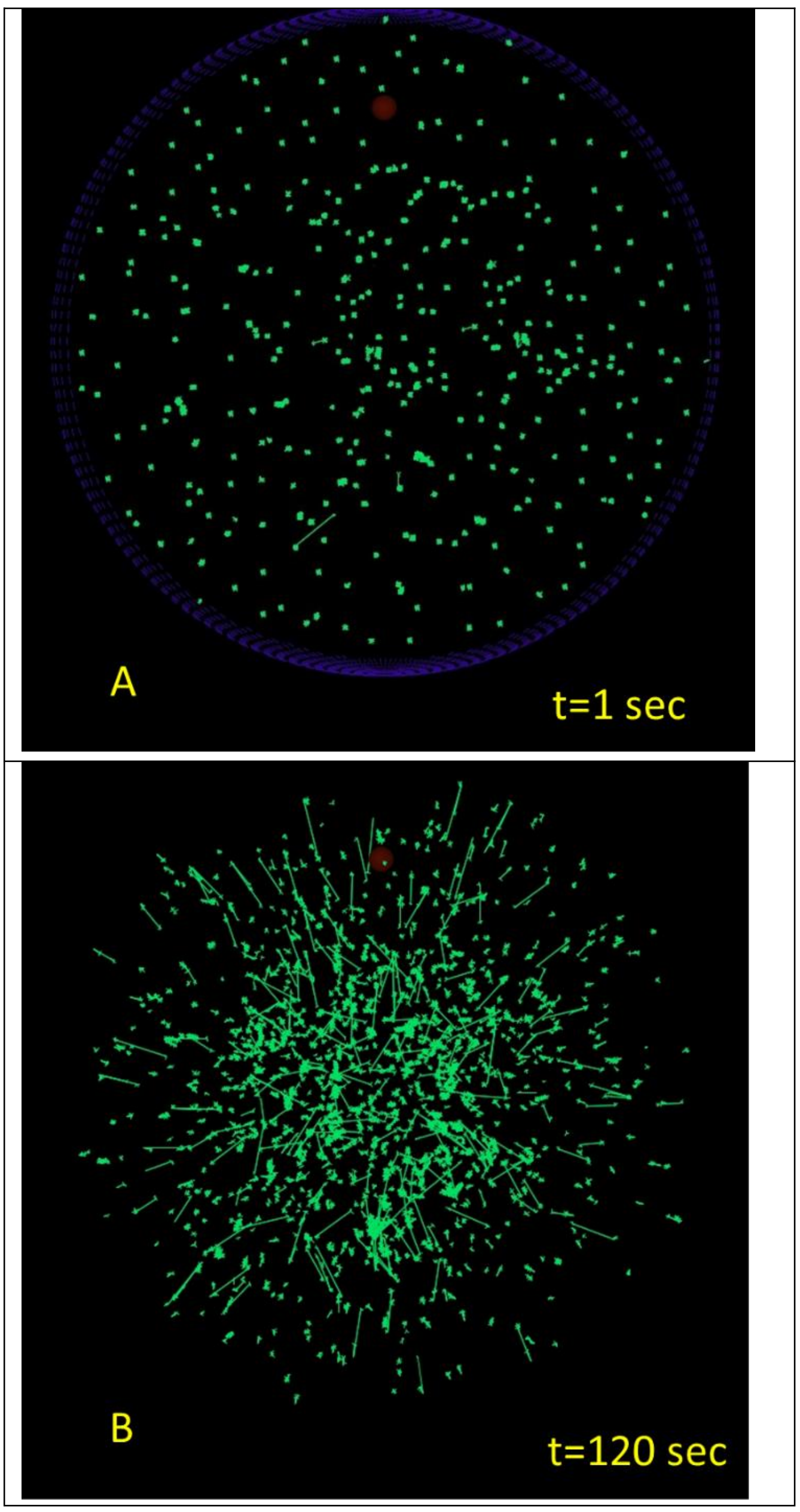

Рисунок 12. *Активный транспорт кинезином актиновых филаментов (МФ) по МТ к плюс-концам приводит к преимущественной локализации этих моторов с их грузом в сердцевине клетки. (A) В начале прогона (1я секунда) моторы, несущие МФ (ярко-зелёные астериски), наблюдаются связанными с МТ по всему объему МТ-сети. (радиус клетки -- 400 мкм; длина МТ – 198 мкм) (B) Со временем (120 сек) моторы, несущие МФ, оказываются преимущественно в сердцевине клетки, куда они добрались со своим грузом.*

С первой же секунды прогона модели молекулы кинезина с заданной небольшой частотой начинают терять молекулы груза. Как следствие, такие свободные молекулы с некоторыми заданными частотами специфически связываются (через соответствующие адапторы) с МФ и затем с МТ. В начале прогона такие моторы, несущие МФ, распределены по всему объему сети МТ и их относительно немного (Рисунок 12А).

Далее, такие моторы с МФ начинают перемещаться по МТ к «плюс»-концу, центростремительно. Со временем такие моторы начинают скапливаются в сердцевине клетки (Рисунок 12В). Такой паттерн явился для нас критерием эффективности этого транспортного процесса. В этом аспекте он сравним со стационарными паттернами в результате другого такого транспортного процесса, как на Рис 11А-В.

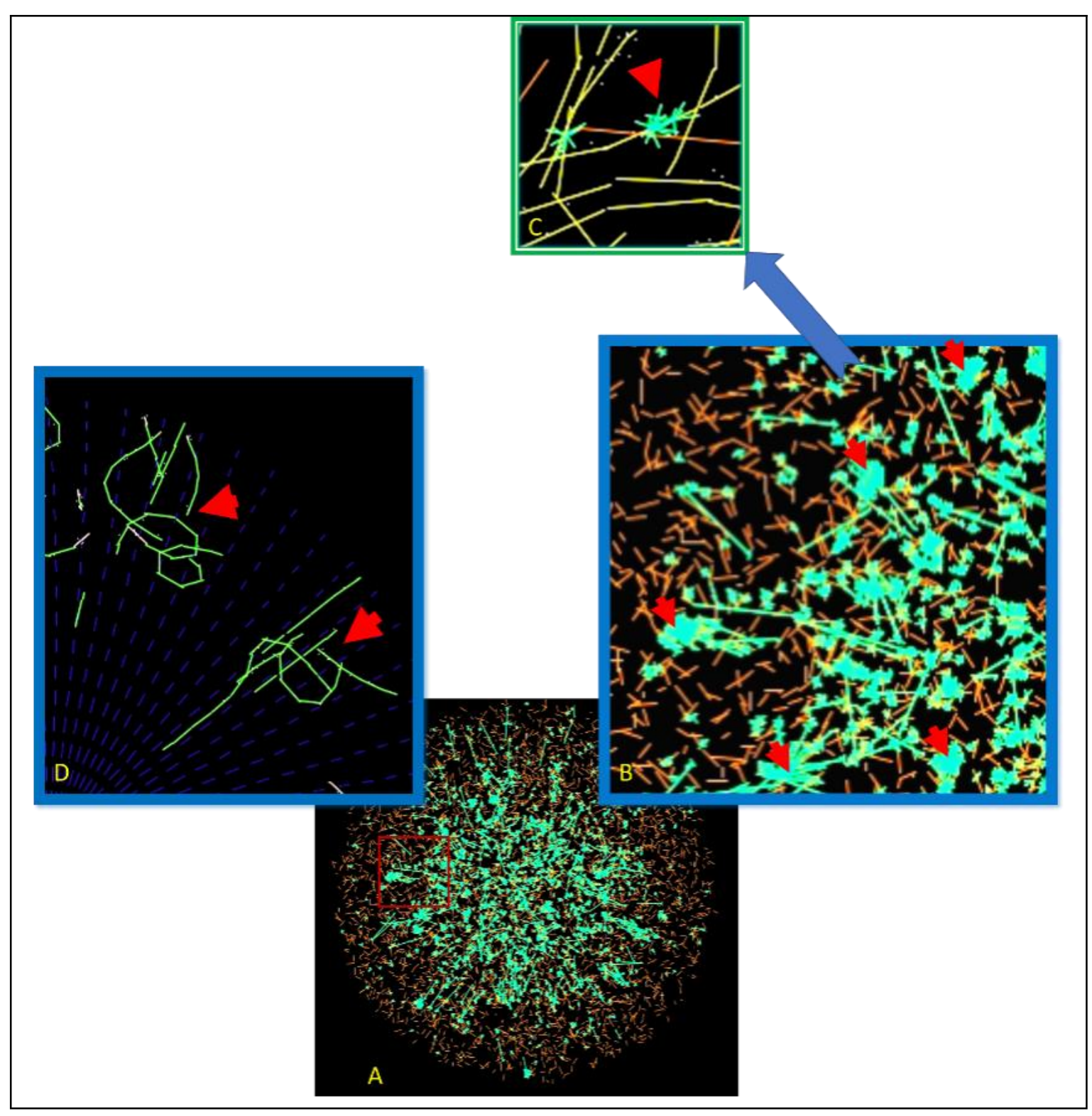

*Рисунок 13. Примеры агрегаций на МТ моторов, несущих филаменты (МФ) для модели Рис 12. (A) Изображение клетки целиком (моторы с филаментами, прикрепленные к МТ, помечены ярко-зеленым). (B) Фрагмент изображения (отмечен красным прямоугольником на имидже (A)) с агрегациями моторов, указанными красными головками стрелок. (C) Фрагмент изображения (B) при еще большем увеличении, так что видны детали скопления моторов на одной нити МТ (красная головка стрелки). (D) Детали петель МТ (красная головка стрелки). МФ – коричневые линии; МТ – зеленые линии.*

Серии прогонов с разными наборами конкретных значений параметров (как описано в разделе Методы) показали робастность этих транспортных процессов. Среди примечательных свойств модели отметим таковое, связанное с параметром MaxMotorsPerActinFilament (пакета Skeledyne). Этот параметр определяет максимально допустимое число моторов, которые могут специфически связаться с одним данным МФ для транспортировки его по микротрубочке.

При условии MaxMotorsPerActinFilament > 1 в ходе прогона модели наблюдается формирование характерных устойчивых компактных агрегаций моторов с актиновыми филаментами на нитях МТ (Рисунок 13B,C). Этим агрегациям соответствуют множественные петли на нитях МТ (Рисунок 13D). Было бы интересно выяснить возможно ли формирование таких петель в реальности (Сравни [Wu et al., 2012[49]]).

Имплементацию транспортных процессов клетки, как на Рис 11 и Рис 12, можно использовать для моделирования действия ядов, разрушающих или МТ или МФ (сравни [Cai et al., 2017]). Наши тесты показали, что нарушение сети МФ ядами сказывается на самой сети и на распределении кинезина, несущего МФ. Если в модели образуются кластеры моторов, связанных с филаментами, и им соответствующие петли из нитей МТ (как на Рис 13), то разрушение МФ приводит к исчезновению и таких кластеров и таких петель. Однако такая обработка не сказывается на транспорте молекулярного груза и достигаемого стационарного распределения в клетке. Нарушение сети МТ ядами сказывается как на транспорте и перераспределении молекулярного груза, так и на транспорте МФ и образовании кластеров и петель.

Отметим в заключение, что вопросы гибкости МТ в физиологии и патологии клетки обсуждались рядом авторов [He et al., 2002[50]; Bicek et al., 2007[51]; Lagomarsino et al., 2007[52]; Diagouraga et al., 2014[53]; Yin et al., 2020[54]]. Согласно опубликованным оценкам МТ способны образовывать петли небольшого радиуса, от нескольких микронов (2-3мкм [Diagouraga et al., 2014]; ~12 мкм [Lagomarsino et al., 2007]). Примечательно, что именно такого порядка (с радиусом примерно 10 мкм) имеют тенденцию образовываться петли в модели Skeledyne при выбранных параметрах моделирования.

## 4.4. Стадия формирования кортикального слоя ядер

Миграции ядер раннего эмбриона, начиная с закономерных перемещений пронуклеусов (упоминавшуюся выше), демонстрируют сложную, но воспроизводимую картину, и давно привлекают внимание исследователей [Baker et al., 1993; Sullivan et al., 1993[55]]. В этом подразделе мы продемонстрируем перспективность использования 3D агентного моделирования конкретно в этом направлении.

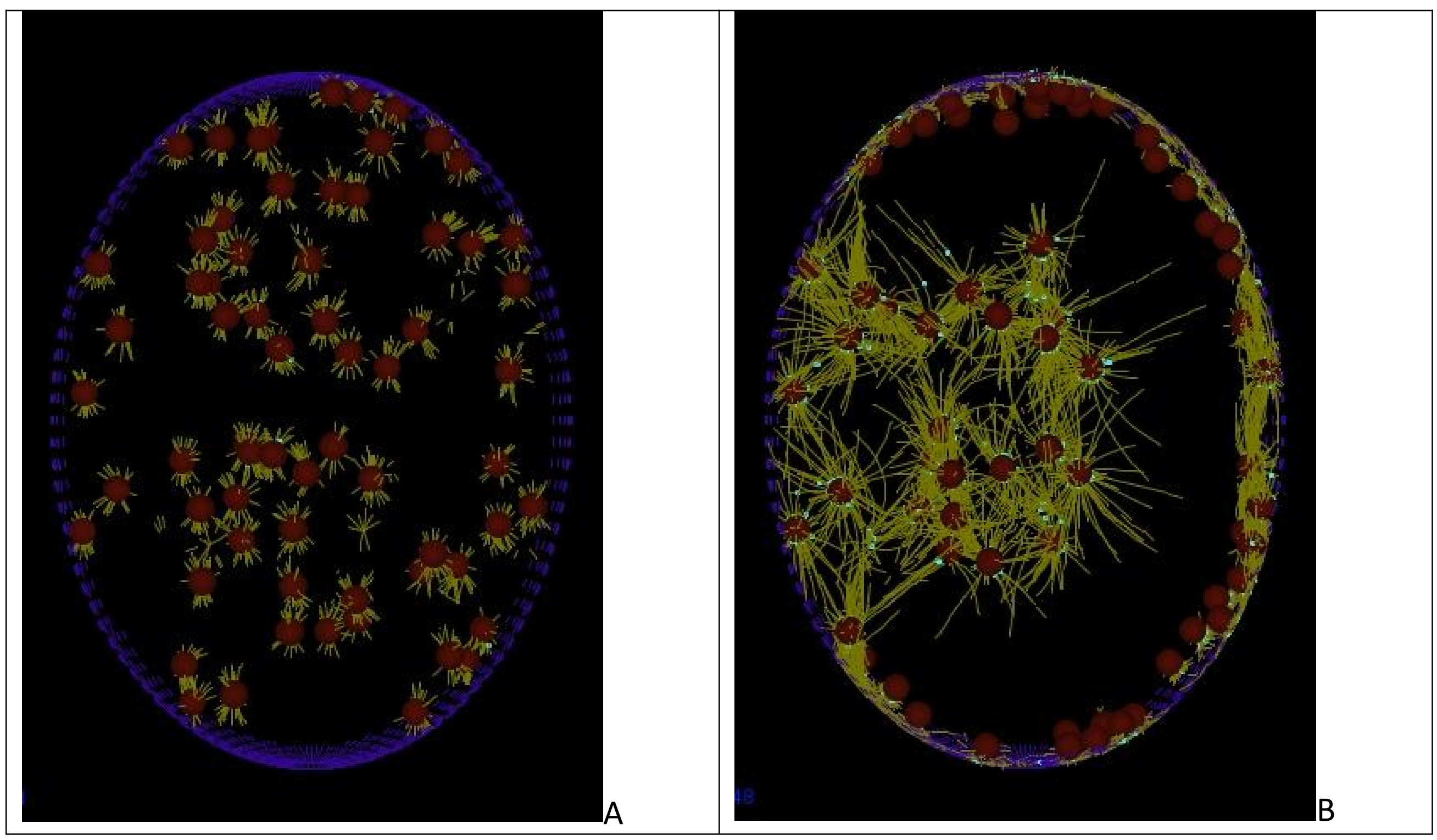

*Рисунок 14. Модель формирования ядерного слоя раннего синцитиального эмбриона дрозофилы. (A) начало процесса с ядрами, хаотично разбросанными в толще цитоплазмы. (B) Значительная часть ядер сместилась в кортекс, с МТ, ориентированными вдоль кортикального слоя. Часть ядер осталась в толще цитоплазмы.*

Известно, что ядра эмбриона на стадии дробления изначально распределены в толще цитоплазмы (например, [Baker et al., 1993; Sullivan et al., 1993; Archambault, Pinson, 2010[56]]). Через некоторое время большая часть ядер оказывается в кортикальном слое, под поверхностью эмбриона. Часть ядер остается в сердцевинной части яйца. Ядра в слое могут контактировать с соседями через пучки МТ.

Пучки МТ имеют тенденцию располагаться ориентированно относительно поверхности яйцеклетки. Полагают, что сердцевинные ядра могут быть связаны посредством МТ с периферией клетки.

Средства моделирования пакета Skeledyne дают возможность моделировать по крайней мере некоторые существенные компоненты этих процессов. Рисунок 14 иллюстрирует поведение ядер (представленных только центриолями), изначально хаотически распределенных в цитоплазме (Рисунок 14А). Через некоторое время ядра преимущественно перемещаются в кортекс так, что МТ растут параллельно поверхности, тогда как оставшиеся ядра скапливаются в сердцевине (Рисунок 14B). Мы рассчитываем, что дальнейшее развитие модели даст возможность промоделировать миграции ядер в более реалистичных деталях.

## 4.5. Транспорт мРНК bcd в синцитиальном эмбрионе

Это тот транспортный процесс, за который ответственна система кортикального транспорта в синцитиальном эмбрионе дрозофилы. Мы в этом разделе приведем некоторые количественные экспериментальные данные этого транспортного процесса, а затем саму модель его.

В самом раннем синцитиальном эмбрионе (первая половина периода дробления, по 6 синхронное деление дробления) локальное плотное и компактное скопление мРНК бикоид в кортикальной цитоплазме головной части перераспределяется так, что несколько распространяется постериорно, формируя достаточно крутой градиент концентрации этой мРНК [Spirov et al., 2009; Little et al., 2011; Fahmy et al., 2014; Ali-Murthy and Kornberg, 2016; Cai et al., 2017]. Этот процесс иллюстрирует Рисунок 15.

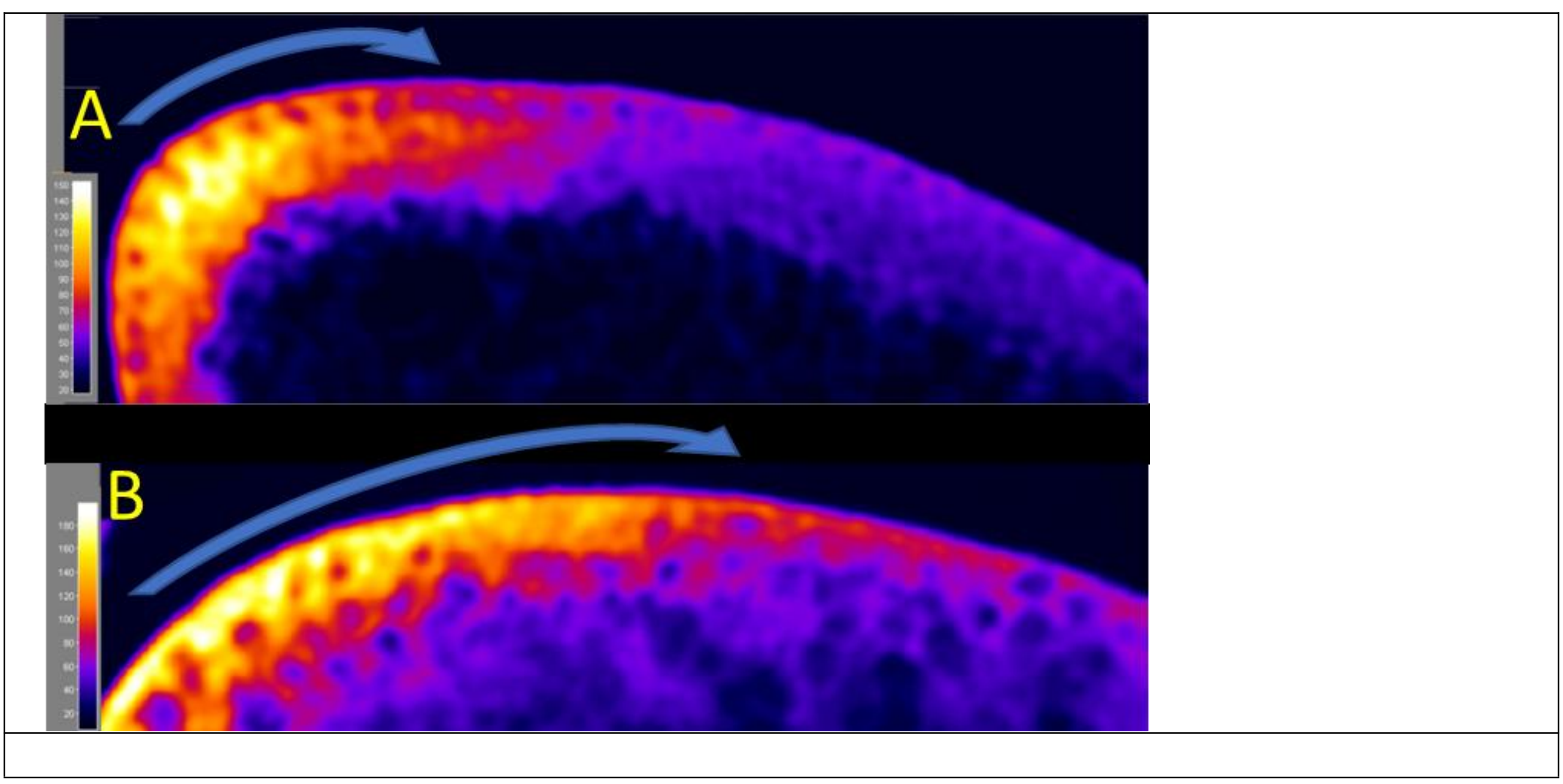

*Рисунок 15. Перераспределение мРНК бикоид постериорно в первой половине стадии делений дробления: получение количественных данных. Локальное скопление мРНК в кортексе головной части самого раннего эмбриона (зиготы) несколько распространяется постериорно, формируя крутой градиент концентрации мРНК. Самый ранний эмбрион (А) в сравнении с немного более поздним (В) (стадия дробления). Полезный сигнал конфокального изображения (иммуноокраска на мРНК бикоида) преобразован в цветовую гамму, так что светло-жёлтый и белый – это самый высокий уровень сигнала. Голубые стрелки передают направление и крутизну наблюдаемых градиентов РНК.*

Публикации лаборатории Баумгартнера (Baumgartner) [Fahmy et al., 2014; Cai et al., 2017] дают основания полагать, что эти перемещения мРНК бикоида выполняются посредством активного транспорта молекулярными моторами по неориентированной сети небольших МТ. Конкретная сеть иллюстрируется Рис 16.

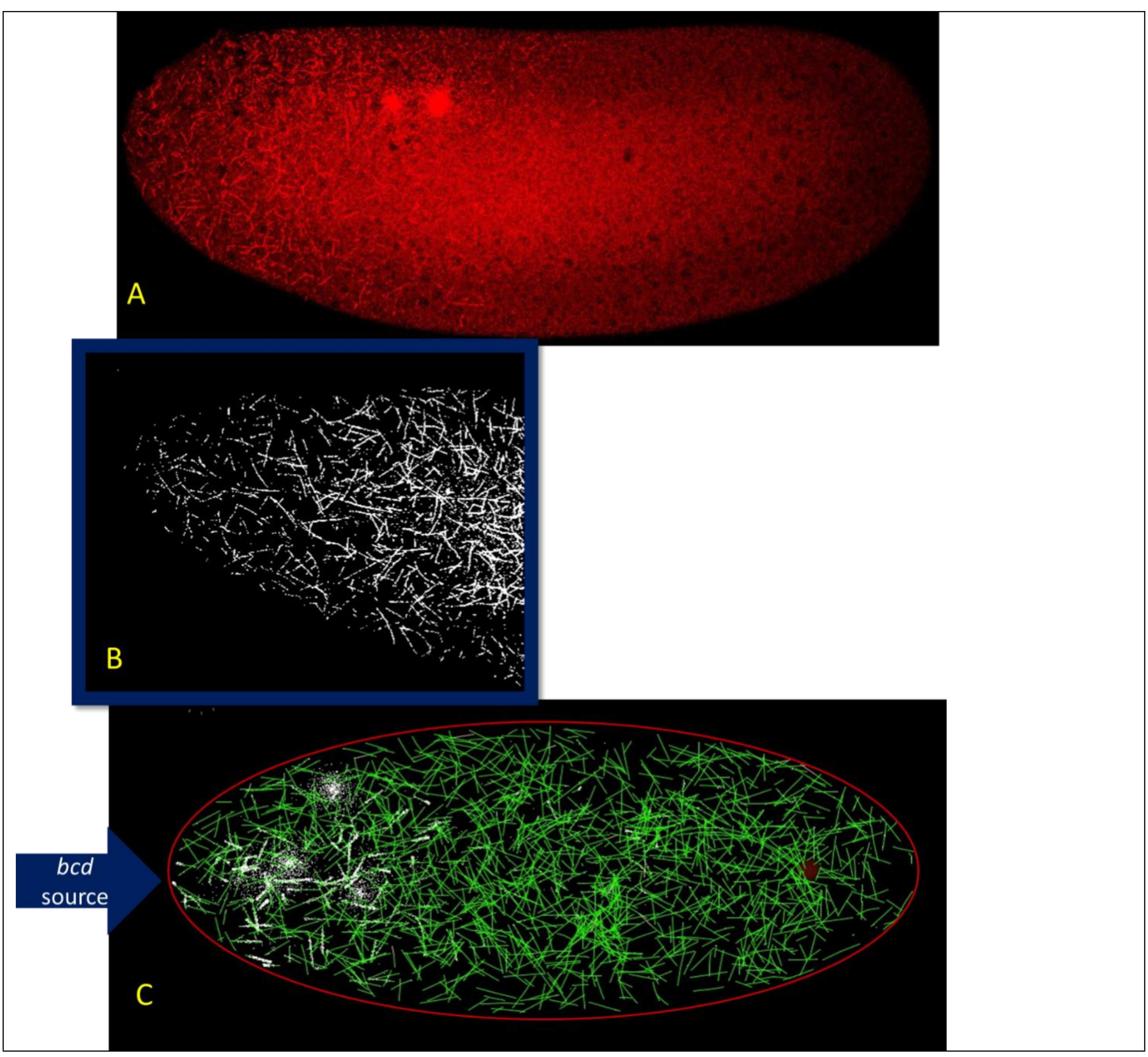

*Рисунок 16. Неориентированная сеть небольших нитей МТ в цитоплазме синцитиального эмбриона дрозофилы. (А) Неориентированная сеть коротких МТ раннего эмбриона дрозофилы, «отсегментированные» средствами программы ImageJ. Для такого процессинга использовано изображение [Fahmy et al., 2014]. (B) Агентная модель такой неориентированной сети коротких МТ (МТ – зеленый, груз – белые точки).*

Как можно видеть (Рисунок 16), сеть состоит из относительно коротких МТ (типично немного десятков мкм длинной), распределенных хаотично в периферийной цитоплазме. Эти характеристики МТ были использованы нами при разработке агентной модели.

### 4.5.1. Моделирование активного транспорта мРНК bcd

Мы моделируем здесь ранний синцитиальный эмбрион, где начально задан груз (мРНК бикоида), локализованный в кортексе головной части эмбриона (Рисунок 17). Изначально в модели заданы 12 тыс таких молекулярных грузов. Груз ничем не закреплен и в дальнейшем будет перераспределяться за счет активного транспорта и свободной диффузии. Параметры транспорта и диффузии задаются изначально и могут быть изменены на лету.

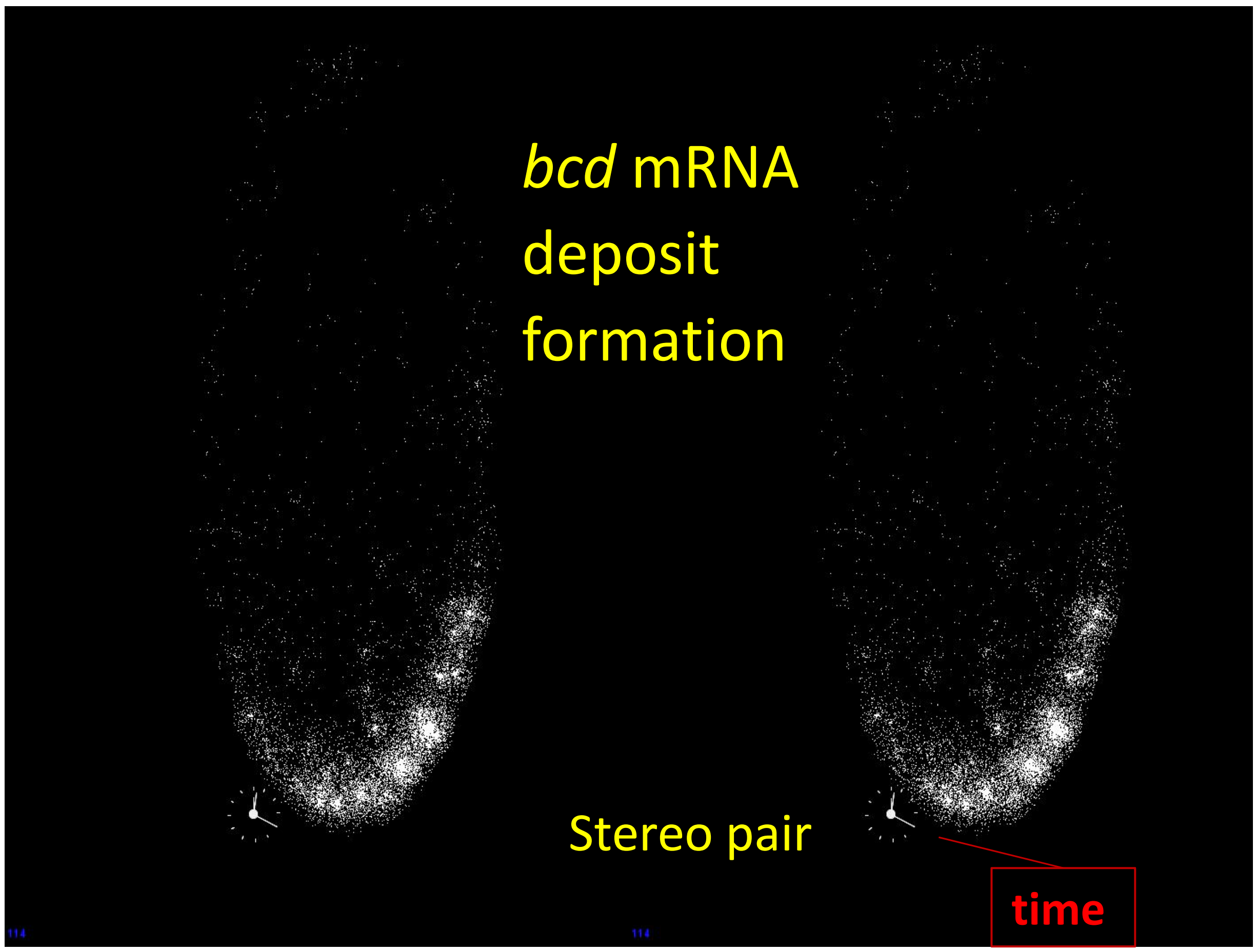

*Рисунок 17. Агентное 3D-моделирование транспортной системы для мРНК бикоид (мелкие белые точки). Ранний синцитиальный эмбрион дрозофилы моделируется как вытянутый эллипсоид с осями в 470 * 200 mkm. Исходное состояние модели, когда груз локализован в кортексе головного кончика самого раннего эмбриона; МТ еще нет.*

Запуск модели сопровождается активацией процесса нуклеации и роста МТ. Процесс формирования все более густой сети МТ устроен следующим образом. В модели на начальной стадии сборка МТ все время превышает деградацию МТ (*Рисунок 18*). Поэтому сеть со временем становится все более плотной. Длина нитей МТ в среднем составляет около 40-50 мкм.

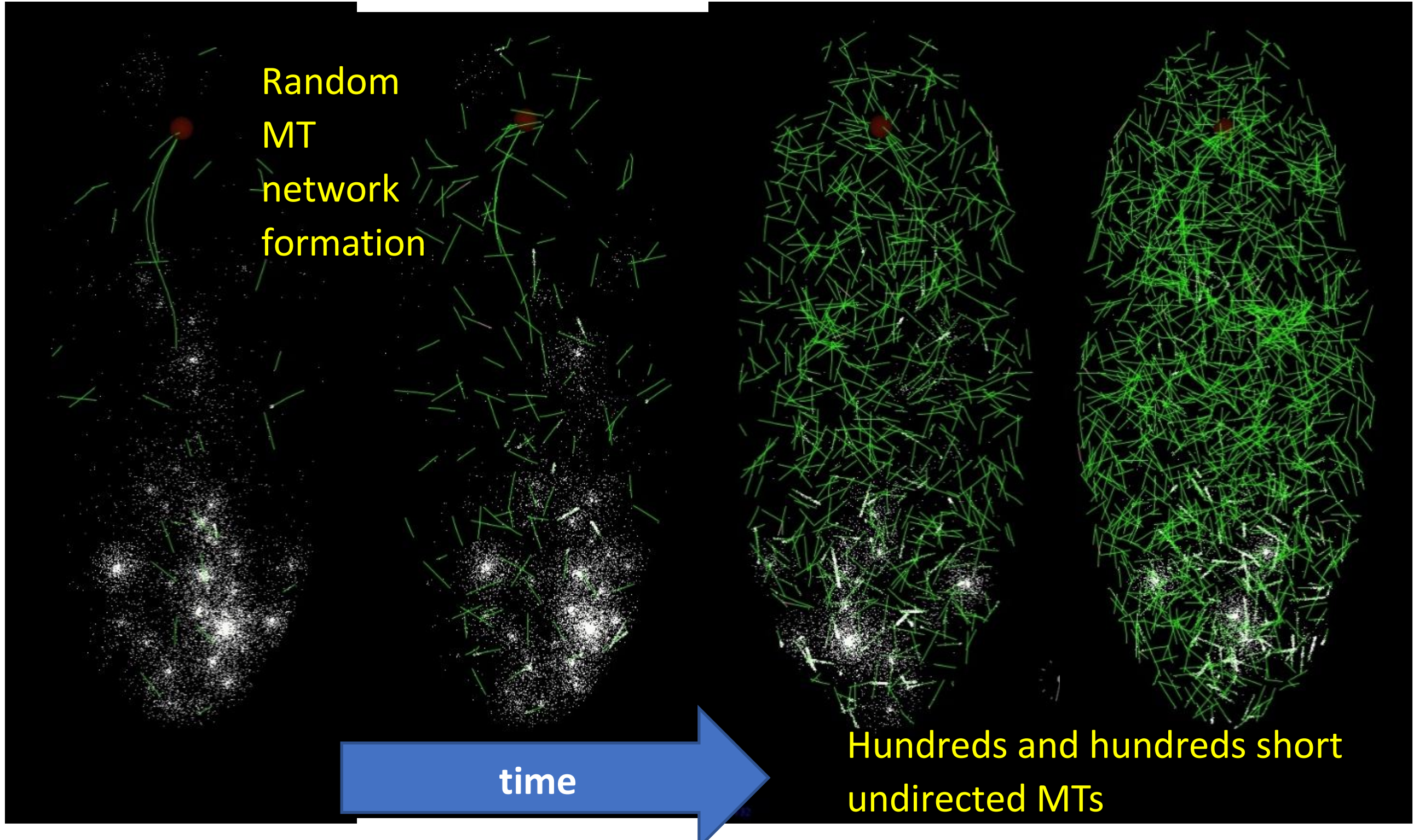

P

*Рисунок 18. Моделирование становления транспортной системы для мРНК bcd. Формирование сети МТ с сотнями коротких ненаправленных нитей МТ (зеленым). мРНК бикоида -- мелкие белые точки.*

Молекула мРНК, молекула адапторного белка и мотор образуют единую частицу. Он моделируется как отдельный объект, способный запрыгивать на нить МТ и спрыгивать с нее, активно перемещаться по нити МТ, как и пассивно диффундировать в цитоплазме (Рисунок 19).

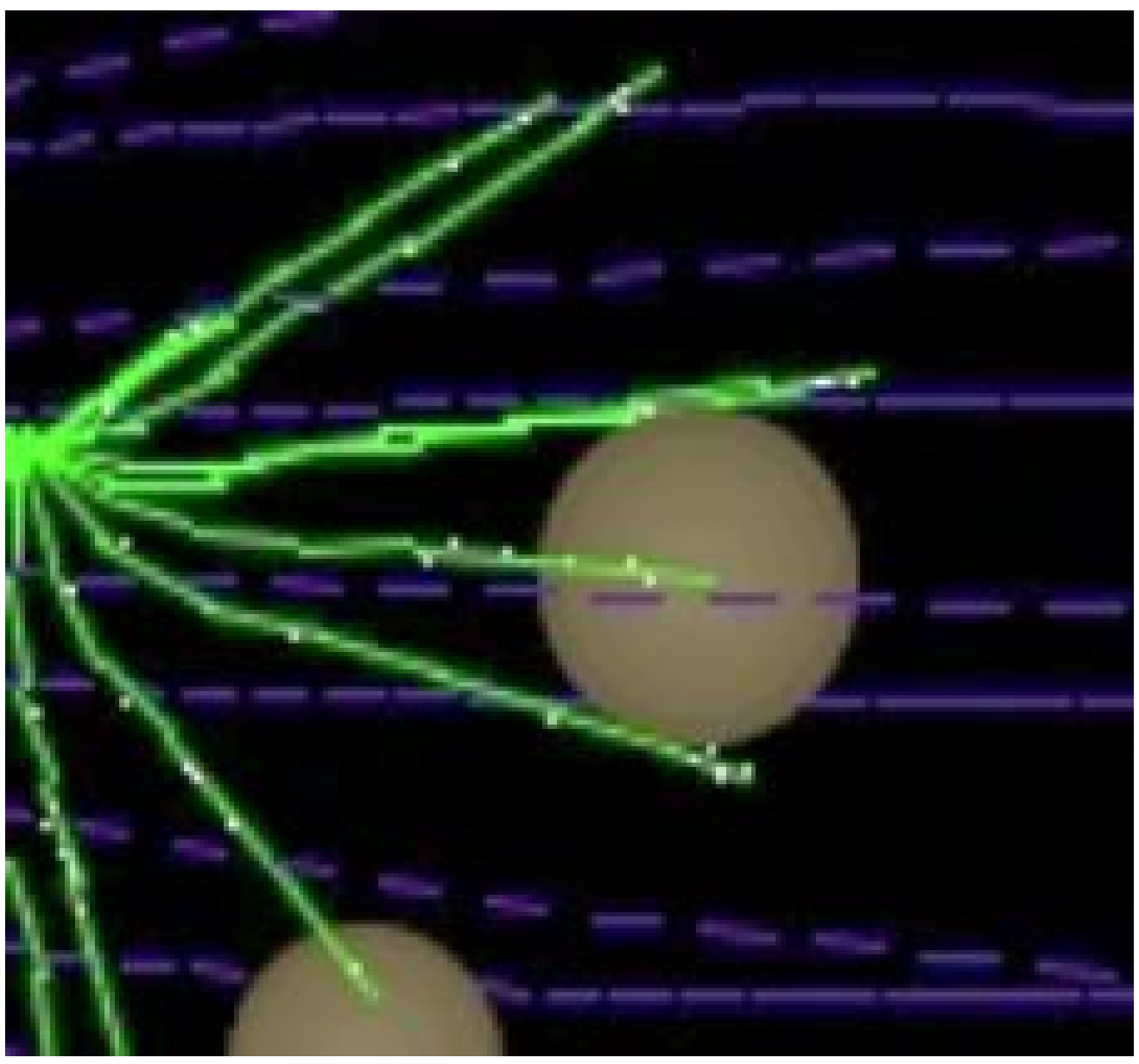

*Рисунок 19. Нити МТ с частицами груза, увеличенное изображение. МТ — зеленые, движущиеся частицы груза — белые точки. Большие светло-коричневые круги – вакуоли.*

Достаточно быстро, за считанные минуты, сеть МТ становится все более неоднородной: отдельные МТ сбиваются в «тяжи», которые становятся все более плотными и массивными (возможно, за счет диффузионных процессов в цитоплазме модели) (Рисунок 20).

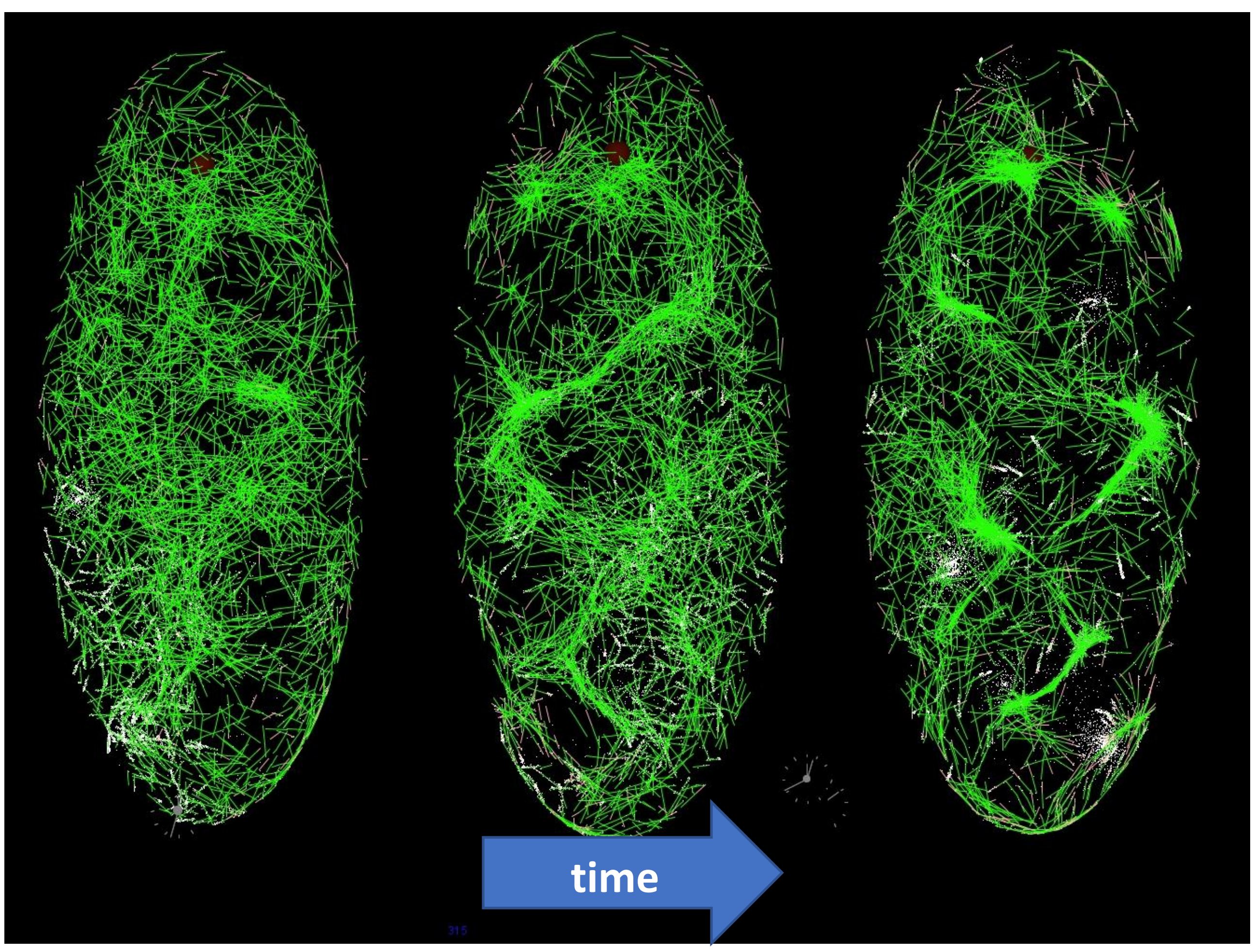

*Рисунок 20. Изменения сети МТ в модели со временем: отдельные нити МТ натыкаются на «тяжи» из нитей, которые становятся более плотными и массивными, и связываются с ними.*

За время прогона модели (примерно 9 минут внутреннего модельного времени) груз значительно перераспределяется от головы к хвосту зиготы (Рисунок 21). При этом формируется и поддерживается крутой пространственный градиент концентрации bcd от головы к хвосту.

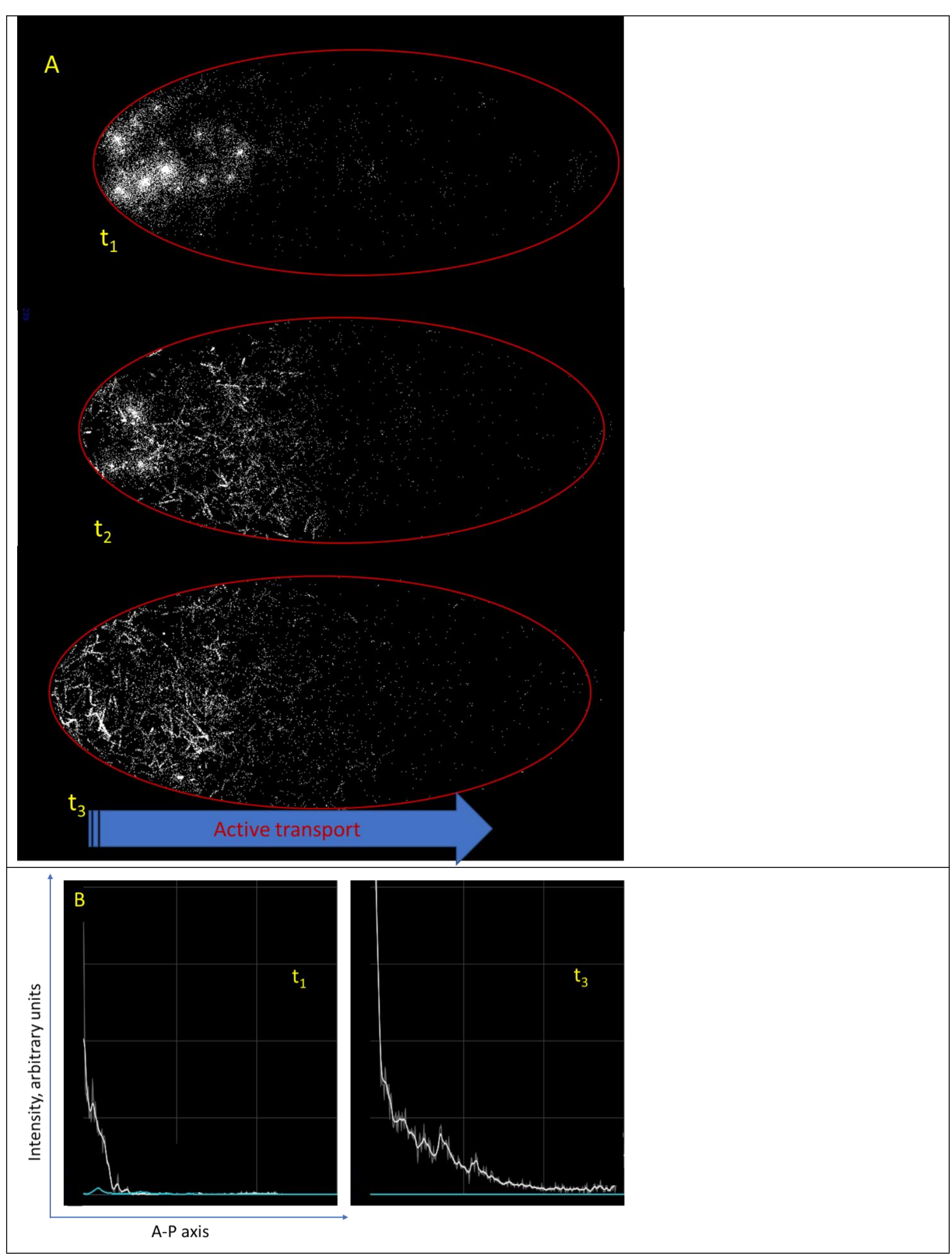

*Рисунок 21. Активный транспорт груза – мРНК бикоид в модели. (A) Результаты моделирования активного транспорта мРНК бикоида молекулярными моторами по неориентированной сети коротких МТ. Динамика перераспределения груза (комплекс этой мРНК с мотором) в начале (t1), в середине (t2) и в конце (t3) моделирования. Формируется крутой пространственный градиент концентрации от головы к хвосту (головной конец – направлен вниз). Визуализированы только молекулярные грузы (белые точки -- мРНК bcd; контуры моделируемой клетки обрисованы красным). (B) Профили*

*концентрации груза (в произвольных единицах) вдоль оси A-P в начале (t1) и в конце (t3) моделирования (исходные данные серым и усредненный профиль белым). В начале мРНК локализована только в головной части, а в результате активного транспорта по сети МТ формирует убывающий передне-задний градиент, близкий к экспоненциальному.*

Таким образом, поведение модели этого раздела в целом напоминает реальные биологические процессы. Однако сеть МТ в модели очень быстро (в считанные минуты модельного времени) становится все более и более неоднородной: отдельные МТ натыкаются на «пряди», которые становятся все более плотными и массивными. Соответственно, груз в модели имеет тенденцию к неравномерному перераспределению в объеме зародыша с локальными неоднородностями (возможно, из-за локальной неоднородности сети МТ).

В продолжение этого подхода хорошо бы выяснить, как уменьшить такую локальную разнородность сети МТ и как добиться более плавного распределения груза. Необходимо воспроизвести плавный градиент плотности МТ от головы к хвосту, наблюдаемый в реальности. Наконец, необходимо воспроизвести процессы активного транспорта при массовом наличии капель желтка в ядре зиготы – раннем зародыше (в пакете Skeledyne имеется такая возможность).

## 4.6. Транспортные процессы в более позднем эмбрионе

Здесь мы рассмотрим возможности агентного моделирования процессов на стадии позднего синцитиального эмбриона и начала процесса целлюляризации.

Следующая стадия, которая наступает примерно через час после процессов перемещения ядер к периферии синцитиального эмбриона (как в предыдущем подразделе), включает быстрый активный транспорт мРНК бикоид базо-апикально.

На стадии целлюляризации центриоли (вблизи ядер) организуют пучки МТ, ориентированные в базо-апикальном направлении. По этим пучкам, в частности, производится активный транспорт мРНК бикоида в базо-апикальном направлении.

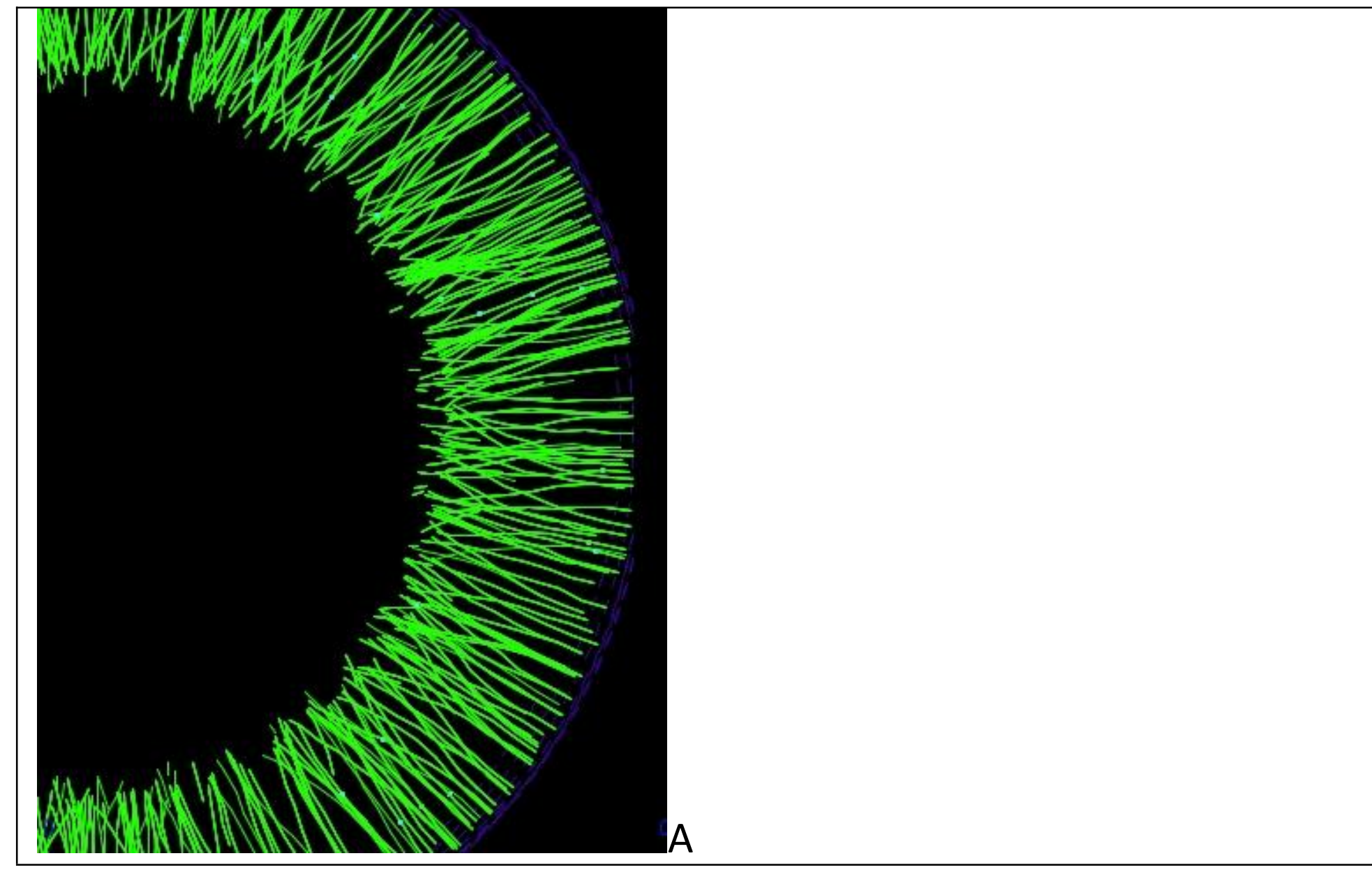
A

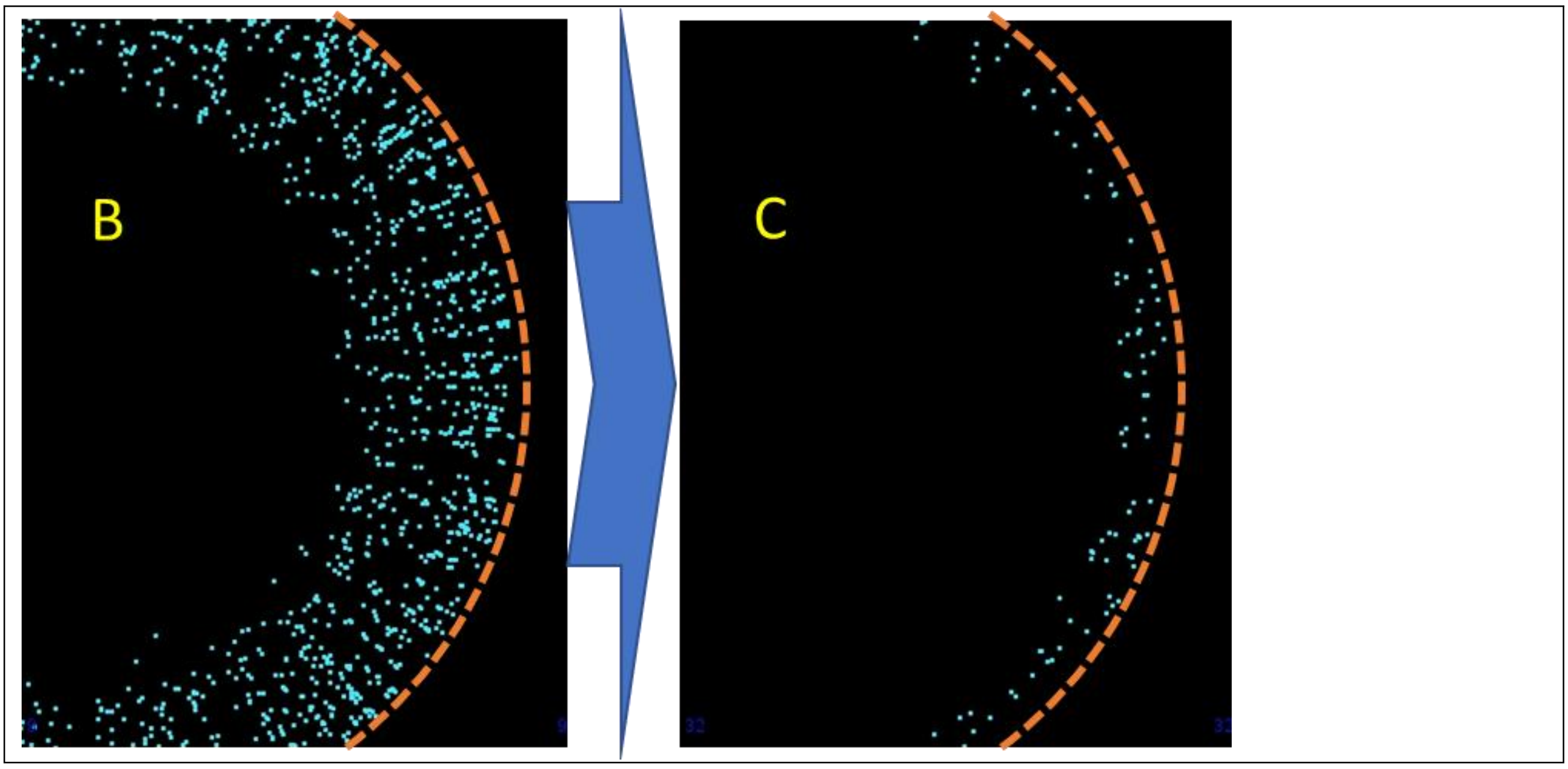

*Рисунок 22. Активный транспорт мРНК бикоида апикально по нитям МТ, ориентированным в базо-апикальном направлении в начале процесса целлюляризации (14й цикл). (А) общая организация ориентированной сети кортикальных МТ агентной модели. (B-C). Груз мРНК (голубые точки) быстро транспортируется по ориентированным МТ апикально. Видны только грузы, присоединенные через адаптор и мотор к микротрубочкам, тогда как в цитоплазме они не видны.*

Моделирование этого процесса иллюстрирует Рисунок 22, когда молекулярный груз быстро транспортируется моторами по сети базо-апикально ориентированных кортикальных МТ. Этот рисунок демонстрирует возможности пакета Skeledyne для дальнейшего развития модели в этом направлении.

## 4.7. Активный транспорт по сетям МТ с фонтанными токами

Токи цитоплазмы играют существенную роль в транспорте мРНК в созревающем ооците [Baker et al. 1993; von-Dassow and Schubiger, 1994[57]; Hecht et al., 2009[58]; Ganguly et al., 2012[59]]. Соответственно, имеются основания ожидать, что токи цитоплазмы играют аналогичные роли в транспорте мРНК и белковых факторов на стадиях раннего эмбриогенеза (зигота и синцитиальный эмбрион). Вместе с тем, детали и молекулярная машинерия этих токов все еще недостаточно изучены. Имея ввиду 3D организацию этих фонтанов и их очевидную динамическую природу, их 3D моделирование могло бы существенно продвинуть нас в исследовании этих процессов. В этом подразделе мы иллюстрируем перспективность развития 3D агентных моделей и в этом направлении.

Начнем мы изложение с некоторых примеров деталей токов цитоплазмы синцитиального эмбриона, послуживших для нас вдохновляющими наблюдениями. На рисунке 23 приведен типичный пример результатов обработки временной серии сагиттальных конфокальных сканов головной части эмбриона gfp-bcd с окончания 13 цикл и по начало 14 цикла, включая митоз. На рисунке видно, как во время митоза флюоресцентный материал (капли желтка) смещается из периплазмы головы вглубь эмбриона, освобождая периплазму, которая становится «прозрачной» на конфокальных изображениях (сравнении рис.23A и рис.C). На рис.23B показано стрелками направление фонтанных токов в ходе митоза, выявленное отслеживанием смещений частиц в цитоплазме.

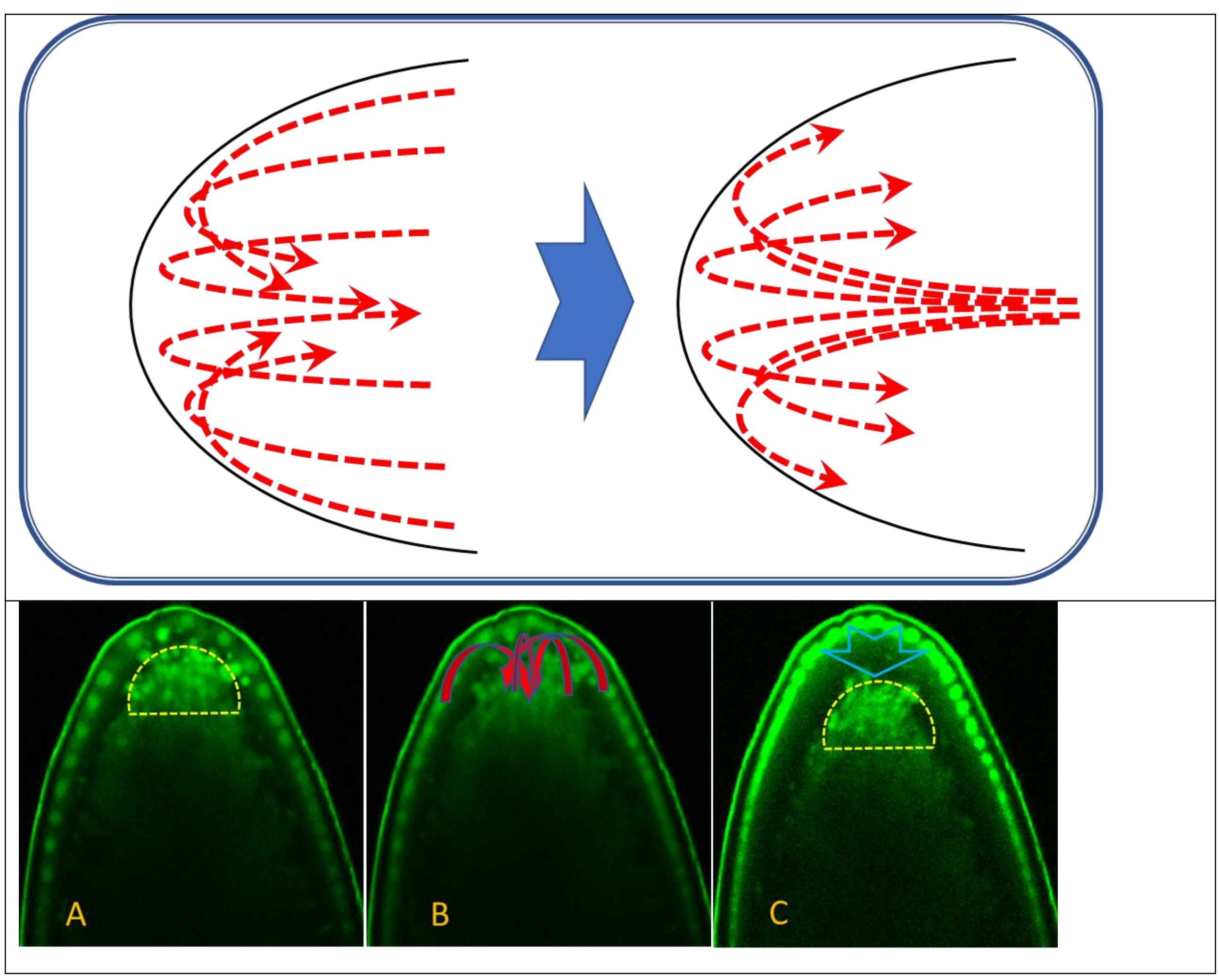

Рисунок 23. Фонтанные токи цитоплазмы в головной части синцитиального эмбриона. (А) Идея прямых и обратных фонтанных токов цитоплазмы в раннем эмбриогенезе дрозофилы. Результат процессинга и анализа временной серии конфокальных изображений с 13 по 14 цикл (включая митоз) живого развивающегося эмбриона gfp-bcd. (A) Усредненное наложением изображение эмбриона в 13 цикле (C) Усредненное наложением изображение эмбриона в начале 14 цикла. (B) «Фонтанные» токи цитоплазмы во время митоза вытесняют массив желтка к сердцевине эмбриона (показаны красными стрелками). Массив флюоресцирующего материала (по-видимому, капли желтка) в головной части эмбриона, смещаемый к вглубь токами цитоплазмы, обведен желтым пунктирным контуром на (А) и (С).

То, что наблюдаемые токи цитоплазмы, имеют более сложную природу (и вариабельны) демонстрирует рисунок 24. Это результат отслеживания траекторий частиц между окончанием 13 и началом 14 цикла для живого развивающегося эмбриона gfp-bcd.

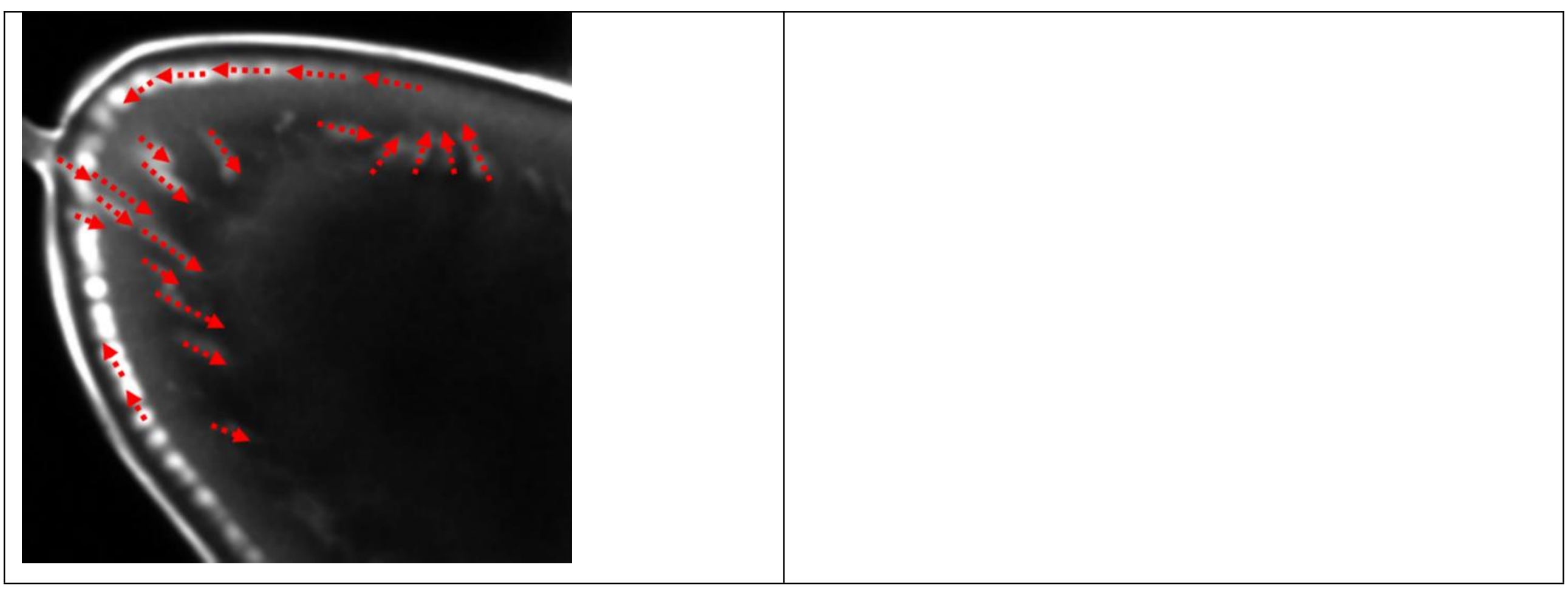

Рисунок 24. Наблюдения деталей фонтанных токов цитоплазмы в головной части эмбриона в период митоза между 13 и 14 циклами живого развивающегося эмбриона gfp-bcd. Локальные направления токов показаны красными пунктирными стрелками.

Видно, что цитоплазма масштабно смещается током от кончика головы вглубь, а затем смещается к периферии и по самым поверхностным слоям возвращается к кончику (рис 24}). То есть, мы наблюдаем в деталях описанные в свое время в классических работах фонтанные токи цитоплазмы. Видно также, что «фонтан» асимметричен.

Мы показали ([Shlemov et al., 2021]), что в это же самое время происходит перераспределение мРНК бикоида в периплазме головной части эмбриона. Сначала материал перемещается из поверхностных слоев глубже в цитоплазму головной, затем материал подвергается обратному перераспределению, обратно перемещаясь в приповерхностные слои цитоплазмы. Поэтому имеются все основания ожидать связанность этих двух процессов.

### 4.7.1. Моделирование роли фонтанных токов в ооците

Здесь мы представим простую общую модель того, как фонтанные токи ориентируют сети МТ и активный транспорт по этой сети для сценариев, напоминающих аналогичные процессы в созревающем ооците (Рисунок 25).

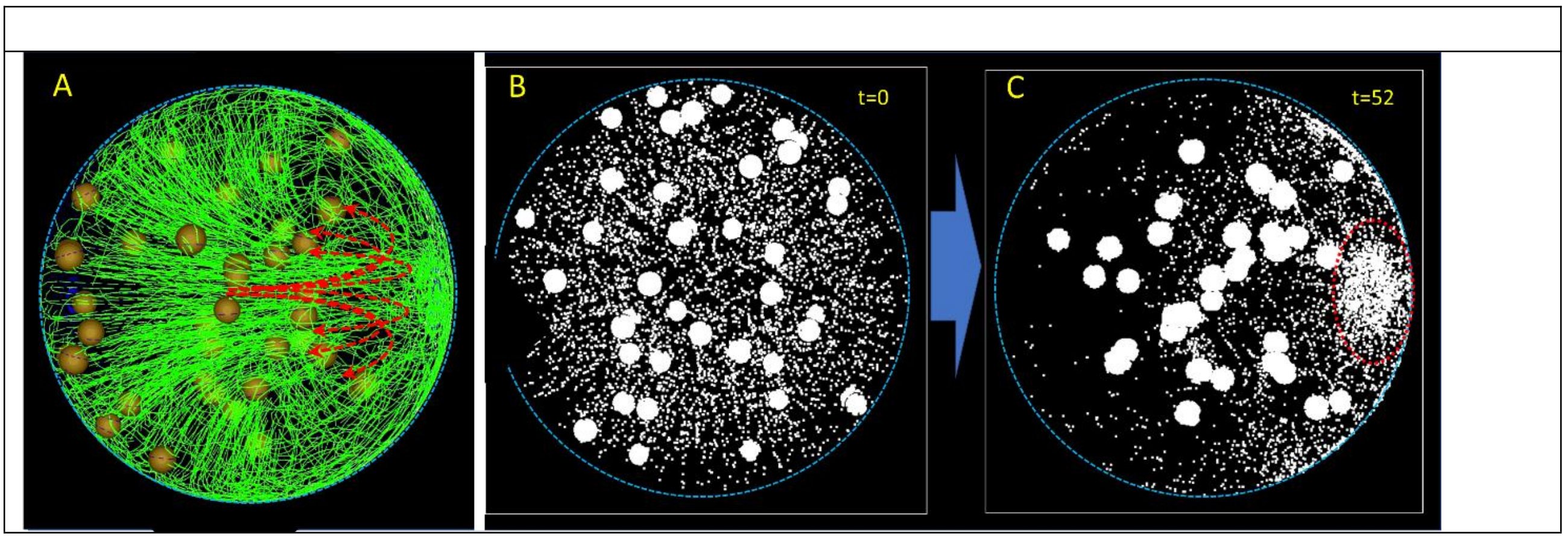

*Рисунок 25. Возможный вклад фонтанных токов цитоплазмы в процессы активного транспорта молекулярного груза. (A) 3D агентая модель клетки, кортекс которой нуклеирует МТ (изначально ориентированы плюс-концом к сердцевине; МТ - зеленым), а кинезин образует комплекс с мРНК оскар и способен транспортировать его по МТ в плюс-направлении. Красными стрелками передан фонтанный ток цитоплазмы: слева направо в толще клетки и растекается к периферии и течет обратно. Фонтанный ток быстро реорганизует систему МТ, придавая ей полярность. (B-C) Перераспределение груза (белые точки) в направлении тока цитоплазмы и удержание его компактным скоплением в кортикальном слое. (B) Изначально груз распределеы в цитоплазме клетки равномерно. (C) Со временем груз перераспределяется  к периферии, в*

*направлении фонтанного тока цитоплазмы клетки и удерживается там активным транспортом (область максимальной плотности груза обведена красным пунктирным овалом; более крупные белые окружности – силуэты вакуолей, добавляемых в модель для регистрации силы и направления тока цитоплазмы).*

Начнем мы с модели, описывающей в общих чертах процесс активного транспорта в ооците, когда существенным составляющим механизмов транспорта могут становиться токи цитоплазмы. В модели задано 5 тыс МТ одинаковой длины (длиной 90% от радиуса клетки) и 6 тыс специфических комплексов кинезина-оскар («грузов»). Кортекс в модели нуклеирует МТ и они изначально ориентированы плюс-концом к сердцевине клетки (например, как на Рис 3B). Изначально комплексы кинезин-оскар распределены в цитоплазме клетки равномерно (Рисунок 25B). Токи цитоплазмы быстро реорганизуют цитоскелет, так что нити МТ следуют направлениям токов цитоплазмы. В результате комплекс кинезин-оскар перераспределяется в клетке к периферии, в направлении фонтанного тока цитоплазмы и удерживается там активным транспортом как компактное скопление (Рисунок 25C).

Меняя параметры такой модели (в частности, начальную организацию сети МТ, радиус клетки и силу тока цитоплазмы) можно найти сценарии более сложной реорганизации системы МТ токами цитоплазмы. Это происходит, в частности, за счет того, что токи «причесывают» нити МТ по избранным направлениям, так и потому, что сильные токи способны отрывать нити МТ от кортикальной поверхности и переносить в другие части клетки, где они образуют компактные скопления.

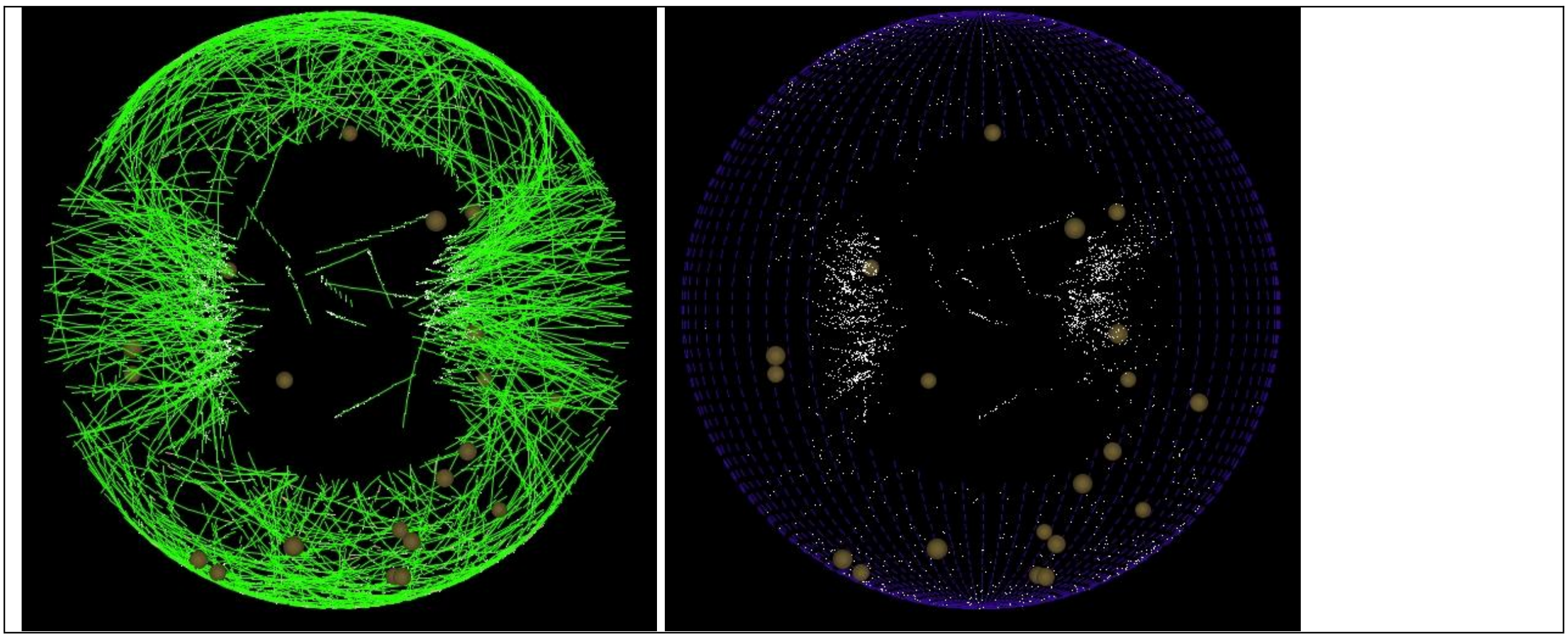

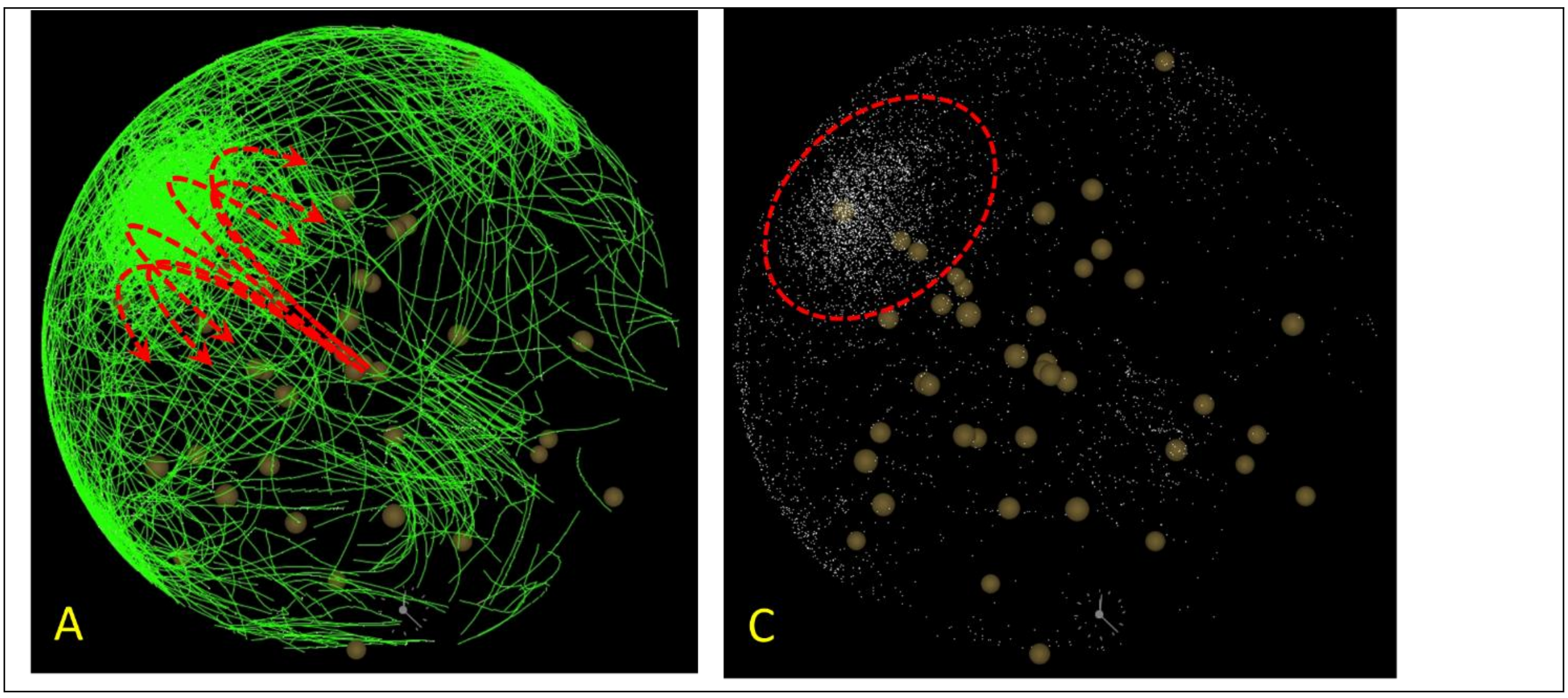

*Рисунок 26. (A) Модель клетки (ооцита) с двумя устойчивыми симметрично ориентированными пучками МТ (окруженные значительно более нерегулярной сетью МТ) и соответствующие устойчивые симметричные скопления грузов на плюс-концах пучков (B). (Никаких фонтанирующих токов!) Поперечное «оптическое» сечение через объем клетки. (C-D) Те же параметры моделирования клетки, что и выше, но с сильными фонтанными токами. Токи не устраняют симметричные пучки полностью, но производят более сложную структуру, включающую начальные пучки. Кроме того, токи создают новую особенность – скопление МТ на полюсе клетки (куда направлен фонтан). Красные изогнутые пунктирные стрелки показывают направление фонтанных токов.*

В продолжение компьютерных тестов мы выбрали версию модели с двумя симметричными пучками МТ (как на Рис 5D-E). Если на сформировавшуюся МТ сеть с такой симметрией {Рис.27A} затем накладывается фонтанный ток, то это приводит к явному усложнению организации сети МТ: фонтанный ток задает «полюс» клетки, где концентрируются нити МТ и грузы на них, изначальная пара пучков так же видоизменяется.

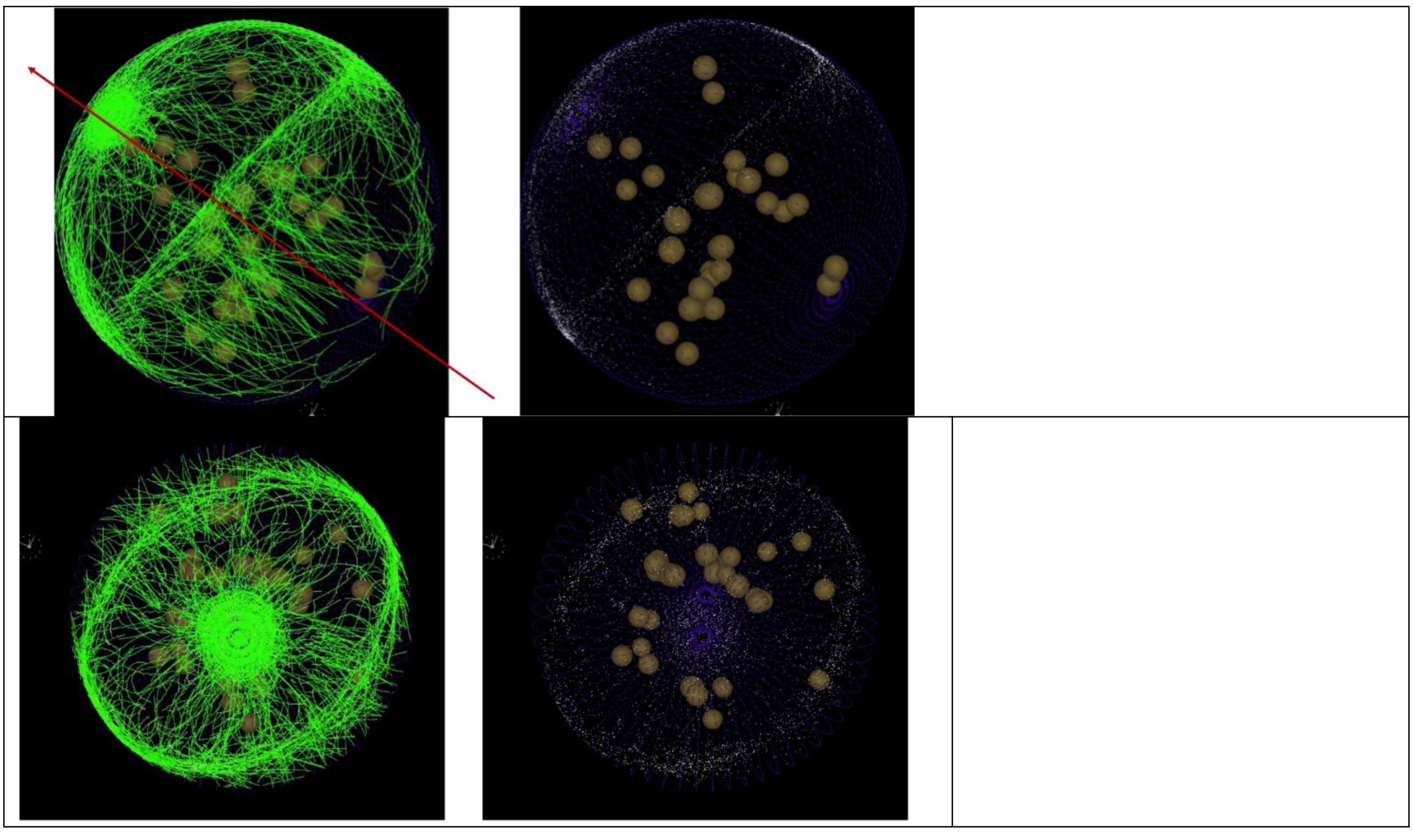

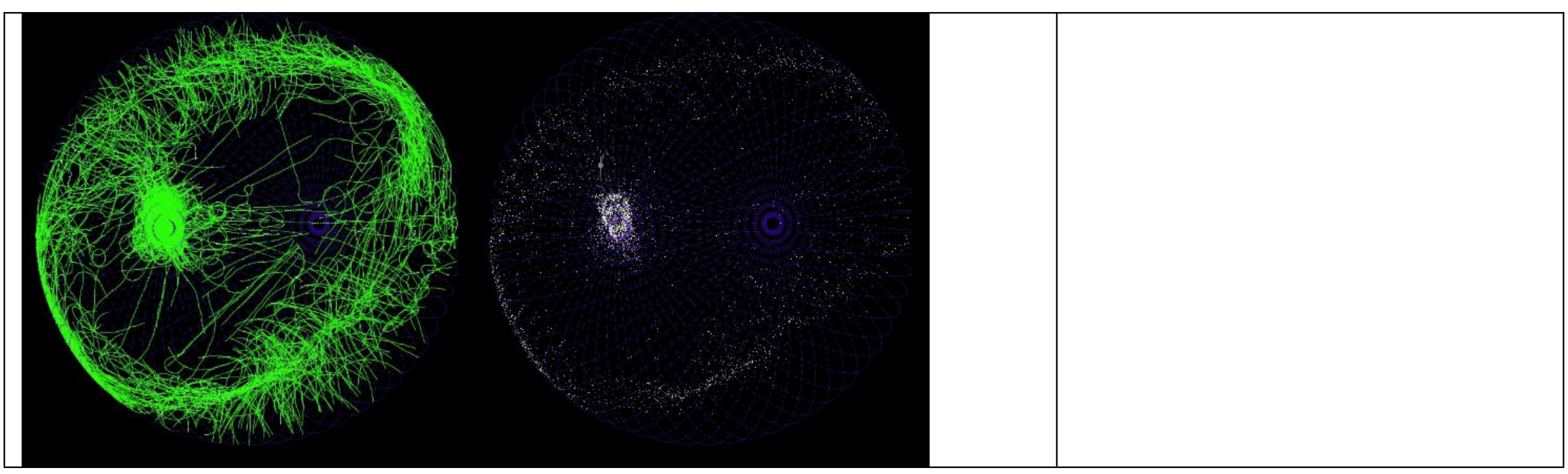

*Рисунок 27. Фонтанные токи (вдоль красной стрелки) поляризуют и реорганизуют всю кортикальную сеть МТ. При дальнейшем изменении параметров формируется еще и «пояс» из нитей МТ (примерно на «экваторе» клетки), который тоже удерживает груз локально. Так что скоплениям нитей МТ соответствуют скопления молекулярного груза соответствующей геометрии. (A-B) Результат существенной реорганизации системы МТ и поляризации распределения груза при дальнейшем изменении основных параметров модели Рис 23. (C-D, E-F) Дальнейшее изменение значений ключевых параметров модели приводит к еще более выраженной реорганизации системы МТ и поляризации распределения груза (C-D). Пояс МТ, созданный цитоплазматическими потоками, включающими остатки ориентированных пучков. Также отчетливо видно накопление МТ на полюсе. Меньшие значения параметра MT segment length и менее сильные фонтанные токи меняют организацию цитоскелета количественно (E-F).*

Изменяя такие ключевые параметры этой модели, как fountain force, MT segment length, MTs length, можно наблюдать дальнейшие реорганизации сети МТ под воздействием фонтанных токов, как на рисунке 27. Появляется пояс из МТ примерно в экваториальной области клетки (он «связывает» между собой зоны начальных пучков). Выраженность этого пояса, как и выраженность скопления на полюсе, зависит от тех ключевых параметров модели.

Как видим, такие модели демонстрируют впечатляющее разнообразие сценариев взаимодействия цитоскелета с фонтанными токами, обосновывая перспективность такого подхода для моделирования разных временнЫх стадий, как и разных аспектов развития ооцита-раннего эмбриона в связи с токами цитоплазмы.

### 4.7.2. Моделирование роли фонтанных токов в синцитиальном эмбрионе

В этом подразделе мы продемонстрируем перспективность исследования роли фонтанных токов в активном транспорте синцитиального эмбриона дрозофилы средствами агентного моделирования. Это эмбрион -- более зрелый, чем в предыдущем подразделе.

Начнем мы с некоторой дальнейшей модификации серии моделей предыдущего подраздела. Если мы, помимо варьирования указанных там параметров, еще предадим клетке форму вытянутого эллипса, то реорганизация начальной сети МТ становится еще более впечатляющей. Пример приведен на Рис 28. Фонтанные токи в условиях этой модели кардинально реорганизуют изначально генерируемую кортексом сеть МТ так, что нити МТ сбиваются в несколько компактных и небольших скоплений. Груз, соответственно, также становится высоко-локализованным.

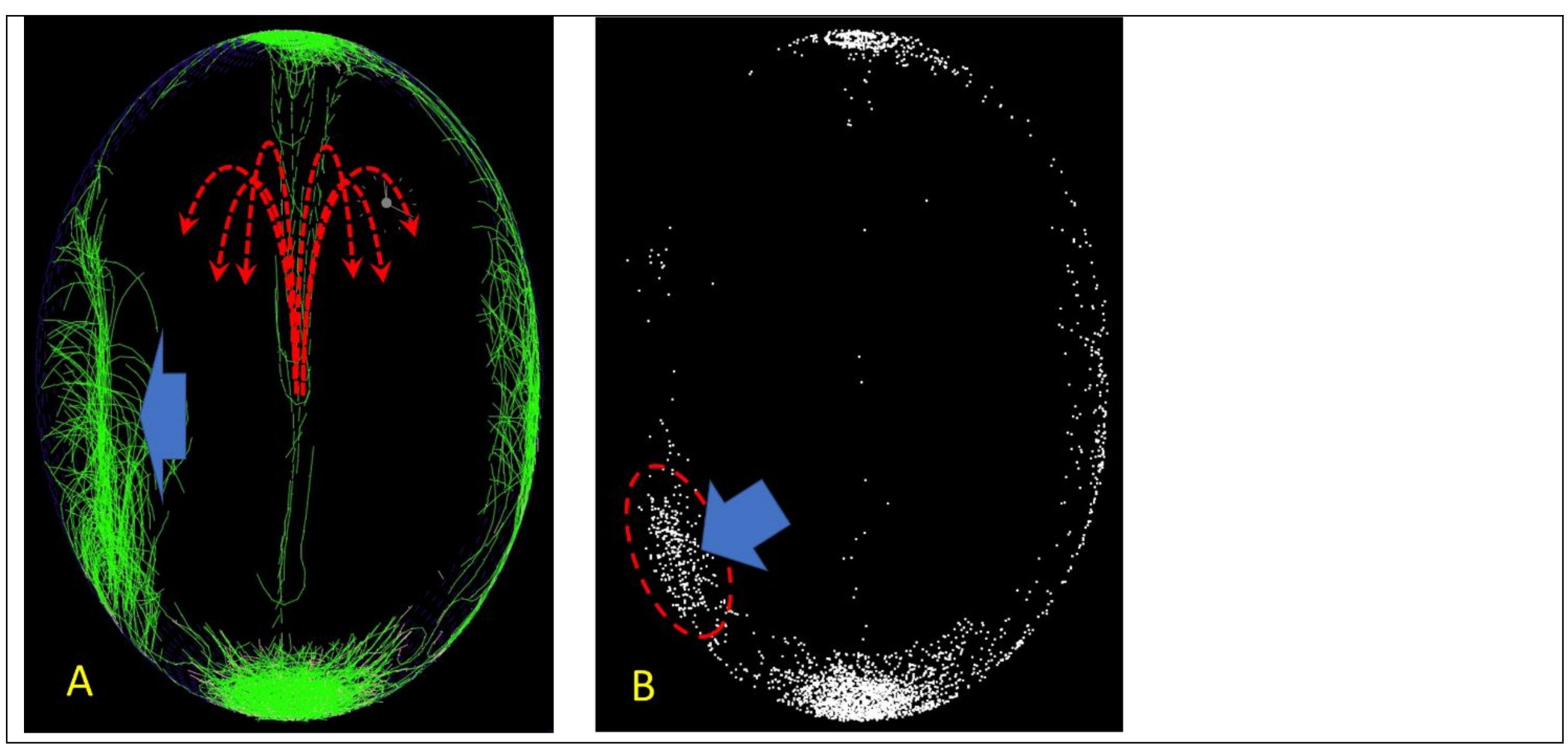

*Рисунок 28. Один из сценариев того, как сильные фонтанные токи кардинально реорганизуют генерируемые кортексом сети МТ. Токи собирают нити МТ в несколько компактных скоплений с высокими концентрациями груза в них.*

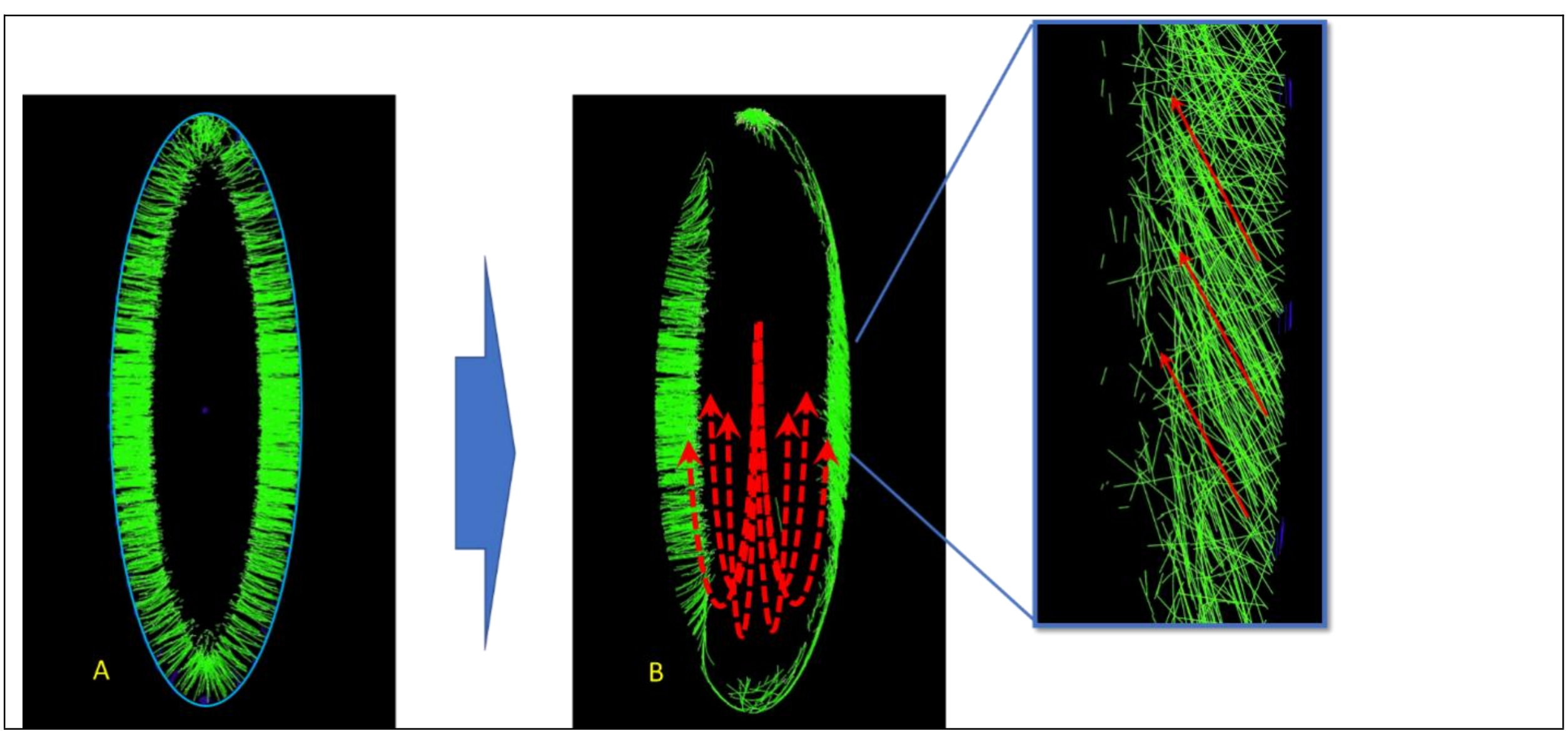

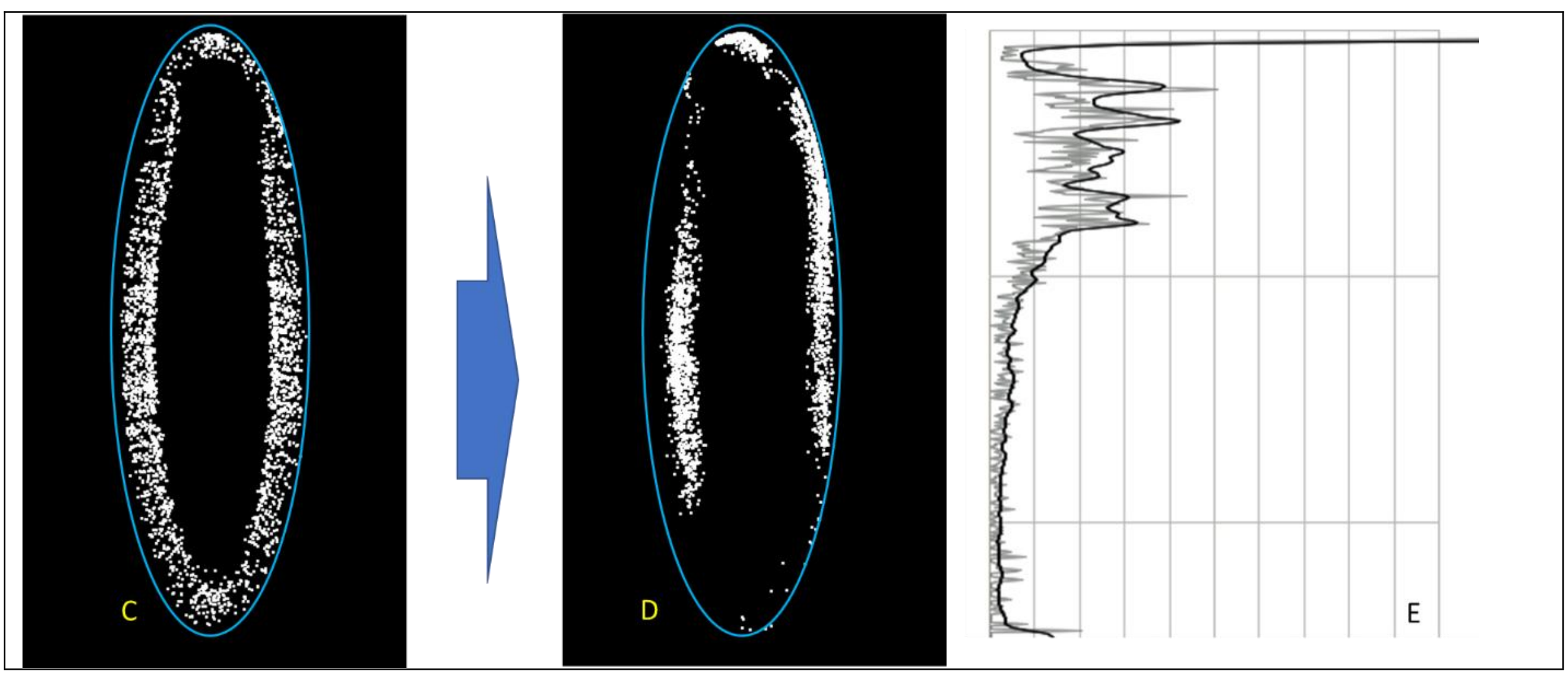

*Рисунок 29.Фонтанные токи цитоплазмы в синцитиальном эмбрионе дают свой вклад в результат активного транспорта молекулярными моторами по ориентированным сетям МТ. (А) Эллипсоидная модель клетки с фонтанным током цитоплазмы вдоль главной оси раннего эмбриона и кортикально нуклеированными МТ (зеленым), ориентированными минус-концом внутрь. МТ транспортируют груз от кортекса к центру клетки. (В) Фонтанные токи цитоплазмы (красные изогнутые пунктирные стрелки) меняют направление свободных концов нитей МТ и частично вырывают их из кортекса и переносят в направлении токов. Врезка передает детали организации сети МТ при большем увеличении: видно, что МТ преимущественно ориентированы в направлении, указанно красными стрелками. (C-D) Картины распределения груза (белые точки) для имиджей (А) и (В)соответственно. (Е) График плотности распределения грузов вдоль главной оси синцитиального эмбриона, соответствующий имиджу (D).*

Предлагается модель клетки (вытянутый эллипсоид по форме) с кортикально нуклеированными и ориентированными «минус» концом внутрь нитями МТ (рис 29А). Запуск модели активирует фонтанный ток цитоплазмы вдоль главной оси раннего эмбриона (рис 29В). Одновременно МТ начинают транспортировать груз от кортекса к центру клетки. При этом, фонтанный ток ориентирует МТ вдоль направления тока и тем самым обуславливает свой вклад в транспорт груза: груз транспортируется не только центростремительно, но и вдоль АР оси (рис 29В и врезка). График показывает, что груз преимущественно локализуется асимметрично и формирует градиент плотности вдоль главной оси эмбриона (рис 29D-E). В контроле же, без фонтанного тока не происходит нарушения симметрии в перераспределении молекулярного груза и модель ведет себя сходно с моделью Рис 22.

Таким образом, представленные в этим разделе результаты прогонов демонстрируют перспективность развития моделей с фонтанными токами цитоплазмы для моделирования развития как ооцита, так и синцитиального эмбриона.

## 5. Заключение

Целая серия разнообразных 3D агентных моделей в этой статье демонстрирует перспективы системно-биологических исследований процессов в ооците – синцитиальном эмбрионе. Это должно быть особенно эффективно, когда имеется возможность использовать количественные экспериментальные данные для разработки и подгонки моделей.

Возможности свести и учесть взаимодействия в одной модели процессы реорганизации цитоскелета, транспорт моторами по МТ и МФ, фонтанные токи цитоплазмы и другие ключевые процессы оогенеза и раннего эмбриогенеза безусловно представляются весьма перспективными.

Вместе с тем, именно такой модельный объект как ооцит—ранний эмбрион дрозофилы предоставляет большой объем не только качественных, но и количественных данных, прекрасно подходящих как для разработки исходных моделей, так и верификации прогнозов этих моделей. Так то это – перспективный объект для моделирования в рамках data driven approach.

## Финансовая поддержка

Работа выполнена при финансовой поддержке Российского фонда фундаментальных исследований (проект № 20-04-01015). This work was supported by the Russian Foundation for Basic Research (project no. 20-04-01015 and Federal Agency for Scientific Organizations of the Russian Federation (project no. 01201351572).

## Список литературы

[1] Medioni C, Mowry K, Besser F. (2012). Principles and roles of mRNA localization in animal development. Development 139(18):3263-76.

[2] Kloc M., Etkin L.D. 2005. RNA localization mechanisms in oocytes. J. Cell Sci. 118:269–282.

[3] Vazquez-Pianzola Paula and Suter Beat, (2012), Conservation of the RNA Transport Machineries and Their Coupling to Translation Control across Eukaryotes, Comparative and Functional Genomics, vol. 2012, Article ID 287852.

[4] St Johnston D, Ahringer J. Cell polarity in eggs and epithelia: parallels and diversity. Cell. 2010 May 28;141(5):757-74.

[5] Medioni C, Mowry K, Besser F. (2012). Principles and roles of mRNA localization in animal development. Development 139(18):3263-76.

[6] Li R, Albertini DF. (2013). The road to maturation: somatic cell interaction and self-organization of the mammalian oocyte. Nat Rev Mol Cell Biol. 14(3):141-52.

[7] Lasko, P. (2012). mRNA localization and translational control in Drosophila oogenesis. Cold Spring Harbor Perspectives in Biology, 4 (10).

[8] Little, S. C., Sinsimer, K. S., Lee, J. J., Wieschaus, E. F., & Gavis, E. R. (2015). Independent and coordinate trafficking of single Drosophila germ plasm mRNAs. Nat Cell Biol. 17(5):558-68.

[9] Castro-González, C., Ledesma-Carbayo, M.J., Peyriéras, N., Santos, A. "Assembling Models of Embryo Development: Image Analysis and the Construction of Digital Atlases". Birth Defects Res. Part C-Embryo Today-Rev., 96(2):109-120. 2012.

[10] Mogilner A., Odde D., (2011). Modeling cellular processes in 3D. Trends Cell Biol., 21, 692-700.

[11] Mogilner A. and Manhart A. (2016). Agent-based modeling: case study in cleavage furrow models. Mol Biol Cell, 27: 3379-84.

[12] Odell G.M. and Foe V.E. (2008). An agent-based model contrasts opposite effects of dynamic and stable microtubules on cleavage furrow positioning. J Cell Biol, 183(3): 471-483.

[13] Wordeman L., Decarreau J., Vicente J.J., Wagenbach M. (2016). Divergent microtubule assembly rates after short- versus long-term loss of end-modulating kinesins. Mol Biol Cell 27(8):1300-9.

[14] Clark, I., Giniger, E., Ruohola-Baker, H., Jan, L.Y., and Jan, Y.N. (1994). Transient posterior localization of a kinesin fusion protein reflects anteroposterior polarity of the Drosophila oocyte. Curr. Biol. 4, 289–300.

[15] Clark I.E., Jan L.Y., Jan Y.N. (1997) Reciprocal localization of Nod and kinesin fusion proteins indicates microtubule polarity in the Drosophila oocyte, epithelium, neuron and muscle, Development 124: 461-470.

[16] Theurkauf, W.E., Hazelrigg, T.I. (1998). In vivo analyses of cytoplasmic transport and cytoskeletal organization during Drosophila oogenesis: characterization of a multi-step anterior localization pathway. Development 125(18): 3655--3666.

[17] Cha, B. J., Koppetsch, B. S. and Theurkauf, W. E. (2001). In vivo analysis of Drosophila bicoid mRNA localization reveals a novel microtubule-dependent axis specification pathway. Cell 106, 35-46.

[18] Cha, B.J., Serbus, L.R., Koppetsch, B.S.,Theurkauf,W.E., 2002. Kinesin I-dependent cortical exclusion restricts pole plasm to the oocyte posterior. Nat. Cell Biol. 4(8):592-8.

[19] Januschke J, Gervais L, Gillet L, Keryer G, Bornens M, Guichet A. The centrosome-nucleus complex and microtubule organization in the Drosophila oocyte. Development. 2006 Jan;133(1):129-39. doi: 10.1242/dev.02179. Epub 2005 Nov 30. PMID: 16319114.

[20] Zimyanin VL, Belaya K, Pecreaux J, Gilchrist MJ, Clark A, Davis I, St Johnston D. 2008. In vivo imaging of oskar mRNA transport reveals the mechanism of posterior localization. Cell 134:843–853.

[21] Lasko P. Patterning the Drosophila embryo: A paradigm for RNA-based developmental genetic regulation. Wiley Interdiscip Rev RNA. 2020 Nov;11(6):e1610. doi: 10.1002/wrna.1610. Epub 2020 Jun 15. PMID: 32543002; PMCID: PMC7583483.

[22] Clark, I., Giniger, E., Ruohola-Baker, H., Jan, L.Y., and Jan, Y.N. (1994). Transient posterior localization of a kinesin fusion protein reflects anteroposterior polarity of the Drosophila oocyte. Curr. Biol. 4, 289–300.
[23] Clark I.E., Jan L.Y., Jan Y.N. (1997) Reciprocal localization of Nod and kinesin fusion proteins indicates microtubule polarity in the Drosophila oocyte, epithelium, neuron and muscle, Development 124: 461-470.
[24] Micklem D.R., Dasgupta,R., Elliott,H., Gergely,F., Davidson,C., Brand,A., González-Reyes,A. and St Johnston,D. (1997) The mago nashi gene is required for the polarisation of the oocyte and the formation of perpendicular axes in Drosophila. Curr. Biol., 7, 468–478.
[25] Serbus LR, Cha BJ, Theurkauf WE, Saxton WM. Dynein and the actin cytoskeleton control kinesin-driven cytoplasmic streaming in Drosophila oocytes. Development. 2005 Aug;132(16):3743-52. doi: 10.1242/dev.01956. PMID: 16077093; PMCID: PMC1534125.
[26] Parton RM, Hamilton RS, Ball G, Yang L, Cullen CF, Lu W, Ohkura H, Davis I. A PAR-1-dependent orientation gradient of dynamic microtubules directs posterior cargo transport in the Drosophila oocyte. J Cell Biol. 2011 Jul 11;194(1):121-35. doi: 10.1083/jcb.201103160. PMID: 21746854; PMCID: PMC3135408.
[27] Alexandrov T, Golyandina N, Holloway D, Shlemov A, Spirov A. (2018). Two-Exponential Models of Gene Expression Patterns for Noisy Experimental Data. J Comput Biol. 25(11):1220-1230.
[28] Little SC, Tkacik G, Kneeland TB, Wieschaus EF, Gregor T (2011). The Formation of the Bicoid Morphogen Gradient Requires Protein Movement from Anteriorly Localized mRNA. PLoS Biol 9(3): e1000596.
[29] Spirov, A., Fahmy, K., Schneider, M., Noll, M., and Baumgartner, S. (2009). Formation of the bicoid morphogen gradient: an mRNA gradient dictates the protein gradient. Development 136, 605-614.
[30] Weil T. T., Xanthakis D, Parton R., Dobbie I., Rabouille C., Gavis E. R., Davis I. (2010). Distinguishing direct from indirect roles for bicoid mRNA localization factors. Development; 137: 169-76.
[31] Fahmy, K., Akber, M., Cai, X., Koul, A., Hayder, A., Baumgartner, S. (2014). αTubulin 67C and Ncd Are Essential for Establishing a Cortical Microtubular Network and Formation of the Bicoid mRNA Gradient in Drosophila. PLoS ONE 9(11): e112053.
[32] Ali-Murthy Z, Kornberg TB. (2016). Bicoid gradient formation and function in the Drosophila pre-syncytial blastoderm. eLife 5 :e13222.
[33] Baker, J., Theurkauf, W.E., Schubiger, G. (1993). Dynamic changes in microtubule configuration correlate with nuclear migration in the preblastoderm Drosophila embryo. J. Cell Biol. 122(1): 113--121.
[34] Farley BM, Ryder SP (2008). Regulation of maternal mRNAs in early development. Crit. Rev Biochem Mol Biol 43:135–162.
[35] Yisraeli, J.K., Sokol, S., and Melton, D.A. 1990. A two-step model for the localization of maternal mRNA in Xenopus oocytes: involvement of microtubules and microfilaments in the translocation and anchoring of Vg1 mRNA. Development 108:289–298.
[36] Cai, X., Spirov, A., Akber, M. and Baumgartner, S. (2017). Cortical movement of Bicoid in early Drosophila embryos is actin- and microtubule-dependent and disagrees with the SDD diffusion model. PLoS One. 12(10):e0185443.
[37] Lucchetta EM, Vincent ME, Ismagilov RF (2008) A Precise Bicoid Gradient Is Nonessential during Cycles 11–13 for Precise Patterning in the Drosophila Blastoderm. PLoS ONE 3(11): e3651.
[38] Cai X, Fahmy K, Baumgartner S. bicoid RNA localization requires the trans-Golgi network. Hereditas. 2019 Sep 10;156:30. doi: 10.1186/s41065-019-0106-8. PMID: 31528161; PMCID: PMC6737670.
[39] Baumgartner S. Revisiting bicoid function: complete inactivation reveals an additional fundamental role in Drosophila egg geometry specification. Hereditas. 2024 Jan 2;161(1):1. doi: 10.1186/s41065-023-00305-9. PMID: 38167241; PMCID: PMC10759373.
[40] Golyandina NE, Holloway DM, Lopes FJ, Spirov AV, Spirova EN, Usevich KD. Measuring gene expression noise in early Drosophila embryos: nucleus-to-nucleus variability. Procedia Comput Sci. 2012 Jan 1;9:373-382. doi: 10.1016/j.procs.2012.04.040. Epub 2012 Jun 2. PMID: 22723811; PMCID: PMC3378315.
[41] http://www.celldynamics.org/celldynamics/downloads/simcode.html
[42] Marat Sabirov, Alexander Spirov, Agent-based 3D modeling of active transport in the oocyte-zygote: problems of scaling, in preparation.
[43] Cohen RS. 2002; Oocyte patterning: dynein and kinesin, inc., Curr Biol. 12(23):R797-9.
[44] St Johnston D. (2005) Moving messages: the intracellular localization of mRNAs. Nature Reviews Molecular Cell Biology 6, 363–375.
[45] Reck-Peterson, S. L., Yildiz, A., Carter, A. P., Gennerich, A., Zhang, N., & Vale, R. D. (2006). Single-molecule analysis of dynein processivity and stepping behavior. Cell, 126(2), 335–348. https://doi.org/10.1016/j.cell.2006.05.046
[46] Goode, B. L., Drubin, D. G. and Barnes, G. (2000). Functional cooperation between the microtubule and actin cytoskeleton. Curr. Opin. Cell Biol. 12, 63-71.
[47] Waterman-Storer, C., Duey, D. Y., Weber, K. L., Keech, J., Cheney, R. E., Salmon, E. D. and Bement, W. M. (2000). Microtubules remodel actomyosin networks in Xenopus egg extracts via two mechanisms of F-actin transport. J. Cell Biol. 150, 361-376.
[48] Foe, V.E., Field, C.M., Odell, G.M. (2000). Microtubules and mitotic cycle phase modulate spatiotemporal distributions of F-actin and myosin II in Drosophila syncytial blastoderm embryos. Development 127(9): 1767-1787.
[49] Wu, Jun & Dickinson, Richard & Lele, Tanmay. (2012). Investigation of in vivo microtubule and stress fiber mechanics with laser ablation. Integrative biology : quantitative biosciences from nano to macro. 4. 471-9. 10.1039/c2ib20015e.
[50] He Y, Yu W, Baas PW. 2002; Microtubule reconfiguration during axonal retraction induced by nitric oxide. J Neurosci. 22(14):5982-5991. doi:10.1523/JNEUROSCI.22-14-05982.2002
[51] Bicek AD, Tüzel E, Kroll DM, Odde DJ. 2007; Analysis of microtubule curvature. Methods Cell Biol. 83:237-268. doi:10.1016/S0091-679X(07)83010-X

[52] Cosentino Lagomarsino, M., Tanase, C., Vos, J. W., Emons, A. M., Mulder, B. M., & Dogterom, M. (2007). Microtubule organization in three-dimensional confined geometries: evaluating the role of elasticity through a combined in vitro and modeling approach. Biophysical journal, 92(3), 1046–1057. https://doi.org/10.1529/biophysj.105.076893

[53] Diagouraga, B., Grichine, A., Fertin, A., Wang, J., Khochbin, S., & Sadoul, K. (2014). Motor-driven marginal band coiling promotes cell shape change during platelet activation. The Journal of cell biology, 204(2), 177–185. https://doi.org/10.1083/jcb.201306085

[54] Yin S, Tien M, Yang H. 2020; Prior-Apprised Unsupervised Learning of Subpixel Curvilinear Features in Low Signal/Noise Images. Biophys J. 118(10):2458-2469. doi:10.1016/j.bpj.2020.04.009

[55] Sullivan W, Fogarty P, Theurkauf W. Mutations affecting the cytoskeletal organization of syncytial Drosophila embryos. Development. 1993 Aug;118(4):1245-54. doi: 10.1242/dev.118.4.1245. PMID: 8269851.

[56] Archambault V, Pinson X. Free centrosomes: where do they all come from? Fly (Austin). 2010 Apr-Jun;4(2):172-7. doi: 10.4161/fly.4.2.11674. Epub 2010 Apr 2. PMID: 20404481.

[57] von Dassow G, Schubiger G. (1994). How an actin network might cause fountain streaming and nuclear migration in the syncytial Drosophila embryo. J Cell Biol. 127:1637-53.

[58] Hecht I, Rappel W. J, Levine H (2009). Determining the scale of the Bicoid morphogen gradient. Proc Natl Acad Sci U S A 106: 1710–1715.

[59] Ganguly S, Williams LS, Palacios IM, Goldstein RE. (2012). Cytoplasmic streaming in Drosophila oocytes varies with kinesin activity and correlates with the microtubule cytoskeleton architecture. Proc Natl Acad Sci U S A. 109(38):15109-14.